\documentclass[prd,twocolumn,nofootinbib,superscriptaddress]{revtex4-2}
\usepackage{lipsum}
\usepackage{bm}
\usepackage{amsmath}
\usepackage{amssymb}
\usepackage{graphicx}
\usepackage{subfigure}
\usepackage{hyperref}
\usepackage{color}
\usepackage{comment}
\usepackage{ulem}
\usepackage{CJK}
\usepackage{longtable}
\hypersetup{
	colorlinks=true,
	linkcolor=red,
	citecolor=blue,
}

\allowdisplaybreaks[2]

\usepackage{graphicx}
\usepackage{dcolumn}
\usepackage{bm}
\begin{document}

\preprint{APS/123-QED}

\title{A complete waveform comparison of post-Newtonian and numerical relativity in eccentric orbits}

\author{Hao Wang}
\email{husthaowang@hust.edu.cn}
\affiliation{Department of Astronomy, School of Physics, Huazhong University of Science and Technology, Wuhan 430074, China}

\author{Yuan-Chuan Zou}
\email{zouyc@hust.edu.cn}
\affiliation{Department of Astronomy, School of Physics, Huazhong University of Science and Technology, Wuhan 430074, China}

\author{Qing-Wen Wu}
\email{qwwu@hust.edu.cn}
\affiliation{Department of Astronomy, School of Physics, Huazhong University of Science and Technology, Wuhan 430074, China}

\author{Xiaolin Liu}
\email{shallyn.liu@foxmail.com}
\affiliation{Instituto de Física Téorica UAM-CSIC, Universidad Autónoma de Madrid, Cantoblanco 28049 Madrid, Spain}

\author{Zhao Li}
\email{lz111301@mail.ustc.edu.cn}
\affiliation{ Department of Astronomy, University of Science and Technology of China, Hefei, Anhui 230026, China}
\affiliation{ School of Astronomy and Space Science, University of Science and Technology of China, Hefei 230026, China}
\affiliation{ Department of Physics, Kyoto University, Kyoto 606-8502, Japan}

\date{\today}

\begin{abstract}
This study presents a thorough comparative analysis between post-Newtonian (PN) and numerically relativistic (NR) waveforms in eccentric orbits, covering nonspinning and spin-aligned configurations. The comparison examines frequency, amplitude, and phase characteristics of various harmonic modes, including $(\ell,m)=(2,2),(2,1),(3,3),(3,2),(4,4),(5,5)$ modes. The study utilizes eccentric PN waveforms based on 3PN quasi-Keplerian parameterization with 3PN radiative reaction, surpassing Newtonian quadrupole moment with higher-order moments.
NR waveforms from RIT and SXS catalogs span mass ratios from 1/4 to 1, eccentricities up to 0.45, and durations exceeding $17000M$ across nonspinning and spin-aligned configurations. Focusing on the (2,2) mode, frequency comparisons between quadrupole and higher-order moments of $\Psi_4^{22}$ and $h^{22}$ were conducted. Amplitude comparisons revealed superior accuracy in quadrupole moments of $\Psi_4^{22}$.
Analysis of total 180 sets of eccentric waveforms showed increasing fitting residuals with rising eccentricity, correlating with smaller mass ratios. Comparisons of initial eccentricity from PN fitting, 3PN quasi-Keplerian parameterization, and RIT/SXS catalogs revealed alignment discrepancies. Frequency, phase, and amplitude comparisons of (2,2) modes show consistent inspiral behavior between PN and NR, with divergences near merger for nonspinning PN and pre-200$M$ for spin-aligned PN.
Average errors of frequency, phase, and amplitude up to $200M$ pre-merger amplify with increasing eccentricity. Average errors for eccentricities of 0-0.2 are below 3\% for frequency, 0.2 for phase, and 6\% for amplitude. For eccentricities of 0.2-0.4, errors increase. The higher-order modes demonstrate consistent trends for frequency and phase, and with increased amplitude errors, underscoring the self-consistency of the PN fitting process.
Fittings on three RIT eccentric waveforms with low mass ratios highlight deviation between PN and NR for such scenarios. Refinements in PN and NR accuracy, especially at higher orders, small mass ratios are essential for precise gravitational wave templates in eccentric orbits, reducing systematic errors in parameter estimation and advancing gravitational wave detection.
\end{abstract}
\maketitle

\section{Introduction}
Gravitational wave astronomy has heralded a new epoch following the inaugural binary black hole (BBH) merger event GW150914 in 2015 \cite{LIGOScientific:2016aoc}. To date, the terrestrial gravitational wave observatories LIGO \cite{LIGOScientific:2014qfs}, Virgo \cite{VIRGO:2014yos}, and KAGRA \cite{KAGRA:2018plz} (LVK) have collectively detected 90 merger events involving binary compact objects during the O1, O2, O3 and O4 observation runs \cite{LIGOcollaboration,LIGOScientific:2024elc}. These events encompass binary black hole mergers, binary neutron star mergers, and black hole-neutron star mergers.

Currently, diverse methodologies are employed to study BBH dynamics, including post-Newtonian (PN) approaches \cite{Blanchet:2013haa}, effective one body (EOB) models \cite{Buonanno:1998gg, Damour:2001tu}, black hole perturbation theory (BHPT) \cite{Teukolsky:1973ha}, and numerical relativity (NR) simulations \cite{Pretorius:2005gq, Campanelli:2005dd, Baker:2005vv}, among others. While the former methods are approximations, NR is recognized for its comprehensive and precise numerical treatment. This study focuses on comparing PN and NR methodologies.
In the PN domain, a tailored approach for low-speed and weak-field scenarios ($v^2 / c^2 \sim G M / c^2 r \ll 1$) characterizes BBH dynamics. It refines Newtonian motion by incorporating high-order corrections through multipole expansions into dynamic quantities.
In contrast, NR addresses BBH dynamics as an initial value problem, necessitating the specification of initial conditions and evolution equations \cite{Baumgarte:1998te}. NR provides a comprehensive solution that includes strong field effects. However, its computational requirements are substantial, often demanding several months on a supercomputer for a single set of simulation.
To establish the consistency between PN and NR results, numerous studies have scrutinized PN and NR waveforms, primarily focusing on circular orbits in nonspinning \cite{Gopakumar:2007vh, Borhanian:2019kxt, Boyle:2007ft, Boyle:2009dg, Pan:2007nw}, spin-aligned \cite{Hannam:2007wf, Hannam:2010ec}, and spin-precession \cite{Campanelli:2008nk} configurations. Notably, research on eccentric orbits is limited, with only a few references comparing the (2,2) mode and higher-order modes of nonspinning waveforms \cite{Habib:2019cui, Chattaraj:2022tay, Hinder:2008kv}.

The predominant emphasis in current research on NR and PN methods on circular orbits can be ascribed to the impact of gravitational wave radiation, which dissipates energy and momentum \cite{Peters:1963ux, Peters:1964zz}. This mechanism results in the circularization of isolated BBHs in the galactic vicinity prior to their merger, even if they possess significant initial eccentricities \cite{Hinder:2007qu}. Consequently, as the binary system transitions into the gravitational wave detection frequency band around 10 Hz, the eccentricity is expected to be negligible.
However, there are some mechanisms through which BBHs can acquire eccentricity before merging \cite{OLeary:2007iqe}. In dense stellar environments like globular clusters \cite{Miller:2002pg, Gultekin:2005fd, OLeary:2005vqo, Rodriguez:2015oxa, Samsing:2017xmd, Rodriguez:2017pec, Rodriguez:2018pss, Park:2017zgj} and galactic nuclei \cite{Gondan:2020svr, Antonini:2012ad, Kocsis:2011jy, Hoang:2017fvh, Gondan:2017wzd, Samsing:2020tda, Tagawa:2020jnc}, interactions such as double-single \cite{Samsing:2013kua, Samsing:2017oij}, double-double interactions \cite{Zevin:2018kzq, Arca-Sedda:2018qgq}, and gravitational capture \cite{Gondan:2020svr, East:2012xq} can induce eccentricity in BBHs. Furthermore, in a three-body system \cite{Naoz:2012bx}, such as binary objects near a supermassive black hole, the eccentricity of the inner binary objects may oscillate due to the Kozai-Lidov mechanism \cite{Naoz:2012bx, VanLandingham:2016ccd, Silsbee:2016djf, Blaes:2002cs, Antognini:2013lpa, Stephan:2016kwj, Katz:2011hn, Seto:2013wwa}, becoming observable upon entering the detection frequency band.
Remarkably, some BBH mergers originating from globular clusters that enter the LIGO sensitive band maintain eccentricities surpassing 0.1 \cite{Rodriguez:2017pec}. The event GW190521 \cite{LIGOScientific:2020iuh} is considered a potential BBH merger with a high eccentricity of $e = 0.69$ \cite{Gayathri:2020coq, Romero-Shaw:2020thy}. With the advancement of detector sensitivity, an increasing number of eccentric BBH mergers are anticipated to be identified in O4 by the LIGO-Virgo-KAGRA (LVK) collaboration or by upcoming ground-based gravitational-wave observatories like the Einstein Telescope (ET) \cite{Punturo:2010zz} or Cosmic Explorer (CE) \cite{Reitze:2019iox}.

In recent decades, several prominent NR collaborations, including Simulating eXtreme Spacetimes (SXS) \cite{Mroue:2013xna, Boyle:2019kee}, Rochester Institute of Technology (RIT) \cite{Healy:2017psd, Healy:2019jyf, Healy:2020vre, Healy:2022wdn}, bi-functional adaptive mesh (BAM) \cite{Hamilton:2023qkv, Bruegmann:2006ulg, Husa:2007hp}, and MAYA \cite{Jani:2016wkt, Ferguson:2023vta}, have conducted extensive simulations of binary compact objects, with their simulation datasets publicly accessible.
In the domain of NR simulations concerning eccentric orbit BBH systems, recent collaborations have delved into a wider parameter space. For example, SXS has released 20 sets of nonspinning eccentric orbit waveforms encompassing mass ratios from 1/3 to 1 and eccentricities from 0 to 0.2, covering the (2,2) mode and higher-order modes (2,1), (3,3), (3,2), (4,4), (4,3), and (5,5) \cite{SXSBBH}. Conversely, RIT has published 824 sets of nonspinning, spin-aligned, and spin-precessed eccentric orbit waveforms with mass ratios ranging from 1/32 to 1 and eccentricities from 0 to 1, also including the (2,2) mode and higher-order modes (2,1), (3,3), (3,2), (4,4), and (4,3) \cite{RITBBH, Healy:2022wdn}.
The extensive collection of NR simulations now accessible provides a rich dataset for thorough exploration and analysis of eccentric orbit BBH mergers.

Substantial progress has been achieved in the investigation of PN waveforms for BBH in eccentric orbits over recent decades. The computation of PN waveforms in eccentric orbit typically involves three crucial components: conservative dynamics, radiative dynamics, and the determination of the quadrupole moment or higher-order moment, also referred to as the amplitude of gravitational waves.
In the realm of conservative dynamics, Memmesheimer et al. \cite{Memmesheimer:2004cv} examined the quasi-Kepler parameterization of the nonspinning configuration up to 3PN under ADM and harmonic coordinates, while recent work by Cho et al. \cite{Cho:2021oai} extended this analysis to the 4PN order. Tessmer et al. \cite{Tessmer:2010hp} concentrated on the quasi-Kepler parameterization of spin-aligned BBHs at 2PN, with a further extension to the 3PN order in the study by Tessmer et al. \cite{Tessmer:2012xr}.
Regarding radiative dynamics, Konigsdorffer and Gopakumar \cite{Konigsdorffer:2006zt, Gopakumar:2001dy} computed the 2PN instantaneous term and 1.5PN hereditary term for nonspinning systems, while Arun et al. \cite{Arun:2009mc, Arun:2007rg, Arun:2007sg, Moore:2016qxz} expanded this investigation to the 3PN order, encompassing instantaneous and hereditary contributions such as the tail and memory effects.
In the context of the quadrupole moment, higher-order moments, or gravitational wave amplitude, Mishra et al., Boetzel et al., and Ebersold et al. \cite{Mishra:2015bqa, Boetzel:2019nfw, Ebersold:2019kdc} calculated the amplitude of the higher-order moment for nonspinning systems at 3PN, including higher-order modes up to the (10,10) mode, analyzing both the instantaneous and hereditary components. Klein et al. \cite{Klein:2010ti} investigated the spin effects of 2PN in the phasing of gravitational waves from binaries on eccentric orbits. Henry et al. \cite{Henry:2023tka} and Paul et al. \cite{Paul:2022xfy} explored radiation dynamics and higher-order moments for spin-aligned waveforms at the 2PN and 3PN level.
These comprehensive computational investigations have improved the accuracy of PN gravitational waveforms for eccentric orbits and significantly broadened the parameter space available for detailed analysis.

The most recent PN and NR discoveries have paved the way for a more in-depth comparison between the two, laying the foundation for the development of more precise hybrid waveforms that blend PN and NR data. Several papers employ surrogate models to generate gravitational waveforms through hybrid approaches \cite{Varma:2019csw,Yoo:2023spi,Varma:2018mmi,Blackman:2017pcm,Blackman:2017dfb,Blackman:2015pia,GramaxoFreitas:2024bpk,Islam:2022laz}. The creation of accurate hybrid waveforms combining PN and NR information for eccentric orbits has been a central focus in gravitational wave detection over the past decade.
While studies by Chattaraj et al., Huerta et al., Hinder et al., Paul et al., and Huerta et al. \cite{Chattaraj:2022tay, Huerta:2016rwp, Hinder:2017sxy, Huerta:2017kez, Paul:2024ujx} have contributed significantly in this area, the parameter space remains notably restricted, particularly in scenarios involving low mass ratios, high eccentricities, and spin-aligned configurations. Therefore, the development of corresponding PN waveforms is crucial to address these limitations.
In this research, we present a comprehensive and detailed comparison of PN and NR waveforms for eccentric orbits, encompassing both nonspinning and spin-aligned setups. Our analysis spans mass ratios from 1/4 to 1 and eccentricities from 0 to 0.45, including waveform harmonic modes (2,2), (3,3), (3,2), (4,4), (4,3), and (5,5), and incorporating the high-order moments of the waveform. Through this meticulous approach, our objective is to advance the generation of precise gravitational wave templates for eccentric orbit binary black holes, signifying progress towards more sophisticated waveform constructions.

This paper is structured as follows: In Section \ref{sec:II:A}, we introduce the foundational principles and waveform construction for BBH systems using PN theory with nonspinning and spin-aligned configurations in eccentric orbits. Section \ref{sec:II:B} delves into the eccentric waveforms obtained from the RIT and SXS catalogs employed in this investigation. Section \ref{sec:II:C} outlines the fitting process of PN and NR waveforms. Section \ref{sec:II:D} provides a comparative analysis of different methods for measuring eccentricity.
Moving to Section \ref{sec:III:A}, we present the residuals resulting from our waveform fitting. Section \ref{sec:III:B} explores the distinctions between various eccentricity measurement methodologies. In Section \ref{sec:III:C}, we present the fitting outcomes and error assessments for the dominant mode (2,2), while Section \ref{sec:III:D} extends these outcomes to higher-order modes (2,1), (3,3), (3,2), (4,4), (4,3), and (5,5). Section \ref{sec:III:E} addresses challenges encountered in fitting waveforms from systems with small mass ratios within the RIT catalog.
Finally, Section \ref{sec:IV} summarizes our discoveries and outlines prospects for future research. Throughout, natural units are utilized, i.e., $G=c=1$, with explicit mention of $G$ and $c$ where necessary for clarity.

\section{methods}\label{sec:II}

\subsection{Eccentric post-Newtonian waveforms}\label{sec:II:A}
In this section, we begin by outlining fundamental concepts concerning a PN system operating in eccentric orbits. This description is notably succinct. For those seeking more comprehensive insights, we recommend consulting the seminal work on PN by Blanchet et al. \cite{Blanchet:2013haa}.
For a nonspinning BBH setup, the constituent black holes possess masses denoted as $m_1$ and $m_2$. The combined mass is represented by $M = m_1 + m_2 =1$, serving as a pivotal mass scale in both the PN and NR domains. The mass ratio is articulated as $q = m_1/m_2$, with $m_1$ being smaller than $m_2$. The reduced mass is calculated as $\mu = m_1 m_2 / M$, while the symmetric mass ratio is defined as $\eta = \mu / M$.

In Newtonian orbits, both energy and angular momentum are conserved, with the constancy of the latter implying that the orbit remains confined to a single plane without precession. As outlined in Ref. \cite{Hinder:2008kv}, we can describe the relative orbit radius $r$ and angular frequency $\dot{\phi}$ (where $\phi$ denotes the relative angular coordinate) as follows:
\begin{equation}\label{eq:1}
r=a(1-e \cos u),
\end{equation}

\begin{equation}\label{eq:2}
\dot{\phi}=\frac{n \sqrt{1-e^2}}{(1-e \cos u)^2},
\end{equation}
where $a$ and $e$ represent the semimajor axis and eccentricity, respectively, keeping constants within Newtonian orbits. The eccentric anomaly $u$ and mean motion $n$ are also introduced, specifically, $n = 2 \pi / P = a^{-3/2} M^{1/2}$, where $P$ signifies the orbital period from one pericenter to another. The eccentric anomaly $u$ obeys Kepler’s equation:
\begin{equation}\label{eq:3}
l=u-e \sin u,
\end{equation}
where the mean anomaly $l$ satisfies $\dot{l} = n$. Given that $n$ remains constant within Newtonian orbits, integration leads to $l = n(t - t_0)$, with $t_0$ denoting the integration constant (representing the initial moment). Eq. (\ref{eq:3}) serves as a transcendental algebraic equation for $u$, which necessitates numerical solutions at each time step. Subsequently, utilizing Eqs. (\ref{eq:1}) and (\ref{eq:2}), we can derive related coordinates $r$ and $\dot{\phi}$ at any given time. By further differentiation and integration, $\dot{r}$ and $\phi$ can be obtained. Through this process, Newtonian orbital dynamics are fully resolved, with each orbit characterized by constants $n$, $e$, $\phi_0 \equiv \phi(t_0)$, and $l_0 \equiv l(t_0)$.

In PN orbits, we can establish conservative dynamical equations akin to those in the Newtonian scenario, albeit with slight modifications. The Newtonian expressions for $r$, $\dot{\phi}$, and $l$ are adjusted by the inclusion of higher-order PN terms. Furthermore, in contrast to the Keplerian parameterization observed in Newtonian dynamics, the quasi-Keplerian parameterization in PN introduces eccentricities $e_t$, $e_r$, and $e_{\phi}$ at distinct coordinates $t$, $r$, and $\phi$. These eccentricities are interconnected and converge to the Newtonian eccentricity $e$ at lower orders.
Typically, in PN literature, $e_t$ is selected as the primary measure of eccentricity. Notably, in PN scenarios, the orbital plane undergoes precession. Owing to this precession effect, the orbital plane rotates by an angle $\delta \phi$ within one orbital period $P$, allowing the definition of the orbital angular frequency as:

\begin{equation}\label{eq:4}
\omega \equiv \frac{2 \pi+\delta \phi}{P}.
\end{equation}

According to Refs. \cite{Hinder:2008kv,Konigsdorffer:2006zt}, two models have been proposed for PN systems in eccentric orbits. The first model, referred to as the $n$ model, employs $n$ and $e_t$ as parameters. The second model, known as the $x$ model, utilizes the common expansion variable of PN, $x\equiv(M \omega)^{2/3}$, along with $e_t$ as parameters. As elucidated in Ref. \cite{Hinder:2008kv}, the latter model more accurately depicts the waveform of eccentric orbits in the PN framework.
In our analysis, we will explore two scenarios of PN eccentric orbital BBH waveforms: one involving nonspinning black holes and the other featuring spin-aligned (non-precessing) black holes, where the directions of orbital angular momentum and spin angular momentum are either parallel or antiparallel. Despite their differences in spin configurations, the dynamics of these two cases exhibit remarkable similarities and can be approached through analogous procedures.

\subsubsection{Nonspinning waveforms}\label{sec:II:A:1}
Citing Refs. \cite{Hinder:2008kv,Memmesheimer:2004cv}, we investigate the conservative orbital dynamics of the 3PN order within modified harmonic coordinates, specifically employing the quasi-Keplerian parametrization. These expressions are dependent on the energy $E$ and the angular momentum $J$ as outlined in Ref. \cite{Memmesheimer:2004cv}. Following the methodology presented in Ref. \cite{Hinder:2008kv}, we reframe these expressions in terms of the variables $x$ and $e_t$. For brevity, we provide condensed versions here, while detailed expressions can be found in Appendix \ref{App:A}.
The PN form of Eq. (\ref{eq:1}) and Eq. (\ref{eq:2}) for nonspinning BBH can be expressed as
\begin{equation}\label{eq:5}
\begin{aligned}
r^\mathrm{NS}= & r_{\mathrm{Newt}}^\mathrm{NS} x^{-1}+r_{1 \mathrm{PN}}^\mathrm{NS}+r_{2 \mathrm{PN}}^\mathrm{NS} x
+r_{3 \mathrm{PN}}^\mathrm{NS} x^2 \\ 
& +\mathcal{O}\left(x^3\right)
\end{aligned}
\end{equation}
and
\begin{equation}\label{eq:6}
\begin{aligned}
\dot{\phi} ^\mathrm{NS}= & \dot{\phi}_{ \mathrm{Newt}}^\mathrm{NS} x^{3 / 2}+\dot{\phi}_{1 \mathrm{PN}}^\mathrm{NS} x^{5 / 2} +\dot{\phi}_{2 \mathrm{PN}}^\mathrm{NS} x^{7 / 2} \\
&+\dot{\phi}_{3 \mathrm{PN}}^\mathrm{NS} x^{9 / 2}+\mathcal{O}\left(x^{11 / 2}\right)
\end{aligned},
\end{equation}
where the superscript NS indicates nonspinning, and $r_{1 \mathrm{PN}}^\mathrm{NS}$, $\dot{\phi}_{1 \mathrm{PN}}^\mathrm{NS}$, etc., denote the PN expansion coefficients, functions of $e_t$ and $u$. Additionally, Eq. (\ref{eq:3}) and $\dot{l} = n$ transform into
\begin{equation}\label{eq:7}
l^\mathrm{NS}=u-e_t \sin u+l_{2 \mathrm{PN}}^\mathrm{NS} x^2+l_{3 \mathrm{PN}}^\mathrm{NS} x^3+\mathcal{O}\left(x^4\right)
\end{equation}
and 
\begin{equation}\label{eq:8}
\begin{aligned}
\dot{l}^\mathrm{NS} & =n^\mathrm{NS} \\
& =x^{3 / 2}+n_{1 \mathrm{PN}}^\mathrm{NS} x^{5 / 2}+n_{2 \mathrm{PN}} ^\mathrm{NS}x^{7 / 2}+n_{3 \mathrm{PN}}^\mathrm{NS} x^{9 / 2}\\
&+\mathcal{O}\left(x^{11 / 2}\right),
\end{aligned}
\end{equation}
where, unlike $r_{1 \mathrm{PN}}^\mathrm{NS}$, $\dot{\phi}_{1 \mathrm{PN}}^\mathrm{NS}$, etc., $l_{1 \mathrm{PN}}^\mathrm{NS}$, $n_{1 \mathrm{PN}}^\mathrm{NS}$, etc., are solely functions of $e_t$. For PN conservative dynamics, the integration of the right-hand side of Eq. (\ref{eq:8}), with respect to the constants $x$ and $e_t$, allows for the direct determination of $l(t)$ in terms of the integration constant $l_0$ at the initial time $t_0$. Subsequently, by substituting the obtained $l(t)$ into Eq. (\ref{eq:7}), we solve for $u$ through numerical methods, then substitute $u$ into Eqs. (\ref{eq:5}) and (\ref{eq:6}) to evaluate $r^\mathrm{NS}$ and $\dot{\phi}^\mathrm{NS}$.

Previously, we focused solely on the conservative dynamics of BBH systems. However, in realistic BBH scenarios, gravitational radiation leads to the loss of energy and angular momentum, causing the parameters \(x\) and \(e_t\) to evolve over time. In the context of PN calculations, this evolution is typically treated using an adiabatic approximation, averaging over one orbital period to compute \(\dot{x}\) and \(\dot{e}_t\). These quantities can be decomposed into instantaneous and nonlinear hereditary terms. The instantaneous terms depend on the retarded time, while the hereditary terms are time integrals, accounting for the system’s entire dynamical history. As discussed in Refs. \cite{Arun:2009mc, Arun:2007sg, Arun:2007rg}, the hereditary contributions include not only tail, tail-of-tail, and tail-square terms (similar to the energy flux) but also a memory contribution at 2.5PN order. Our 3PN radiative dynamics calculations are based on Ref. \cite{Arun:2009mc} and are expressed as follows: 

\begin{equation}\label{eq:9}
\dot{x}^\mathrm{NS}= \dot{x}_{\mathrm{inst}}^\mathrm{NS}+\dot{x}_{\mathrm{hered}}^\mathrm{NS}
\end{equation}
and
\begin{equation}\label{eq:10}
\dot{e_t}^\mathrm{NS}= \dot{e_t}_{\mathrm{inst}}^\mathrm{NS}+\dot{e_t}_{\mathrm{hered}}^\mathrm{NS},
\end{equation}
where the subscripts \textit{inst} and \textit{hered} mean \textit{instantaneous} and \textit{hereditary}. Their specific expressions are
\begin{equation}\label{eq:11}
\begin{aligned}
\dot{x}_{\mathrm{inst}}^\mathrm{NS}= & \frac{2 c^3 \eta}{3 G M} x^5 \left(\dot{x}_{\mathrm{Newt}}^\mathrm{NS}+\dot{x}_{1 \mathrm{PN}}^\mathrm{NS} x+\dot{x}_{2 \mathrm{PN}}^\mathrm{NS} x^2 \right. \\
& \left. +\dot{x}_{3 \mathrm{PN}}^\mathrm{NS} x^3\right)
\end{aligned},
\end{equation}

\begin{equation}\label{eq:12}
\begin{aligned}
\dot{e_t}_{\mathrm{inst}}^\mathrm{NS}= & -\frac{c^3 \eta}{G M} e_t x^4 \left(\dot{e_t}_{\mathrm{Newt}}^\mathrm{NS}+\dot{e_t}_{1 \mathrm{PN}}^\mathrm{NS} x+\dot{e_t}_{2 \mathrm{PN}}^\mathrm{NS} x^2 \right. \\
& \left. +\dot{e_t}_{3 \mathrm{PN}}^\mathrm{NS} x^3 \right)
\end{aligned},
\end{equation}

\begin{equation}\label{eq:13}
\begin{aligned}
\dot{x}_{\mathrm{hered}}^\mathrm{NS}= & \frac{64c^3 \eta}{5 G M} x^4 \left(\dot{x}_{1.5\mathrm{PN}}^\mathrm{NS} x^{3/2}+\dot{x}_{2.5 \mathrm{PN}}^\mathrm{NS} x^{5/2} \right. \\
&\left. +\dot{x}_{3 \mathrm{PN}}^\mathrm{NS} x^3 \right)
\end{aligned}
\end{equation}
and 

\begin{equation}\label{eq:14}
\begin{aligned}
\dot{e_t}_{\mathrm{hered}}^\mathrm{NS}= & \frac{32c^3 \eta}{5 G M} e_t x^4 \left(\dot{e_t}_{1.5\mathrm{PN}}^\mathrm{NS} x^{3/2}+\dot{e_t}_{2.5 \mathrm{PN}}^\mathrm{NS} x^{5/2} \right. \\
&\left. +\dot{e_t}_{3 \mathrm{PN}}^\mathrm{NS} x^3 \right)
\end{aligned},
\end{equation}
respectively.
Ref. \cite{Hinder:2008kv} addresses the correction for radiative reaction at 2PN order, with the hereditary term included only at 1.5PN order, which introduces certain limitations. To obtain a more accurate PN waveform, we adopt the 3PN results from Ref. \cite{Arun:2009mc} for the specific expressions in Eqs. (\ref{eq:11}), (\ref{eq:12}), (\ref{eq:13}), and (\ref{eq:14}), while confirming consistency with their 2PN counterparts. However, the results in Ref. \cite{Arun:2009mc} are provided in ADM (Arnowitt-Deser-Misner) coordinates, requiring the transformation of the time eccentricity \(e_t\) to harmonic coordinates. The parameter \(x\), being coordinate-invariant, does not require transformation and can be used directly. For brevity, the detailed derivations of \(\dot{x}^\mathrm{NS}\) and \(\dot{e_t}^\mathrm{NS}\) are relegated to Appendix \ref{App:A}. The adiabatic evolution equations governing \(\dot{x}^\mathrm{NS}\) and \(\dot{e_t}^\mathrm{NS}\) form an autonomous system, solvable independently of Kepler’s equation. By providing initial conditions \(x_0\) and \(e_{t0}\), we can numerically solve the system of ordinary differential equations to obtain \(\dot{x}^\mathrm{NS}(t)\) and \(\dot{e_t}^\mathrm{NS}(t)\). From this point, the steps of the conservative dynamics, as outlined previously, are followed: integrating Eq. (\ref{eq:8}), solving the root of Eq. (\ref{eq:7}), and substituting the resulting \(u\) into Eqs. (\ref{eq:5}) and (\ref{eq:6}). Finally, by differentiating and integrating \(r^\mathrm{NS}\) and \(\dot{\phi}^\mathrm{NS}\), we obtain \(\dot{r}^\mathrm{NS}\) and \(\phi^\mathrm{NS}\). At this stage, we have fully resolved the dynamics of BBHs in eccentric orbits with gravitational radiation reaction. 

In most studies, the quadrupole moment, which represents the leading-order Newtonian contribution, forms the basis for constructing PN waveforms. This method, known as the restricted waveform approximation, is a crucial component in the development of waveforms. The gravitational wave strain $h$ due to the quadrupole moment can be expressed as follows \cite{Hinder:2008kv}:
\begin{equation}\label{eq:15}
h=h_{+}-i h_{\times},
\end{equation}

\begin{equation}\label{eq:16}
\begin{aligned}
h_{+}= & -\frac{M \eta}{R}\left\{( \operatorname { c o s } ^ { 2 } \theta + 1 ) \left[\cos 2 \phi^{\prime}\left(-\dot{r}^2+r^2 \dot{\phi}^2+\frac{M}{r}\right)\right.\right. \\
& \left.\left.+2 r \dot{r} \dot{\phi} \sin 2 \phi^{\prime}\right]+\left(-\dot{r}^2-r^2 \dot{\phi}^2+\frac{M}{r}\right) \sin ^2 \theta\right\},
\end{aligned}
\end{equation}

\begin{equation}\label{eq:17}
\begin{aligned}
h_{\times}= & -\frac{2 M \eta}{R} \cos \theta\left\{\left(-\dot{r}^2+r^2 \dot{\phi}^2+\frac{M}{r}\right) \sin 2 \phi^{\prime}\right. \\
& \left.-2 r \cos 2 \phi^{\prime} \dot{r} \dot{\phi}\right\},
\end{aligned}
\end{equation}
where $\phi^{\prime} \equiv \phi-\varphi$, and $\theta$ and \(\varphi\) are the spherical polar angles. The gravitational wave strain \(h\) can be expressed as the leading (2,2) mode \(h^{22}\), using the spin-weighted spherical harmonics \({ }_{-2} Y_2^2(\theta, \varphi) = \frac{1}{2} e^{2 i \varphi} \sqrt{5 / \pi} \cos ^4(\theta / 2)\) for \(\ell = 2\), \(m = 2\), 
\begin{equation}\label{eq:18}
\begin{gathered}
h^{22}=\int_{-2} Y_2^{2 *}(\theta, \varphi) h(\theta, \varphi) d \Omega, \\
=-\frac{4 M \eta e^{-2 i \phi}}{R} \sqrt{\frac{\pi}{5}}\left(\frac{M}{r}+(\dot{\phi} r+i \dot{r})^2\right),
\end{gathered}
\end{equation}
where \(Y_2^{2 *}(\theta, \varphi)\) represents the complex conjugate of \({ }_{-2} Y_2^2(\theta, \varphi)\). This gives us the dominant gravitational wave mode \(h^{22}\). The negative and positive modes of $\ell$ are in agreement, thus our focus in this study is solely on the positive modes. However, due to symmetry, the quadrupole moment alone is insufficient to derive higher modes such as \(h^{32}\), \(h^{43}\), and others. To capture the full complexity of the waveform, we must go beyond the quadrupole moment in Eq. (\ref{eq:18}) and incorporate higher-order moments. 

Refs. \cite{Mishra:2015bqa,Boetzel:2019nfw,Ebersold:2019kdc} have thoroughly investigated the amplitudes of eccentric orbits, deriving both the instantaneous and hereditary terms of the waveform as functions of \(e_t\) and \(x\). However, the waveforms in these references are expressed as low-eccentricity expansions of \(e_t\), valid only for eccentricities less than 0.1, and are not applicable to the high-eccentricity regime considered in this work. Consequently, we rely solely on the instantaneous waveforms for general orbits provided in Ref. \cite{Mishra:2015bqa}. The derivation of the 3PN waveforms for general orbits utilizes the MPM-PN (multipolar post-Minkowskian and post-Newtonian) formalism. The gravitational waveform is then expressed using spin-weighted spherical harmonics as shown in Eq. (\ref{eq:18}). For brevity, the detailed expressions for the various waveform modes are relegated to Appendix \ref{App:A}. Ref. \cite{Mishra:2015bqa} represents the nonspinning instantaneous waveform as: 


\begin{equation}\label{eq:19}
h^{\ell m, \mathrm{NS}}=\frac{4 G M \eta}{c^4 R} \sqrt{\frac{\pi}{5}} e^{-i m \phi} H^{\ell m, \mathrm{NS}}.
\end{equation}

Each mode originates from a different PN order, and thus, they cannot be expressed simultaneously. As an illustration, we consider the $h^{22}$ mode, which can be written as
\begin{equation}\label{eq:20}
h^{22, \mathrm{NS}}=\frac{4 G M \eta }{c^4 R} \sqrt{\frac{\pi}{5}} e^{-2 i \phi} H^{22, \mathrm{NS}},
\end{equation}
where the amplitude $H^{22, \mathrm{NS}}$ is expressed as a sum of terms from different PN orders: \begin{equation}\label{eq:21}
H^{22, \mathrm{NS}}=H_{\text {Newt }}^{22, \mathrm{NS}}+H_{1 \mathrm{PN}}^{22, \mathrm{NS}}+H_{2 \mathrm{PN}}^{22, \mathrm{NS}}+H_{2.5 \mathrm{PN}}^{22, \mathrm{NS}}+H_{3 \mathrm{PN}}^{22, \mathrm{NS}}.
\end{equation}
Focusing on the leading Newtonian order, we have: \begin{equation}\label{eq:22}
H^{22, \mathrm{NS}}_{\text {Newt }}=\frac{G M}{r}+r^2 \dot{\phi}^2+2 i r \dot{r} \dot{\phi}-\dot{r}^2,
\end{equation}
which reduces to the quadrupole moment expression in Eq. (\ref{eq:18}). Thus, we obtain the harmonic waveform modes for non-spinning BBH in eccentric orbits. Although the hereditary term contributions have been omitted, the accuracy obtained still falls within an acceptable range as demonstrated by the fitting outcomes of PN and NR in Sec. \ref{sec:II:C}. The exploration of amplitudes incorporating hereditary terms will be deferred to future investigations. 

\subsubsection{Spin-aligned waveforms}\label{sec:II:A:2}
For a spin-aligned or non-precessing BBH system in eccentric orbits, Ref. \cite{Henry:2023tka} provides a comprehensive analysis and formulation of gravitational wave modes and fluxes up to 3PN order. In this work, spin effects are incorporated into the quasi-Keplerian parameterization up to 3PN in harmonic coordinates, using the covariant Tulczyjew-Dixon SSC (Newton-Wigner spin-supplementary condition) \cite{Newton:1949cq}. Key results from Ref. \cite{Henry:2023tka} include next-to-leading order instantaneous spin-orbit and spin-spin contributions to the waveform modes, as well as the inclusion of hereditary (tail and memory) contributions to both modes and fluxes in eccentric orbits. Here, we introduce several new concepts: the antisymmetric mass ratio $\delta=(m_1-m_2)/M$, and the dimensionless spin magnitudes $\chi_1 \equiv \left|\boldsymbol{S}_1\right| /\left( m_1^2\right)$ and $\chi_2 \equiv \left|\boldsymbol{S}_2\right| /\left( m_2^2\right)$. Both $\chi_1$ and $\chi_2$ are aligned with the $z$-axis of the coordinate system, and their values range from [-1,1]; negative values indicate that the spin angular momentum is antiparallel to the orbital angular momentum, while positive values indicate alignment. Here, $|\boldsymbol{S}_1|$ and $|\boldsymbol{S}_2|$ represent the spin magnitudes. Following the approach of the non-spinning case, we first introduce the 3PN quasi-Keplerian parameterization for the spin-aligned case \cite{Henry:2023tka,Tessmer:2010hp}. Since we remain within the same harmonic coordinate system, the radial coordinate $r^{\mathrm{SP}}$ can be expressed as: 
\begin{equation}\label{eq:23}
r^{\mathrm{SP}}=r^{\mathrm{NS}} + r^{\mathrm{SO}} + r^{\mathrm{SS}},
\end{equation}
where $r^{\mathrm{NS}}$ denotes the non-spinning component of $r$, as given by Eq. (\ref{eq:5}). The superscript $\mathrm{SP}$ indicates the \textit{spin} contribution, $\mathrm{SO}$ denotes spin-orbit coupling, which represents the first-order term in spin, and $\mathrm{SS}$ refers to spin-spin coupling, which is the quadratic term in spin. The detailed expressions for $r^{\mathrm{SO}}$ and $r^{\mathrm{SS}}$ are (provided in Appendix \ref{App:A}): 
\begin{equation}\label{eq:24}
r^{\mathrm{SO}}=r_{1.5 \mathrm{PN}}^{\mathrm{SO}} + r_{2.5 \mathrm{PN}}^{\mathrm{SO}},
\end{equation}
and 
\begin{equation}\label{eq:25}
r^{\mathrm{SS}}=r_{2 \mathrm{PN}}^{\mathrm{SS}} + r_{3 \mathrm{PN}}^{\mathrm{SS}}.
\end{equation}
Similarly, for the orbital angular velocity $\dot{\phi}^{\mathrm{SP}}$ , we have
\begin{equation}\label{eq:26}
\dot{\phi}^{\mathrm{SP}}=\dot{\phi}^{\mathrm{NS}} + \dot{\phi}^{\mathrm{SO}} + \dot{\phi}^{\mathrm{SS}},
\end{equation}
where the abbreviations for $\dot{\phi}^{\mathrm{SO}}$ and $\dot{\phi}^{\mathrm{SS}}$ are 
\begin{equation}\label{eq:27}
\dot{\phi}^{\mathrm{SO}}=\dot{\phi}_{1.5 \mathrm{PN}}^{\mathrm{SO}} + \dot{\phi}_{2.5 \mathrm{PN}}^{\mathrm{SO}},
\end{equation}
and 
\begin{equation}\label{eq:28}
\dot{\phi}^{\mathrm{SS}}=\dot{\phi}_{2 \mathrm{PN}}^{\mathrm{SS}} + \dot{\phi}_{3 \mathrm{PN}}^{\mathrm{SS}}.
\end{equation}
The Kepler's equation for the spin-aligned BBH is
\begin{equation}\label{eq:29}
l^{\mathrm{SP}}=l^{\mathrm{NS}} + l^{\mathrm{SO}} + l^{\mathrm{SS}},
\end{equation}
where the abbreviations for $l^{\mathrm{SO}}$ and $l^{\mathrm{SS}}$ are 
\begin{equation}\label{eq:30}
l^{\mathrm{SO}}=l_{2.5 \mathrm{PN}}^{\mathrm{SO}},
\end{equation}
and 
\begin{equation}\label{eq:31}
l^{\mathrm{SS}}=l_{3 \mathrm{PN}}^{\mathrm{SS}}.
\end{equation}
Similarly, for the mean motion $n$, it can be expressed as
\begin{equation}\label{eq:32}
\dot{l}^\mathrm{SP} = n^{\mathrm{SP}}=n^{\mathrm{NS}} + n^{\mathrm{SO}} + n^{\mathrm{SS}},
\end{equation}
where the abbreviations for $n^{\mathrm{SO}}$ and $n^{\mathrm{SS}}$ are
\begin{equation}\label{eq:33}
n^{\mathrm{SO}}=n_{1.5 \mathrm{PN}}^{\mathrm{SO}} + n_{2.5 \mathrm{PN}}^{\mathrm{SO}},
\end{equation}
and 
\begin{equation}\label{eq:34}
n^{\mathrm{SS}}=n_{2 \mathrm{PN}}^{\mathrm{SS}} + n_{3 \mathrm{PN}}^{\mathrm{SS}}.
\end{equation}
The above content provides the quasi-Keplerian parameterization at 3PN order for the dynamics of spin-aligned eccentric orbits in a BBH system. For further details, refer to Refs. \cite{Henry:2023tka,Tessmer:2010hp}. Next, we consider the evolution of $x$ and $e_t$ due to the energy and angular momentum carried away by gravitational radiation. Analogous to the non-spinning case, $\dot{x}^\mathrm{SP}$ and $\dot{e_t}^\mathrm{SP}$ can be expressed as 
\begin{equation}\label{eq:35}
\dot{x}^{\mathrm{SP}}=\dot{x}^{\mathrm{NS}} + \dot{x}^{\mathrm{SO}} + \dot{x}^{\mathrm{SS}},
\end{equation}
and 
\begin{equation}\label{eq:36}
\dot{e_t}^{\mathrm{SP}}=\dot{e_t}^{\mathrm{NS}} + \dot{e_t}^{\mathrm{SO}} + \dot{e_t}^{\mathrm{SS}}.
\end{equation}
As in the nonspinning case, the spin-aligned scenario also includes contributions from both the instantaneous and hereditary terms. The abbreviations for $\dot{x}^{\mathrm{SO}}$, $\dot{x}^{\mathrm{SS}}$, $\dot{e_t}^{\mathrm{SO}}$, and $\dot{e_t}^{\mathrm{SS}}$ are
\begin{equation}\label{eq:37}
\dot{x}^{\mathrm{SO}}= \dot{x}_{\mathrm{inst}}^\mathrm{SO}+\dot{x}_{\mathrm{hered}}^\mathrm{SO},
\end{equation}

\begin{equation}\label{eq:38}
\dot{x}^{\mathrm{SS}}= \dot{x}_{\mathrm{inst}}^\mathrm{SS},
\end{equation}

\begin{equation}\label{eq:39}
\dot{e_t}^{\mathrm{SO}}= \dot{e_t}_{\mathrm{inst}}^\mathrm{SO}+\dot{e_t}_{\mathrm{hered}}^\mathrm{SO},
\end{equation}

\begin{equation}\label{eq:40}
\dot{e_t}^{\mathrm{SS}}= \dot{e_t}_{\mathrm{inst}}^\mathrm{SS}.
\end{equation}
It is worth mentioning that the hereditary term of spin-spin coupling comes from beyond the 3PN order, so it is absent here. The specific expressions of the PN order of the above equations are
\begin{equation}\label{eq:41}
\dot{x}_{\mathrm{inst}}^\mathrm{SO}= \dot{x}_{1.5 \mathrm{PN}}^\mathrm{SO}+\dot{x}_{2.5 \mathrm{PN}}^\mathrm{SO},
\end{equation}

\begin{equation}\label{eq:42}
\dot{x}_{\mathrm{hered}}^\mathrm{SO}= \dot{x}_{3 \mathrm{PN}}^\mathrm{SO},
\end{equation}

\begin{equation}\label{eq:43}
\dot{x}_{\mathrm{inst}}^\mathrm{SS}= \dot{x}_{2 \mathrm{PN}}^\mathrm{SS}+\dot{x}_{3 \mathrm{PN}}^\mathrm{SS},
\end{equation}

\begin{equation}\label{eq:44}
\dot{e_t}_{\mathrm{inst}}^\mathrm{SO}= \dot{e_t}_{1.5 \mathrm{PN}}^\mathrm{SO}+\dot{e_t}_{2.5 \mathrm{PN}}^\mathrm{SO},
\end{equation}

\begin{equation}\label{eq:45}
\dot{e_t}_{\mathrm{hered}}^\mathrm{SO}= \dot{e_t}_{3 \mathrm{PN}}^\mathrm{SO},
\end{equation}

\begin{equation}\label{eq:46}
\dot{e_t}_{\mathrm{inst}}^\mathrm{SS}= \dot{e_t}_{2 \mathrm{PN}}^\mathrm{SS}+\dot{e_t}_{3 \mathrm{PN}}^\mathrm{SS}.
\end{equation}
Similar to the nonspinning case, we substitute the radiation equation into the conservation equation to obtain $r^{\mathrm{SO}}$ and $\dot{\phi}^{\mathrm{SP}}$, and we obtain the BBH evolution dynamics of the spin-aligned eccentric orbit. 
For the spin-aligned case, as in the previous nonspinning case, we only use the instantaneous terms expression of the general orbit, because the expression with hereditary terms in Ref. \cite{Henry:2023tka} is only expansions for low eccentricity and is not applicable to the cases of medium and high eccentricity, and is therefore not applicable to the content of our study. Each higher mode of spin-aligned waveform can be expressed as
\begin{equation}\label{eq:47}
h^{\ell m,\mathrm{SP}}=\frac{4 G M \eta}{c^4 R} \sqrt{\frac{\pi}{5}} e^{-i m \phi} H^{\ell m,\mathrm{SP}}.
\end{equation}
We take the 22 mode as an example to give its specific PN order expression:
\begin{equation}\label{eq:48}
h^{22,\mathrm{SP}}=\frac{4 G M \eta }{c^4 R} \sqrt{\frac{\pi}{5}} e^{-2 i \phi} H^{22,\mathrm{SP}},
\end{equation}
where the amplitude $H^{22,\mathrm{SP}}$ can be expressed as
\begin{equation}\label{eq:49}
H^{22,\mathrm{SP}}=H^{22,\mathrm{NS}} + H^{22,\mathrm{SO}} + H^{22,\mathrm{SS}},
\end{equation}
where $H^{22,\mathrm{NS}}$ is the nonspinning portion of the aforementioned waveform. And the abbreviations of $H^{22,\mathrm{SO}}$ and $H^{22,\mathrm{SO}}$ are
\begin{equation}\label{eq:50}
H^{22,\mathrm{SO}}=H_{1.5 \mathrm{PN}}^{22,\mathrm{SO}} + H_{2.5 \mathrm{PN}}^{22,\mathrm{SO}} + H_{3 \mathrm{PN}}^{22,\mathrm{SO}},
\end{equation}
and 
\begin{equation}\label{eq:51}
H^{22,\mathrm{SS}}=H_{2 \mathrm{PN}}^{22,\mathrm{SS}} + H_{3 \mathrm{PN}}^{22,\mathrm{SS}}.
\end{equation}
So far, we can obtain any desired spin-aligned PN waveform in an eccentric orbit. All the abbreviations mentioned above can be found in Appendix \ref{App:A} for detailed.

\subsection{Eccentric Numerical Simulations}\label{sec:II:B}
Several NR collaborations have conducted extensive simulations of BBH mergers in quasicircular orbits. However, publicly accessible simulations involving eccentric orbits remain relatively scarce. The NR simulations of eccentric orbits used in this study were sourced from the RIT (Rochester Institute of Technology) \cite{RITBBH} and SXS (Simulating eXtreme Spacetimes) \cite{SXSBBH} catalogs. The simulations in the RIT catalog were performed using the LazEv code \cite{Zlochower:2005bj}, which employs the moving puncture approach \cite{Campanelli:2005dd} and the BSSNOK (Baumgarte-Shapiro-Shibata-Nakamura-Oohara-Kojima) formalism for the evolution system \cite{Nakamura:1987zz,Shibata:1995we,Baumgarte:1998te}. The LazEv code operates within the Cactus \cite{Schnetter:2003rb}/Carpet \cite{Zlochower:2012fk}/Einstein Toolkit \cite{Loffler:2011ay} framework. The fourth release of the RIT catalog included 824 eccentric BBH NR simulations, covering a range of configurations including nonspinning, spin-aligned, and spin-precessing systems—with eccentricities ranging from 0 to 1 \cite{Healy:2022wdn}. The eccentric waveforms utilized in this work are primarily sourced from the non-spinning and spin-aligned waveforms in the RIT catalog. The second source is the SXS Collaboration, which employs a multi-domain spectral method \cite{Lindblom:2005qh,Szilagyi:2009qz,Kidder:1999fv,Scheel:2008rj} in conjunction with a first-order version of the generalized harmonic formulation \cite{Hemberger:2012jz,Pretorius:2004jg,Samary:2012bw,Garfinkle:2001ni} of Einstein’s equations with constraint damping to evolve the initial data. The Spectral Einstein Code (SpEC) \cite{SXSBBH} is used for these NR simulations.

In NR, gravitational wave waveforms are extracted by computing the Newman-Penrose scalar or Weyl scalar  $\Psi_4$ at a finite radius and then extrapolating it to null infinity \cite{Boyle:2009vi,Bishop:2016lgv,Reisswig:2009rx}. It is noteworthy that $h$ can also be extrapolated directly. Specifically, the SXS $h$ waveforms are directly derived from the metric without the utilization of $\Psi_4$ \cite{Mroue:2013xna, Boyle:2019kee}. The scalar $\Psi_4$ can be expanded using spin-weighted spherical harmonic functions, similar to Eq. (\ref{eq:18}), as follows: \begin{equation}\label{eq:52}
r \Psi_4=\sum_{\ell, m} r \Psi_4^{\ell m} {}_{-2}Y_{\ell, m}(\theta, \phi),
\end{equation} 
where $r$ is the extraction radius, and $\Psi_4^{\ell m}$ represents the expansion coefficient or harmonic mode. As $r$ approaches infinity, the relationship between $h$ and $\Psi_4$ is given by: \begin{equation}\label{eq:53}
\Psi_4=\frac{\partial^2}{\partial t^2} h.
\end{equation}
The expansion coefficients $h^{\ell m}$ and $\Psi_4^{\ell m}$ follow the same relationship. Both $\Psi_4^{\ell m}$ and $h^{\ell m}$ data can be accessed from the RIT and SXS catalog databases. These data are available as harmonic modes, with up to (4,4) modes for RIT and (5,5) modes for SXS. The RIT and SXS catalogs also include detailed metadata for each simulated waveform, such as mass ratio, spin, initial orbital angular momentum, initial ADM energy, and the properties of the final remnants. This metadata provides comprehensive information for analysis.

To streamline the representation of the parameter space and aid in research, we introduce the concept of the effective spin in the $z$-direction, which aligns with the orbital angular momentum $L$. This effective spin is defined as
\begin{equation}\label{eq:54}
\chi_{\mathrm{eff}}=\frac{m_1 \chi_1+m_2 \chi_1}{m_1+m_2}.
\end{equation}

\begin{figure}[htbp!]
\centering
\includegraphics[width=7cm,height=7cm]{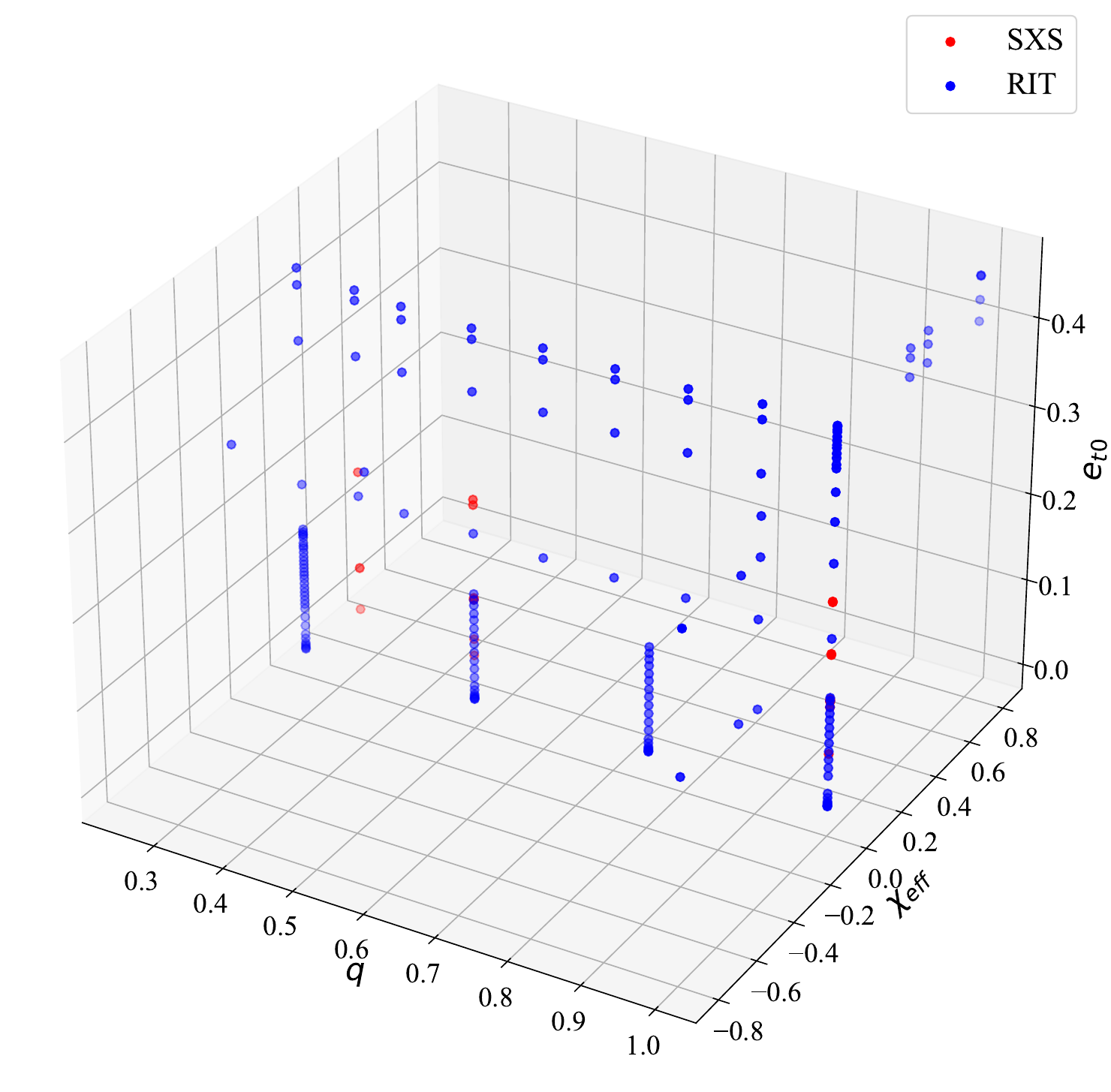}
\caption{\label{FIG:1}The parameters used in our study include three configurations no spin, spin alignment, which cover the parameter space mass ratio $q$ from 1/4 to 1, and the initial eccentricity ${e_t}_0$ from 0 to 0.45. The red and blue markers represent the SXS and RIT catalogs, respectively.}
\end {figure}

During waveform processing, we initially excluded the first $300M$ and $100M$ of the SXS and RIT waveforms, respectively, to eliminate the effects of transient junk radiation. Following this, we identified the peak value of the gravitational wave amplitude of $h^{22}$, which was used as the reference for time alignment, setting this moment as the new time zero for the waveform. Table \ref{table:1} in Appendix \ref{App:B} provides a detailed set of parameters for the waveforms used in the comparison between PN and NR simulations. Additionally, the parameter ranges for the SXS and RIT waveforms are illustrated in FIG. \ref{FIG:1}. In FIG. \ref{FIG:1}, the initial eccentricity ${e_t}_0$ is determined through PN fitting to (2,2) mode frequency of higher-order moments after removing transient junk radiation, a concept that will be further elaborated. RIT catalog provides a broad range of waveform parameters, the lowest mass ratio available for eccentric orbit waveforms is 1/32. However, we have restricted our analysis to waveforms with mass ratios ranging from 1/4 to 1. The SXS catalog includes 20 nonspinning eccentric waveforms with mass ratios $q=1, 1/2, 1/3$ and initial eccentricities ${e_t}_0 \in[0,0.2]$. In total, 180 waveforms were analyzed, including 160 nonspinning waveforms and 20 spin-aligned waveforms from the SXS and RIT catalogs. The rationale for this selection will be discussed in the following sections. 

\subsection{Fitting the post-Newtonian waveforms to numerical
relativity waveforms}\label{sec:II:C}
NR simulations are the most accurate method for modeling the dynamics of BBH systems. They precisely capture not only the inspiral phase but also the merger and ringdown phases, including complex strong-field dynamic effects. In contrast, the PN model provides high-order corrections to the speed of light to incrementally approximate the gravitational wave waveform. However, it is limited to cases where the black holes are widely separated, and their velocities remain well below the speed of light. Following the approach outlined in Ref. \cite{Hinder:2008kv}, we adopt a similar methodology: we use the NR waveform as the reference, fit the PN waveform to the NR data, and thereby extract a set of PN fitting parameters. 

Before performing the fitting, we must first determine which quantities to compare, as we have two key elements: the gravitational wave strain $h$ and the Newman-Penrose scalar $\Psi_4$. Ref. \cite{Hinder:2008kv} provides the correct framework for fitting $\Psi_4$. However, many studies directly use $h$ when constructing hybrid waveforms that combine PN and NR data. In this study of eccentric waveforms, we explore fitting both $h$ and $\Psi_4$ to evaluate which one better aligns with the fitting expectations. For ease of comparison, we decompose both $h$ and $\Psi_4$ into their amplitude and phase components as follows: 
\begin{equation}\label{eq:55}
h^{\ell m}=\mathcal{A}^{\ell m}(t) \exp \left[i \Phi^{\ell m}(t)\right],
\end{equation}
\begin{equation}\label{eq:56}
\Psi_4^{\ell m}=A^{\ell m}(t) \exp \left[i \varphi^{\ell m}(t)\right],
\end{equation}
and the amplitude, phase and frequency of $h^{l m}$ can be obtained using the following equations:
\begin{equation}\label{eq:57}
\mathcal{A}^{\ell m}=\left|h^{\ell m}\right|,
\end{equation}
\begin{equation}\label{eq:58}
\Phi^{\ell m}=\mathrm{arg}(h^{\ell m}),
\end{equation}
\begin{equation}\label{eq:59}
\omega^{\ell m}=\frac{d \Phi^{\ell m}}{d t}.
\end{equation}
The same is true for $\Psi_4^{\ell m}$, we can get the amplitude, phase and frequency of it by
\begin{equation}\label{eq:60}
A^{\ell m}=\left|\Psi_4^{\ell m}\right|,
\end{equation}
\begin{equation}\label{eq:61}
\varphi^{\ell m}=\mathrm{arg}(\Psi_4^{\ell m}),
\end{equation}
\begin{equation}\label{eq:62}
\varpi^{\ell m}=\frac{d \varphi^{\ell m}}{d t}.
\end{equation}

Following the methodology outlined in Ref. \cite{Hinder:2008kv}, we use the gravitational wave frequency of the (2,2) mode as the parameter for PN fitting to NR waveforms, as it is a coordinate-invariant quantity and represents the dominant mode. A time interval $I = \left[t_1, t_2\right]$ is selected, and least squares fitting is applied to obtain the optimal initial parameters for the PN fit to the NR waveform. In this context, $t_1$ represents the time following the truncation of each waveform from the junk radiation, while $t_2$ denotes a time point before the merger. The specific $[t_1, t_2]$ values for each waveform are provided in the Table \ref{table:1} located in Appendix \ref{App:A}. As described in Sec. \ref{sec:II:A:1}, determining the strain $h^{22}$ requires first establishing the initial parameters ${e_t}_0$ and ${x}_0$ for ${e_t}$ and ${x}$, followed by the integration constant $l_0$ and the initial phase parameter $\Phi^{22}_0$. For $\Psi_4^{22}$, the same operation is performed after taking two time derivatives of $h$. In the context of NR data, the $\Psi_4^{22}$ component can either be directly sourced from the catalog or computed by taking the second time derivative of $h^{22}$. Given the significant data noise in the $\Psi_4^{22}$ dataset from the RIT catalog, this study adopts the latter approach, computing the second time derivative of $h^{22}$ to ensure data accuracy and reliability. We take $\Psi_4^{22}$ as an example to illustrate the fitting process. When performing frequency fitting only, the initial phase $\varphi^{22}_0$ need not be considered. To conduct the least squares fit, we minimize the residual 
\begin{equation}\label{eq:63}
Q\left(y_0\right) \equiv \frac{1}{N} \sum_{t \in I}\left[\varpi^{22}_{\mathrm{PN}}\left(t ; y_0\right)-\varpi^{22}_{\mathrm{NR}}(t)\right]^2, 
\end{equation}
where subscripts $\mathrm{PN}$ and $\mathrm{NR}$ indicate that the waveforms come from PN and NR, respectively, $N$ is the number of steps, and $y_0$ denotes the initial parameters: \begin{equation}\label{eq:64}
y_0 \equiv\left[x_0, {e_t}_0, l_0\right]. 
\end{equation}
We uniformly sample the time interval $[t_1, t_2]$ with 1000 points, denoted as $N=1000$, and subsequently employ Eq. (\ref{eq:63}) to globally fit the $\varpi^{22}$ from PN and NR waveforms. This fitting approach aims to enhance the consistency between the PN and NR waveforms to the fullest extent possible.
For clarity, we omit the superscript 22 for frequency, amplitude, and phase, assuming their dominance in the (2,2) mode. Once the optimal initial parameters for the frequency are obtained, least squares fitting is applied to match $\varphi_{\mathrm{PN}}$ and $\varphi_{\mathrm{NR}}$, determining the initial phase $\varphi_0$. According to Ref. \cite{Hinder:2008kv}, if the fitting interval exceeds at least $200M$, the result will stabilize, becoming independent of the interval length. Otherwise, errors can occur. Hence, the fitting interval during the inspiral phase should be as long as possible to obtain the most accurate initial parameters $y_0$.

\begin{figure*}[htbp!]
\centering
\includegraphics[width=15cm,height=15cm]{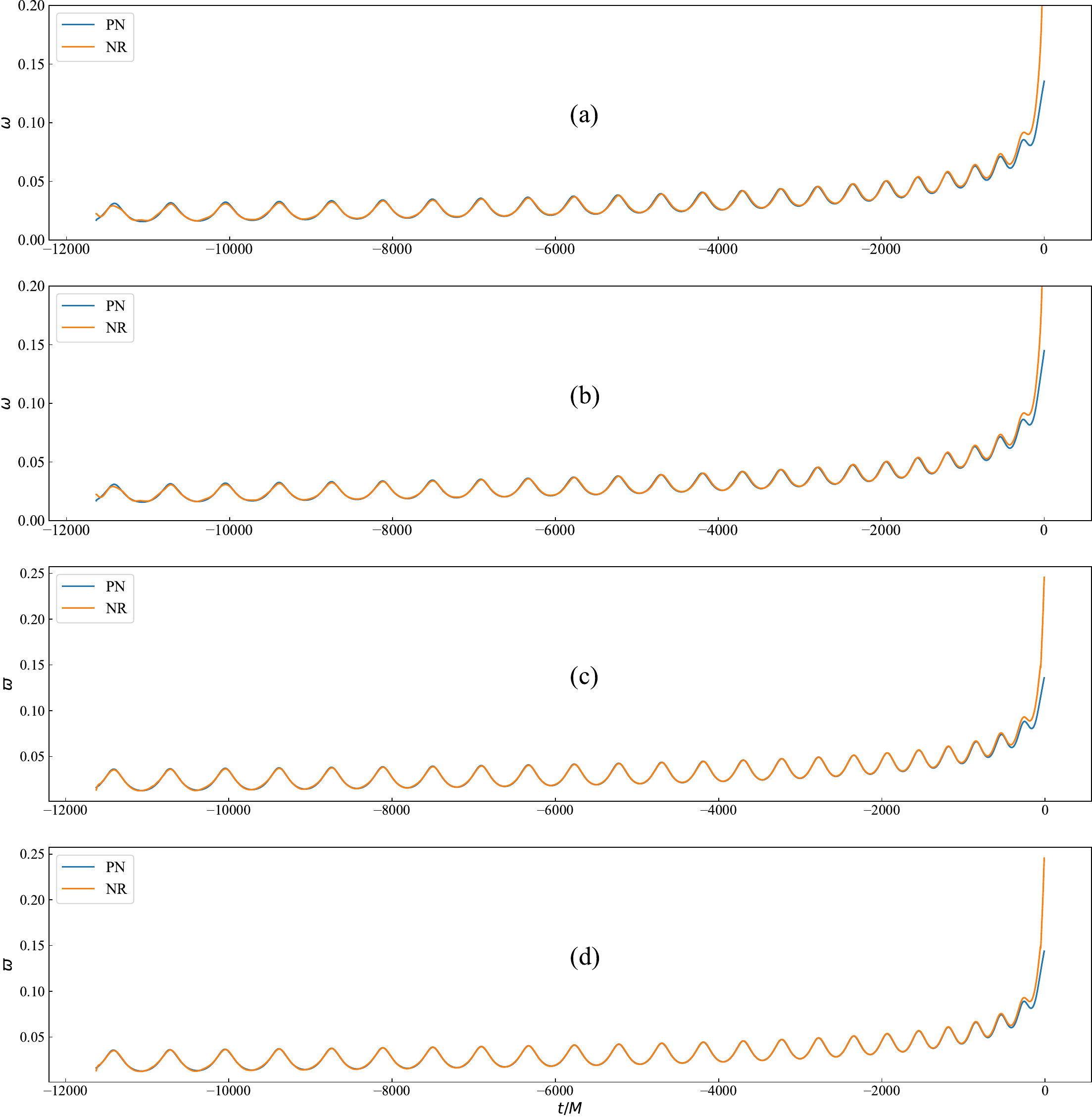}
\caption{\label{FIG:2}Frequency fitting of the (2,2) mode across four distinct cases for waveform RIT:eBBH:1282: the quadrupole moment of $h^{22}$ (panel (a)), the higher-order moment of $h^{22}$ (panel (b)), the quadrupole moment of $\Psi_4^{22}$ (panel (c)), and the higher-order moment of $\Psi_4^{22}$ (panel (d)).}
\end {figure*}

When performing the fitting of PN to NR waveforms, we encounter four distinct scenarios: the choice between fitting the frequency of $h^{22}$ or $\Psi_4^{22}$. Additionally, for both $h^{22}$ and $\Psi_4^{22}$, we can either fit the frequency of their quadrupole moment (Eq. (\ref{eq:18})) or consider the higher-order moments (Eq. (\ref{eq:20})). In many studies \cite{Chattaraj:2022tay, Huerta:2016rwp, Hinder:2017sxy, Huerta:2017kez}, the quadrupole moment of $h^{22}$ is emphasized as the primary component for constructing waveforms. We use the waveform RIT:eBBH:1282 as a case study to demonstrate the outcomes across various fitting scenarios. Notably, there is a scarcity of previous studies comparing extended PN and NR waveforms for eccentric orbits, making this work relatively novel and indicative of broader trends. The characteristics of other waveforms closely align with this exemplar. In FIG. \ref{FIG:2}, we examine the frequency fitting of the (2,2) mode across four different cases: the quadrupole moment of $h^{22}$ (panel (a) of FIG. \ref{FIG:2}), the higher-order moment of $h^{22}$ (panel (b) of FIG. \ref{FIG:2}), the quadrupole moment of $\Psi_4^{22}$ (panel (c) of FIG. \ref{FIG:2}), and the higher-order moment of $\Psi_4^{22}$ (panel (d) of FIG. \ref{FIG:2}). The waveform fitting interval extends from $100M$ after the waveform's onset to $1000M$ before the merger. As FIG. \ref{FIG:2} illustrates, in all four cases, we observe a strong agreement between PN and NR waveforms during the inspiral phase, with discrepancies only becoming apparent near the merger. This highlights the exceptional accuracy of PN methods in capturing the inspiral phase of extended waveforms, an observation not fully addressed in prior literature. The fitting results for the four cases, including residuals $Q$ (we use 1000 steps; the residuals show little dependence on the number of steps), $l_0$, $x_0$, and ${e_t}_0$, are presented in Table \ref{tab:I}. In accordance with the findings in Ref. \cite{Zlochower:2012fk}, the magnitude of the $Q$ value in this context aligns closely with the $L_2$ norm (approximately $10^{-6}$) of the Hamiltonian constraint within the bulk for sixth and eighth order finite difference implementations. Based on these results, the magnitude of the residuals is on the order of $10^{-7}$, indicating that all four cases provide a good fit to the NR waveform frequency. In fact, for shorter waveforms and lower eccentricities, the residual $Q$ can be one to two orders of magnitude smaller. The similarity in residuals in Table \ref{tab:I} suggests that any of the four cases will yield accurate results. However, among the four cases, (d) consistently has the smallest residual, a trend that applies across all other waveforms after extensive fitting. It is important to note that there are non-physical fluctuations, representing junk radiation in the early part of the frequency of $h^{22}$ (as seen in panels (a) and (b) of FIG. \ref{FIG:2}). These fluctuations necessitate global fitting for $h^{22}$, which means using as much of the waveform as possible, including the region close to the merger, rather than only part of the waveform. These non-physical fluctuations are not present in $\Psi_4^{22}$. For parameters such as $l_0$, $x_0$, and ${e_t}_0$, while we have stringent accuracy requirements for waveform fitting parameters (as will be explained later), the fitting results across the four cases do not maintain a consistent level of accuracy. As shown in Table \ref{tab:I}, $l_0$ can only be accurate to 0.1, $x_0$ to 0.001, and ${e_t}_0$ to 0.01 across the four cases. We attribute this to numerical errors in the NR waveform and the differences between the higher-order and quadrupole moments in the PN expansion. There is no absolute right or wrong in these four scenarios, and any of them can be reasonably chosen as the fitting basis. For these reasons, we select the frequency of the 22 mode from the higher-order moment of $\Psi_4^{22}$ as the fitting target. 
\begin{table}[!ht]
\caption{\label{tab:I}Fitting results of the four cases in FIG. \ref{FIG:2}, including the residuals $Q$ (we set the number of steps to 1000, the residual results have little dependence on the number of steps), $l_0$, $x_0$, ${e_t}_0$.}
    \centering
    \begin{tabular}{|l|l|l|l|l|}
    \hline
        ~ & $Q$ & $l_0$ & $x_0$ & ${e_t}_0$  \\ \hline
        (a) & $3.74 \times 10^{-7}$ & 4.27440580  & 0.04831607  & 0.19214287   \\ \hline
        (b) & $2.40 \times 10^{-7}$ & 4.31218357  & 0.04826974  & 0.19500954   \\ \hline
        (c) & $1.51 \times 10^{-7}$ & 4.26107246  & 0.04828178  & 0.19464287   \\ \hline
        (d) & $1.17 \times 10^{-7}$ & 4.29085024  & 0.04824280  & 0.19711954  \\ \hline
    \end{tabular}
\end{table}

\begin{figure*}[htbp!]
\centering
\includegraphics[width=15cm,height=15cm]{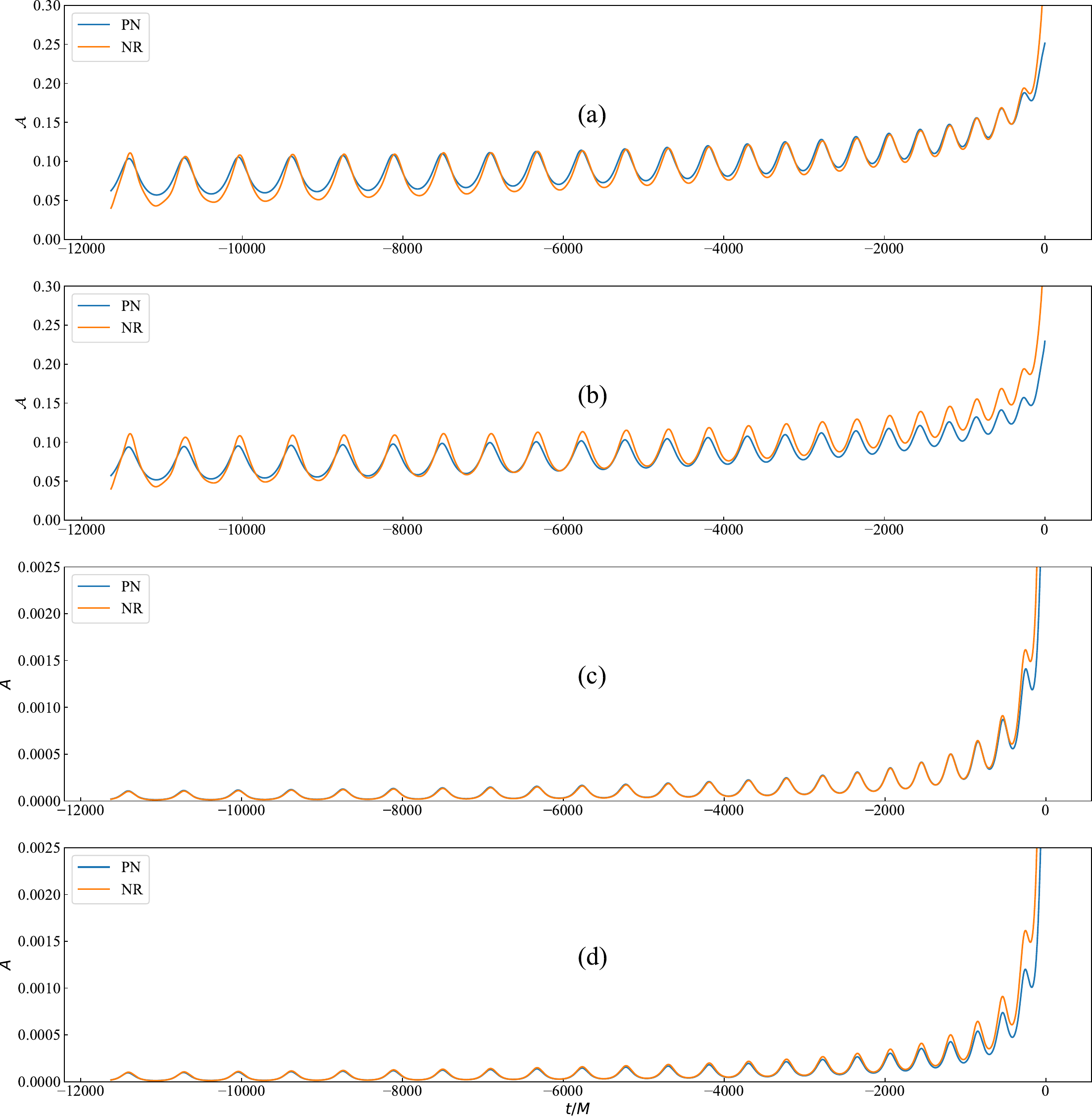}
\caption{\label{FIG:3}The amplitude of the waveform RIT:eBBH:1282 in the four cases of FIG. \ref{FIG:2}: the quadrupole moment of $h^{22}$ (panel (a)), the higher-order moment of $h^{22}$ (panel (b)), the quadrupole moment of $\Psi_4^{22}$ (panel (c)), and the higher-order moment of $\Psi_4^{22}$ (panel (d)).}
\end {figure*}

Following the discussion on frequency, it is crucial to address another key aspect of the waveform: the amplitude. Unlike frequency, amplitude is not necessarily gauge-invariant, which can cause it to behave differently. In FIG. \ref{FIG:3}, we present the amplitude of the waveform RIT:eBBH:1282, corresponding to the four cases in FIG. \ref{FIG:2}: the quadrupole moment of $h^{22}$ (panel (a)), the higher-order moment of $h^{22}$ (panel (b)), the quadrupole moment of $\Psi_4^{22}$ (panel (c)), and the higher-order moment of $\Psi_4^{22}$ (panel (d)). In panel (a) of FIG. \ref{FIG:3}, we observe that the amplitude agrees well with the NR waveform's amplitude during a brief period prior to the merger, with increasing deviation as we move further from the merger. Panel (b) shows that the amplitude of the PN waveform deviates significantly from the NR amplitude throughout the entire time period. Panel (c) reveals the best agreement between the amplitude of PN and NR waveforms, with nearly no deviation in the inspiral phase and only deviations near the merger, consistent with the well-known limitations of PN approximations near the merger phase. In panel (d), the amplitude exhibits some consistency with the NR waveform during the inspiral, but visible deviations occur and grow larger as the merger approaches. From FIG. \ref{FIG:3}, we conclude that the amplitude of the quadrupole moment of $\Psi_4^{22}$ shows the best agreement with the NR waveform, while the other cases exhibit varying degrees of deviation. These amplitude deviations are not unique to waveform RIT:eBBH:1282 but are present across all waveforms, particularly those that are relatively long (over $3000M$). This phenomenon occurs in both RIT and SXS waveforms. This observation supports the approach taken in many previous studies that rely solely on the quadrupole moment as a waveform approximation. While this method may work for short waveforms, it becomes less effective for longer waveforms.

\begin{figure*}[htbp!]
\centering
\includegraphics[width=15cm,height=7.5cm]{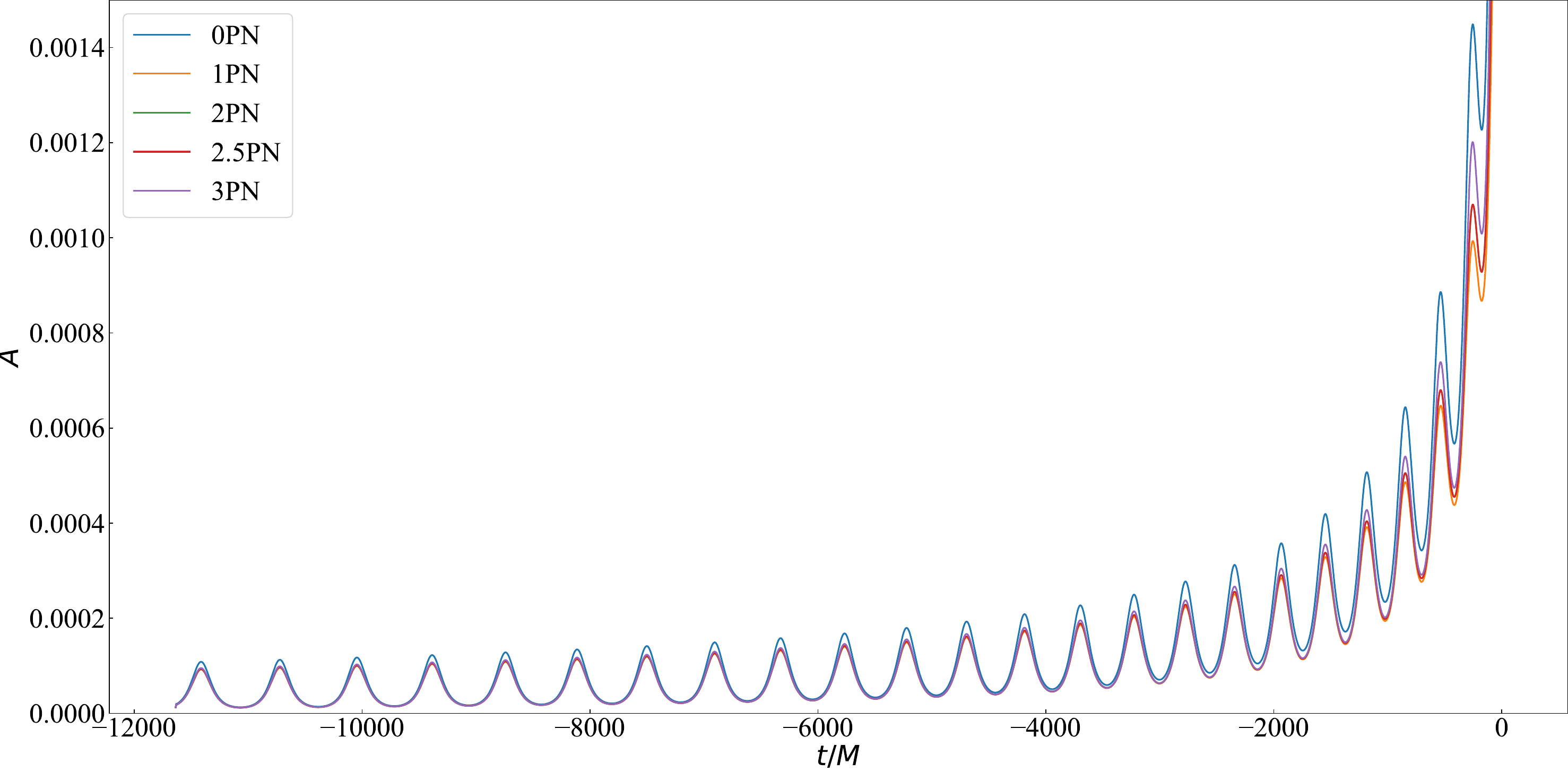}
\caption{\label{FIG:4}Various PN orders amplitudes of Eq. (\ref{eq:21}) for waveform RIT:eBBH:1282. For instance, selecting the 2PN order entails the inclusion of the preceding Newtonian order, the 1PN order, and so forth.}
\end {figure*}

Our goal here is not to investigate the underlying causes of these discrepancies but to focus on their phenomenological implications. As such, this paper will not delve into the amplitude of \(h^{22}\) due to its significant deviations. Instead, we concentrate on analyzing the amplitude of \(\Psi_4^{22}\) across different PN orders. In FIG. \ref{FIG:4}, we present waveform RIT:eBBH:1282, computed at various PN orders of Eq. (\ref{eq:21}), to illustrate the contributions of different PN approximations. For instance, selecting the 2PN order includes the preceding Newtonian, 1PN, and other relevant terms. The results in FIG. \ref{FIG:4} reveal distinct contributions from various PN orders to the amplitude, particularly in the discrepancies that emerge near the merger. During the inspiral phase, the Newtonian order contributes most significantly to the amplitude, surpassing the contributions of higher-order PN terms. As noted earlier, the close agreement between the Newtonian order's amplitude and that of the NR waveform reinforces their consistency. In contrast, the amplitudes associated with higher PN orders exhibit varying degrees of deviation from the NR amplitude. These findings indicate that within the range of 3PN order, the Newtonian term provides the most accurate amplitude results. This observation also applies to higher harmonic modes, a point we will expand upon in later sections of this paper. 

There is a vast array of RIT and SXS waveforms available for analysis. A total of 180 sets of waveforms have been carefully selected for direct fitting, covering both nonspinning and spin-aligned configurations. Consequently, this study necessitates a substantial amount of waveform fitting.
For each waveform, three free parameters $y_0 \equiv [x_0, {e_t}_0, l_0]$ are essential for fitting, with each parameter requiring precise determination within specified accuracy bounds. ${e_t}_0$ must be determined with a precision of at least 0.001, $x_0$ must be resolved to a minimum of 0.0001, and $l_0$ must be ascertained with an accuracy threshold of at least 0.01 to ensure the required precision. This precision is directly linked to the sensitivity of parameters in the PN fitting process, with $x_0$ exhibiting the highest sensitivity, followed by ${e_t}_0$, and $l_0$ at a lower sensitivity level. These observations have been derived from extensive fitting exercises conducted in our research.
The utilization of the standard parameter range of $y_0$ for NR waveform fitting often results in extensive computations and issues, frequently leading to infinite or invalid values under the square root. To optimize computational efficiency, we have developed some techniques to constrain parameter ranges effectively.

In light of the fitting process for waveform RIT:eBBH:1282, it is crucial to revisit FIG. \ref{FIG:1} in Sec. \ref{sec:II:B} to understand why only 180 sets of waveforms are retained for fitting, instead of encompassing all the eccentric waveforms from RIT. The waveforms chosen for fitting must possess sufficient length. While we previously indicated that a correctly fitted waveform should have a minimum duration of $200M$, practical fitting requires a waveform duration exceeding $350M$ in NR simulations to achieve an accurate match. Furthermore, the waveform frequency should exhibit at least two peaks, signifying two periastron passages. This criterion leads to the exclusion of numerous highly eccentric waveforms from RIT. Accommodating a highly eccentric waveform necessitates substantial initial separation and heightened initial eccentricity in NR simulations.

\begin{figure*}[htbp!]
\centering
\includegraphics[width=15cm,height=7.5cm]{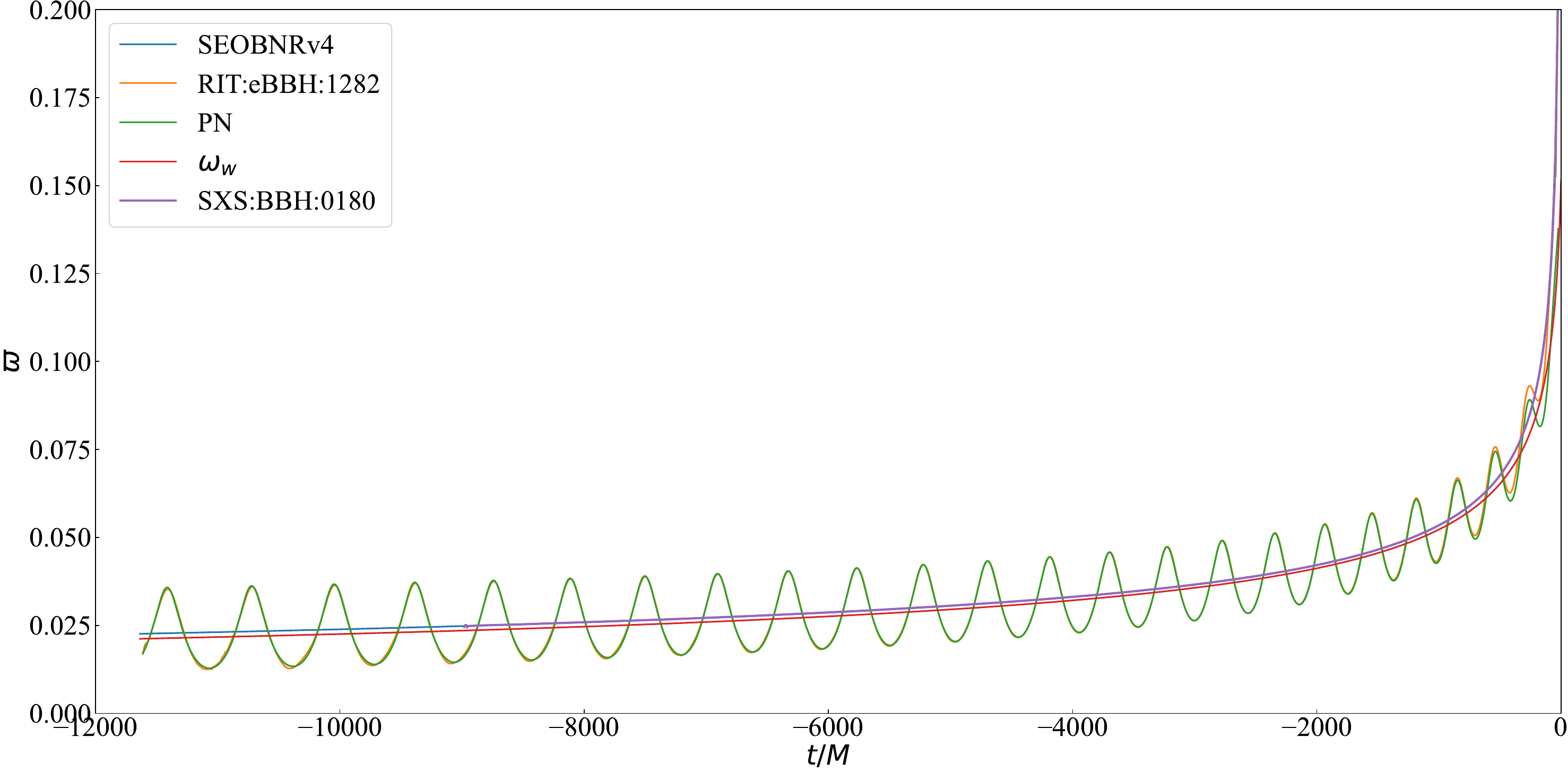}
\caption{\label{FIG:5} Frequency of the waveform RIT:eBBH:1282 and its PN fitting, average orbital frequency $\omega_w$, alongside a corresponding circular orbit waveform with $q=1$ generated by the SEOBNRv4 code in Pycbc. To further validate the accuracy of the SEOBNRv4 waveform, we also include the circular orbit waveform SXS:BBH:0180.}
\end {figure*}

In the process of determining $x_0$, directly setting a fitting interval can introduce challenges in precisely defining the range, potentially leading to issues such as encountering infinite values during fitting and obtaining imprecise estimates. To overcome these challenges and enhance precision in derivation, we propose approximating the evolution of $x(t)$ based on the frequency evolution of circular orbit waveforms with identical mass ratios during the fitting of NR waveforms. The relationship $x = (M {\omega}_o)^{2/3} = (M {\omega}_w/2)^{2/3} \approx (M {\omega}_c/2)^{2/3}$ provides a means to derive $x$ and, conversely, ${\omega}_o$ (the orbital average frequency), ${\omega}_w$ (waveform average frequency), and ${\omega}_c$ (circular orbit waveform frequency).
While slight differences exist between $\omega_{w}$ and the circular orbit frequency $\omega_c$, the approximation remains accurate within a margin of $\pm0.001$ (for RIT:eBBH:1282) throughout the waveform's frequency evolution. As a result, the evolution of $x$ can be approximated using the frequency of circular orbit waveforms. It is important to note that as the waveform length increases and eccentricity gradually rises, the effectiveness of this approximation diminishes, necessitating an expansion of the parameter $x$ guess range.
In FIG. \ref{FIG:5}, we illustrate this phenomenon using the waveform RIT:eBBH:1282, spanning approximately $12000M$, alongside a corresponding circular orbit waveform with $q=1$ generated by the SEOBNRv4 code in Pycbc \cite{Biwer:2018osg}. This comparison underscores the constraints of the parameter $x$ guess range. FIG. \ref{FIG:5} showcases the RIT waveform and its PN fit, along with the average frequency $\omega_w$ derived from $x$, and the SEOBNRv4 waveform for a mass ratio of $q=1$ from Pycbc. To further validate the accuracy of the SEOBNRv4 waveform, we also include the circular orbit waveform SXS:BBH:0180, albeit shorter in duration than SEOBNRv4, provided here solely for comparison. When utilizing the SEOBNRv4 waveforms from Pycbc, we convert their units from the International System of Units (SI) to the Natural System of Units.
As depicted in FIG. \ref{FIG:5}, although the frequency of the circular orbit waveform and the orbit-averaged frequency for the same mass ratio are not perfectly aligned, the discrepancy is minimal. This slight distinction is why some studies \cite{Mroue:2010re,Setyawati:2021gom} directly utilize the circular orbit waveform as an approximation for the average frequency of eccentric waveforms.
The methodology of estimating the orbital average frequency using a circular orbit waveform is pertinent not only for nonspinning waveforms but also for spin-aligned waveforms, provided waveforms with the same mass ratio and spin characteristics are employed, as detailed in our prior study \cite{Wang:2023ueg}.

Moreover, in the case of ${e_t}_0$, it is not feasible to simply assign a range of [0, 1] due to the potential occurrence of infinite or negative values under the square root at high eccentricities in the process of PN fitting. Both RIT and SXS have developed proprietary methods for assessing eccentricity within their datasets, providing a valuable reference point. By leveraging these eccentricities, typically falling within a range of approximately $\pm 0.1$, we can establish an initial guess range for eccentricity. This strategy helps alleviate issues linked to unbounded ranges and enhances the robustness of the fitting process.

Additionally, for $l_0$, we can consider the range [0, 6.5], recognizing that due to periastron precession, the value of $l_0$ may exceed $2\pi$. Furthermore, we can refine the range of $l_0$ based on the variation in frequency $\varpi$ at periastron and apastron.

When determining the initial guess parameter range, a strategy akin to Newton's bisection method can be utilized. Initially, we can select 10 initial guess parameters, for example, ${e_t}_0 \in [0.1, 0.2, 10]$, where 10 denotes the number of eccentricity divisions, and progressively fine-tune the guess range for the initial parameters. Following an iteration, this process might yield ${e_t}_0 \in [0.19, 0.20, 10]$. By iteratively adjusting the guess parameter range three or four times, we can derive initial parameters with sufficient accuracy while significantly reducing computational expenses. This iterative approach can be similarly applied to all three parameters: ${e_t}_0$, $l_0$, $x_0$. Once these parameters are determined, we utilize the least squares fitting method to determine the initial phase $\varphi_0$, thereby fully defining all PN parameters. The results of fitting 180 sets of eccentric waveforms from RIT and SXS are elaborated in Table \ref{table:1} in Appendix \ref{App:B}.

\subsection{Different methods of measuring initial eccentricity}\label{sec:II:D}
Upon determining the PN fitting parameter ${e_t}_0$ from the NR waveform, we establish a method to quantify the eccentricity of the NR waveform. The accuracy of this approach relies on the radiation and conservation dynamics at the 3PN order. It is noteworthy that within general relativity, a gauge-invariant definition of eccentricity is lacking; nevertheless, various references \cite{Mroue:2010re,Healy:2017zqj,Ireland:2019tao,Ramos-Buades:2018azo} have introduced diverse methodologies to gauge the orbital eccentricity of BBH in eccentric orbits.
In NR simulations, two primary techniques are utilized to determine the orbital eccentricity of BBH systems. One method involves analyzing the waveform's features, such as frequency oscillations, phase variations, and amplitude changes, to infer the eccentricity. However, this method is constrained in capturing only local waveform oscillations, primarily at lower PN and Newtonian orders, thus leading to notable errors in eccentricity estimation. Conversely, the second method relies on assessing the orbital energy and angular momentum to ascertain the initial orbital eccentricity of the BBH system.
In this study, we concentrate solely on the latter method of eccentricity calculation, which entails deriving the initial eccentricity of a BBH system through its orbital energy and angular momentum. By contrasting the PN fit parameter ${e_t}_0$ with the initial eccentricities obtained from this method and the existing initial eccentricity data from the RIT and SXS catalogs, we aim to discern any discrepancies between them.

For nonspinning systems, following the methodology outlined in Refs. \cite{Wang:2023vka} and \cite{Radia:2021hjs}, we utilize the generalized 3PN quasi-Keplerian parameterization to estimate the initial eccentricity.
The initial eccentricity ${e_t}_0$ can be determined utilizing the expression given in Eq. (25e) of Ref. \cite{Memmesheimer:2004cv} under harmonic coordinates. For conciseness, Eq. (25e) is provided in Appendix \ref{App:A}. The key parameters necessary for computing eccentricity in Eq. (25e) are the initial binding energy $E_{\mathrm{b}}$ and the initial angular momentum $L$, both of which are available in the metadata of the RIT and SXS catalogs. The binding energy can be calculated as
\begin{equation}\label{eq:65}
E_{\mathrm{b}}=M_{\mathrm{ADM}}-M,
\end{equation}
where $M_{\mathrm{ADM}}$ denotes the ADM mass. 

For spin-aligned systems, we utilize Eq. (75) from Ref. \cite{Tessmer:2010hp}, detailed in Appendix \ref{App:A}, which provides an expression precise up to the 2PN order in ADM coordinates.
While this equation is accurate up to the 2PN order in ADM coordinates, the similarity of eccentricity values computed in ADM coordinates using the Eq. (75) with those determined in harmonic coordinates based on the fitting outcomes of the nonspinning waveform from Ref. \cite{Wang:2023vka} and the PN fitting in this study indicates minimal discrepancies. The difference between the 2PN and 3PN orders is also slight, with errors in both cases being only 0.001, mirroring the precision of our eccentricity estimation through PN fitting. Notably, Ref. \cite{Tessmer:2012xr} offers a calculation of the eccentricity ${e_t}_0$ at the 3PN order; however, due to potential inaccuracies in its calculation formula, achieving the correct result is unattainable, thereby rendering its findings unsuitable for this study.

Both the SXS and RIT catalogs present their independently measured eccentricities; however, the precision of these values remains uncertain due to the differing methodologies employed by each catalog. The eccentricity measurement technique introduced by RIT, elaborated in Ref. \cite{Healy:2022wdn}, is notably straightforward, with its accuracy extensively scrutinized therein.
The process of generating eccentric waveforms and measuring eccentricity unfolds as follows: A novel parameter $\epsilon$ within the range of 0 to 1 is introduced, adjusting the tangential linear momentum as $p_t = p_{t,qc}(1-\epsilon)$, where $p_{t,qc}$ denotes the tangential linear momentum in a quasicircular orbit. In this approach, the initial positions of the binary black hole remain fixed at apastron, while the initial orbital eccentricity gradually increases during the simulations, transitioning from a quasi-circular orbit ($e=0$) towards the head-on collision limit ($e=1$). The corresponding initial orbital frequency (including the 22-mode of gravitational waves) decreases by the factor ${\omega}_{o,e} = {\omega_{o,qc}} (1-\epsilon)$. Consequently, the initial eccentricity of the orbit can be approximated as $e=2\epsilon-\epsilon^2$, providing a second-order approximation with respect to $\epsilon$.
The SXS catalog does not explicitly outline its eccentricity measurement method in Refs. \cite{Mroue:2013xna,Boyle:2019kee}.

\section{results}\label{sec:III}

\subsection{Fitting residuals} \label{sec:III:A}
The residual fitting between PN and NR data can potentially be influenced by various factors such as mass ratio, spin, eccentricity, waveform length, and more. Our fitting insights suggest the following outcome: The fitting residual is primarily linked to the eccentricity of the system, with mass ratio or spin playing a secondary role, and showing no significant correlation with other factors. A decrease in eccentricity results in a reduced fitting residual, while an increase in eccentricity leads to an increased fitting residual. This implies that a very long waveform with low eccentricity, like RIT:eBBH:1282, exhibits a small fitting residual due to the consistency between PN and NR data. On the other hand, an increase in the fitting residual indicates a decrease in the accuracy of PN approximations compared to NR. This loss in accuracy can be attributed to the heightened impact of strong-field effects in NR simulations as eccentricity increases, exacerbating the limitations of PN approximations.

\begin{figure*}[htbp!]
\centering
\includegraphics[width=15cm,height=12cm]{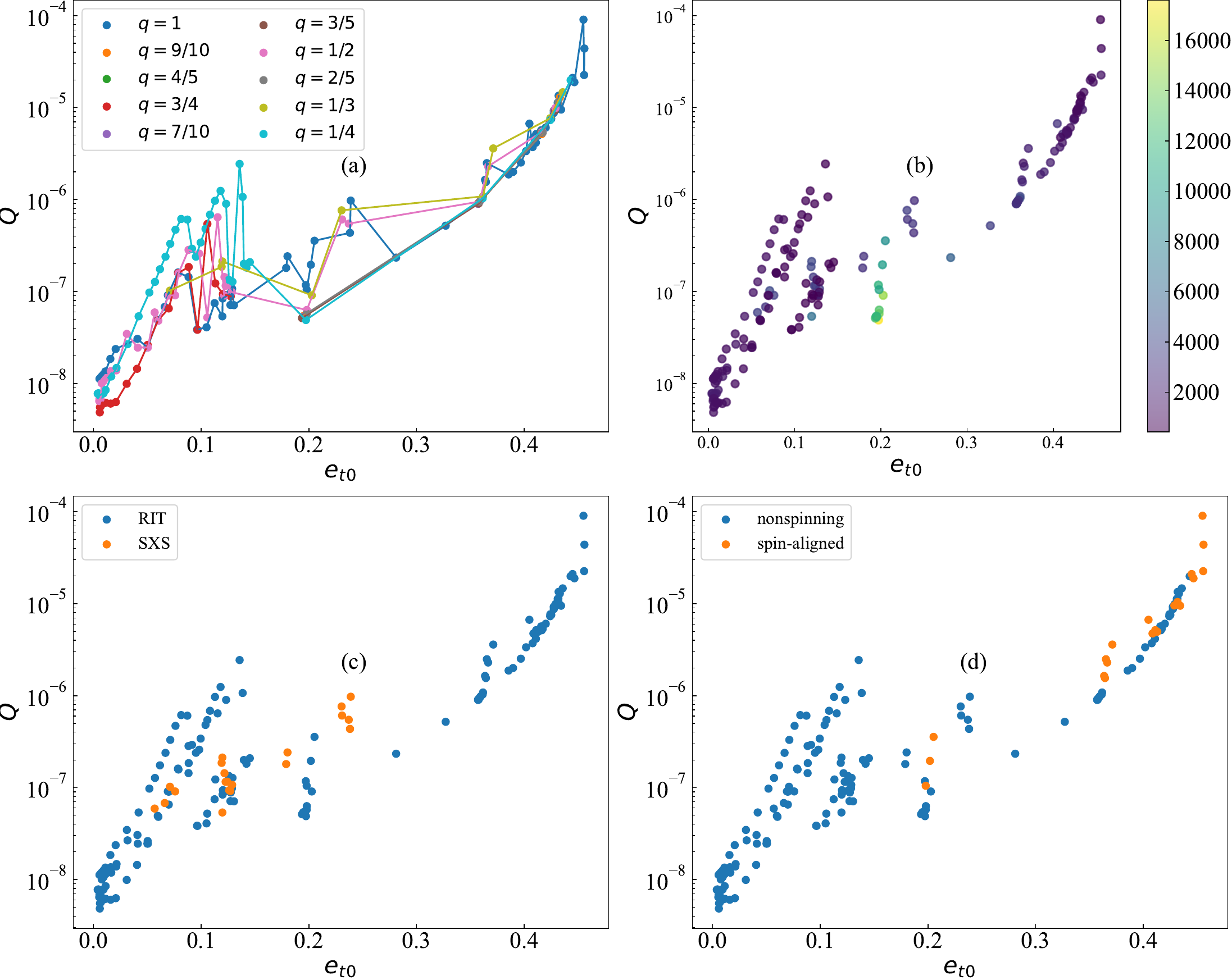}
\caption{\label{FIG:6}Residuals obtained from PN fitting of all the eccentric waveforms of RIT and SXS in FIG. \ref{FIG:1}, encompassing waveforms with varying mass ratios, different eccentricities, different waveform lengths and both nonspinning and spin-aligned configurations.}
\end {figure*}

In FIG. \ref{FIG:6}, we present the residuals from PN fitting of all the eccentric waveforms from RIT and SXS shown in FIG. \ref{FIG:1}, encompassing waveforms with varying mass ratios, different eccentricities, and both nonspinning and spin-aligned configurations. In FIG. \ref{FIG:6}, although the graphs appear in four distinct panels, they illustrate the same data, with variations reflecting the different research contexts. 

In panel (a) of FIG. \ref{FIG:6}, we illustrate the variation of the fitting residual with the mass ratio $q$. It may seem puzzling why the fitting residual correlates with the mass ratio, as the accuracy of the waveform ideally should remain unaffected by the mass ratio, whether in PN or NR simulations. Historically, NR simulated waveforms have been regarded as the most accurate compared to PN and EOB approximations. However, NR simulations may not be entirely error-free and could contain distinct inaccuracies, such as numerical errors.
In panel (a), it becomes evident that as the mass ratio decreases, so does the fitting residual. This trend holds true for both the RIT and SXS catalogs. For specific values, please consult Appendix \ref{App:A}. Essentially, the software codes EinsteinToolkit and SpEC, commonly used for numerical simulations, exhibit the highest accuracy when the mass ratio is unity during the simulation of BBH mergers in eccentric orbits. Conversely, as the mass ratio decreases, the margin of error widens.
In NR simulations, a lower mass ratio results in a longer waveform under identical initial conditions, necessitating a finer numerical simulation grid for increased accuracy. The increased fitting residual in panel (a) suggests that NR waveforms incur a more significant deviation relative to PN waveforms in scenarios involving small mass ratios, as supported by findings in Ref. \cite{Habib:2019cui}, which also conducts a comparison of PN and NR involving small mass ratios. In Sec. \ref{sec:III:E}, we will analyze waveforms with small mass ratios from RIT as case studies to elucidate the potential challenges and implications of PN fitting in scenarios with small mass ratios.

In panel (b) of FIG. \ref{FIG:6}, we have depicted the impact of waveform length on the PN fitting residuals. The color bar in panel (b) represents the duration of the waveform, spanning from the initial time to the merger point. One might expect that the PN fitting residual correlates with the length of the waveform. Surprisingly, this is not the case. It is clear that the length of the waveform has minimal influence on the fitting residuals. The majority of waveforms we analyze are short, typically ranging from approximately 1000M to 2000M in length. Some exhibit significant eccentricities, resulting in larger fitting residuals, while others with lower eccentricities yield smaller residuals. Among a few elongated waveforms, the longest extends beyond 16000M. Despite this extended duration, its fitting residual remains moderate due to its intermediate eccentricity level. This observation highlights that PN fitting residuals are not tied to waveform length, or any correlation is very weak.

In panel (c) of FIG. \ref{FIG:6}, we showcase the impact of different catalogs on PN fitting residuals. The trends in the fitting residuals of RIT and SXS are notably similar, suggesting that distinct catalogs or computational codes have minimal effects on PN fitting. RIT and SXS have previously cross-validated their simulated waveforms in earlier studies, leading to a high level of agreement between them.

In panel (d) of FIG. \ref{FIG:6}, we explore the effect of spin alignment and no spin on the PN fitting residual. The scatter points in both scenarios display a similar overall trend, with a slight elevation observed in the scatter points for spin alignment compared to those without spin. This discrepancy can be attributed to the influence of spin on waveform characteristics. The noted difference does not substantially affect the upward trend of scatters in relation to eccentricity throughout the figure, as there is no significant divergence between the trend of spin-aligned scatters and that of nonspinning scatters. Further investigations involving a broader range of spin-aligned eccentric waveforms will provide a more thorough analysis in future research. Thus, it can be concluded that the presence or absence of nonprecessing spin has a negligible effect on the PN fitting residual. For precise values, please consult the detailed PN fitting data in Appendix \ref{App:B}.

In summary, the analysis of FIG. \ref{FIG:6} leads us to the conclusion that the PN fitting residual is primarily linked to the eccentricity of the waveforms. A clear trend emerges indicating that as eccentricity increases, so does the fitting residual. This increase in the fitting residual indicates a decrease in the accuracy of PN approximations compared to NR. This decline can be attributed to the heightened impact of strong-field effects in NR simulations as eccentricity rises, exacerbating the limitations of the PN approximation. While the mass ratio does have a minor influence on the fitting residual, this effect may be mainly attributed to the intrinsic limitations of the NR or PN simulations for small mass ratios. Additionally, our findings suggest that variations in numerical simulation codes and spin configurations do not significantly impact the fitting residual.

\subsection{Initial eccentricity} \label{sec:III:B}
In the preceding Section \ref{sec:II:D}, we introduced the nonspinning quasi-Keplerian parameterization at 3PN orders and the spin-aligned quasi-Keplerian parameterization at 2PN orders for quantifying the initial eccentricity based on the initial energy and orbital angular momentum. Furthermore, we emphasized that both RIT and SXS have autonomously derived their individual initial eccentricity values.

As previously discussed, there is no definitive standard for measuring eccentricity, leading to the absence of a universally accepted method. Consequently, it is difficult to claim the superiority of any specific approach for determining eccentricity. Generally, as eccentricity increases, so does the margin of difference in its measurement. In this study, we focus on evaluating the relative discrepancies between two eccentricity measurement methods and the PN fitting eccentricity, \({e_t}_0\). The initial eccentricity, determined using the quasi-Keplerian parameterization based on initial energy and orbital angular momentum, is denoted as \({{{e_t}_{0}}_{\mathrm{KP}}}\). The difference of it with respect to \({e_t}_0\) is expressed as: 
\begin{equation}\label{eq:66}
{\varepsilon}_{\mathrm{KP}}= |{{{e_t}_{0}}_{\mathrm{KP}}}-{e_t}_0|. 
\end{equation}
We chose not to calculate the relative difference by dividing by \({e_t}_0\) because of the potential for significant discrepancies between \({{{e_t}_{0}}_{\mathrm{KP}}}\) and \({e_t}_0\), which could yield large numerical differences that might obscure meaningful interpretation in a plot. For detailed numerical values of \({{{e_t}_{0}}_{\mathrm{KP}}}\) and \({e_t}_0\), please refer to Appendix \ref{App:B}. The initial eccentricity measured by the RIT catalog is denoted as \({{{e_t}_{0}}_{\mathrm{RIT}}}\), and the difference compared to \({e_t}_0\) is given by: \begin{equation}\label{eq:67}
{\varepsilon}_{\mathrm{RIT}}= |{{{e_t}_{0}}_{\mathrm{RIT}}}-{e_t}_0|. 
\end{equation}
Similarly, the initial eccentricity measured by the SXS catalog is denoted as \({{{e_t}_{0}}_{\mathrm{SXS}}}\), with the difference relative to \({e_t}_0\) defined as: \begin{equation}\label{eq:68}
{\varepsilon}_{\mathrm{SXS}}= |{{{e_t}_{0}}_{\mathrm{SXS}}}-{e_t}_0|. 
\end{equation}

\begin{figure*}[htbp!]
\centering
\includegraphics[width=15cm,height=6.5cm]{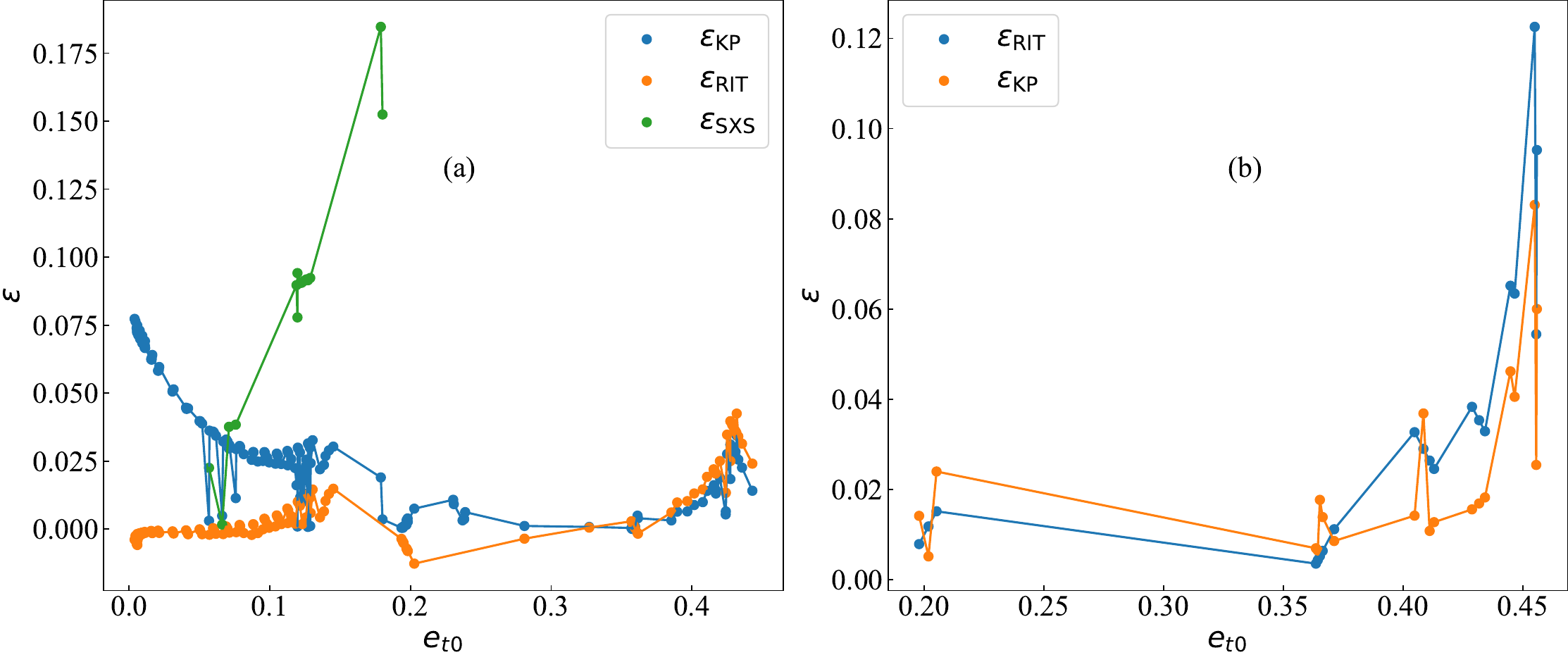}
\caption{\label{FIG:7}Differences between the eccentricities determined by quasi-Keplerian parameterization, RIT and SXS catalogs and those obtained through PN fitting. Panel (a) in corresponds to the nonspinning configuration, while panel (b) pertains to the spin-aligned configuration.}
\end {figure*}

In FIG. \ref{FIG:7}, we present the differences between the eccentricities determined by the two aforementioned methods and those obtained through PN fitting. Panel (a) corresponds to the nonspinning configuration, while panel (b) pertains to the spin-aligned configuration. 

From panel (a), which focuses on the nonspinning configuration, we observe that the differences of all three measurement methods increase with eccentricity \({e_t}_0\). Among these methods, the eccentricity measured by RIT, \({{e_t}_0}_{\mathrm{RIT}}\), exhibits the closest agreement with PN fitting, followed by the quasi-Keplerian parameterization, \({{e_t}_0}_{\mathrm{KP}}\), while the SXS measurements, \({{e_t}_0}_{\mathrm{SXS}}\), show the largest deviations. Detailed datas from the RIT catalog and Appendix \ref{App:B} reveal that the RIT eccentricity measurements exhibit smooth, continuous behavior without abrupt fluctuations, maintaining close alignment with PN fitting results across a wide range of eccentricities \cite{RITBBH}. For \({{{e_t}_{0}}_{\mathrm{RIT}}}\), the difference with respect to PN fitting remains within 0.025 for low and moderate eccentricities (0-0.4) and within 0.05 for intermediate eccentricities (0.4-0.45). In contrast, measurements using the \({{e_t}_0}_{\mathrm{KP}}\) method display significant differences at very low eccentricities, a known limitation of the approach \cite{Sperhake:2007gu}. This method may produce eccentricity that deviate by more than 1 at both very low (around 0.01) and very high eccentricities (around 0.9), making it unreliable in these extreme cases. For \({{e_t}_0}_{\mathrm{KP}}\), the difference relative to PN fitting in the low eccentricity range (0-0.2) is approximately 0.025-0.075, though in other eccentricity ranges, the \({{e_t}_0}_{\mathrm{KP}}\) results align more closely with RIT and PN measurements. Regarding the SXS measurements, \({{{e_t}_{0}}_{\mathrm{SXS}}}\), the difference for low eccentricities (0-0.1) ranges between 0 and 0.04, but for higher eccentricities, the differences become too significant to be considered reliable. Moreover, the SXS catalog lacks complete measurement data for some waveforms, further contributing to the uncertainties in these assessments. 

In panel (b), which examines the spin-aligned configuration, the trends in differences mirror those observed for the nonspinning case. However, the differences are more pronounced for medium eccentricities in the spin-aligned configuration. Specifically, for \({{e_t}_0}_{\mathrm{KP}}\), the maximum difference exceeds 0.08, while for \({{e_t}_0}_{\mathrm{RIT}}\), the maximum difference surpasses 0.12. These findings highlight an increasing lack of reliability in all three measurement methods as spin effects become more pronounced, further underscoring the influence of spin on eccentricity assessments. 

\subsection{Dominant mode}\label{sec:III:C}
In this section, we present the PN fitting results for the dominant (2,2) mode, with higher modes discussed in the following section. We use three sets of nonspinning and three sets of spin-aligned waveforms with increasing initial eccentricity to illustrate the PN fitting results for frequency, phase, and amplitude. A noteworthy feature of these waveform sets is that they differ only in their initial eccentricities, while all other parameters, such as mass ratio, initial separation, and spin remain identical and constant. The six waveforms are nonspinning cases RIT:eBBH:1330, RIT:eBBH:1331, RIT:eBBH:1332, and spin-aligned cases RIT:eBBH:1899, RIT:eBBH:1900, RIT:eBBH:1901. Detailed fitting parameters for these waveforms can be found in Appendix \ref{App:A}.

\begin{figure*}[htbp!]
\centering
\includegraphics[width=15cm,height=21cm]{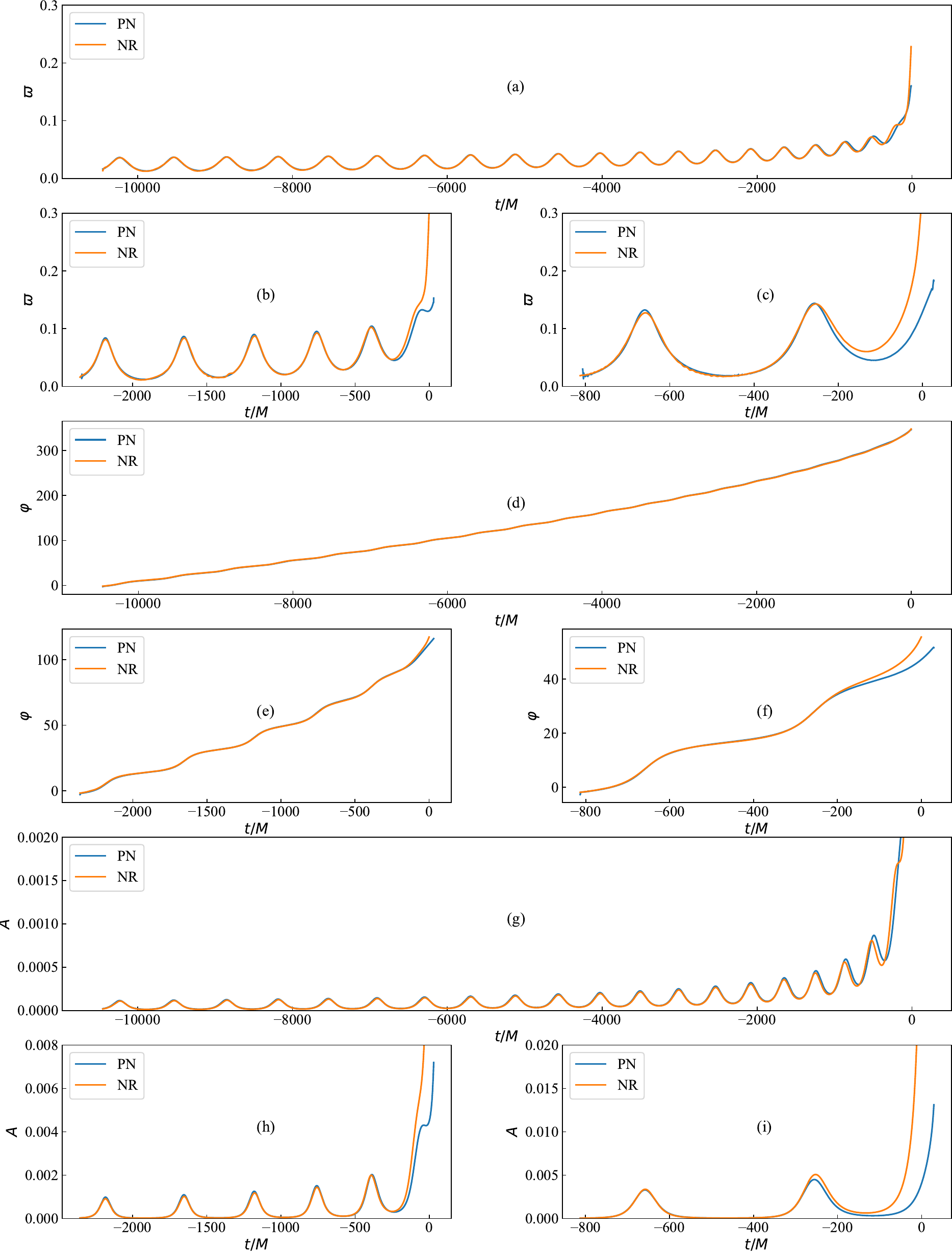}
\caption{\label{FIG:9}Frequency, phase, and leading Newtonian order amplitude of three sets of spin-aligned waveforms RIT:eBBH:1899 (panel (a), (d), (g)), RIT:eBBH:1900 (panel (b), (e), (h)), RIT:eBBH:1901 (panel (c), (f), (i)), also accompanied by their PN fitting.}
\end {figure*}

In FIG. \ref{FIG:8}, we display the frequency, phase, and amplitude of three nonspinning waveforms: RIT:eBBH:1330 (panels (a), (d), (g)), RIT:eBBH:1331 (panels (b), (e), (h)), and RIT:eBBH:1332 (panels (c), (f), (i)), along with their corresponding PN fits. Similarly, FIG. \ref{FIG:9} presents the frequency, phase, and amplitude of three spin-aligned waveforms: RIT:eBBH:1899 (panels (a), (d), (g)), RIT:eBBH:1900 (panels (b), (e), (h)), and RIT:eBBH:1901 (panels (c), (f), (i)), also with their PN fits. In both figures, we consider only the leading Newtonian order for the amplitude, as discussed in Sec. \ref{sec:II:C}, which yields the most accurate fits compared to higher-order moments. Both FIGs. \ref{FIG:8} and \ref{FIG:9} reveal a clear trend: as eccentricity increases, the quality of the fitting decreases, as indicated by the larger fitting residuals in FIG. \ref{FIG:6}. Higher eccentricities result in more pronounced deviations near periastron, where local maxima are observed in the figures, highlighting the significant influence of the strong-field effects in NR within this region. Another consistent observation across all panels of FIGs. \ref{FIG:8} and \ref{FIG:9}, regardless of eccentricity, is the notable deviation between PN and NR starting approximately $200M$ before the merger. This divergence is inherent to the nature of PN and NR, with the strong-field effects of NR dominating as PN gradually loses its accuracy approaching the merger. 

\begin{figure*}[htbp!]
\centering
\includegraphics[width=15cm,height=21cm]{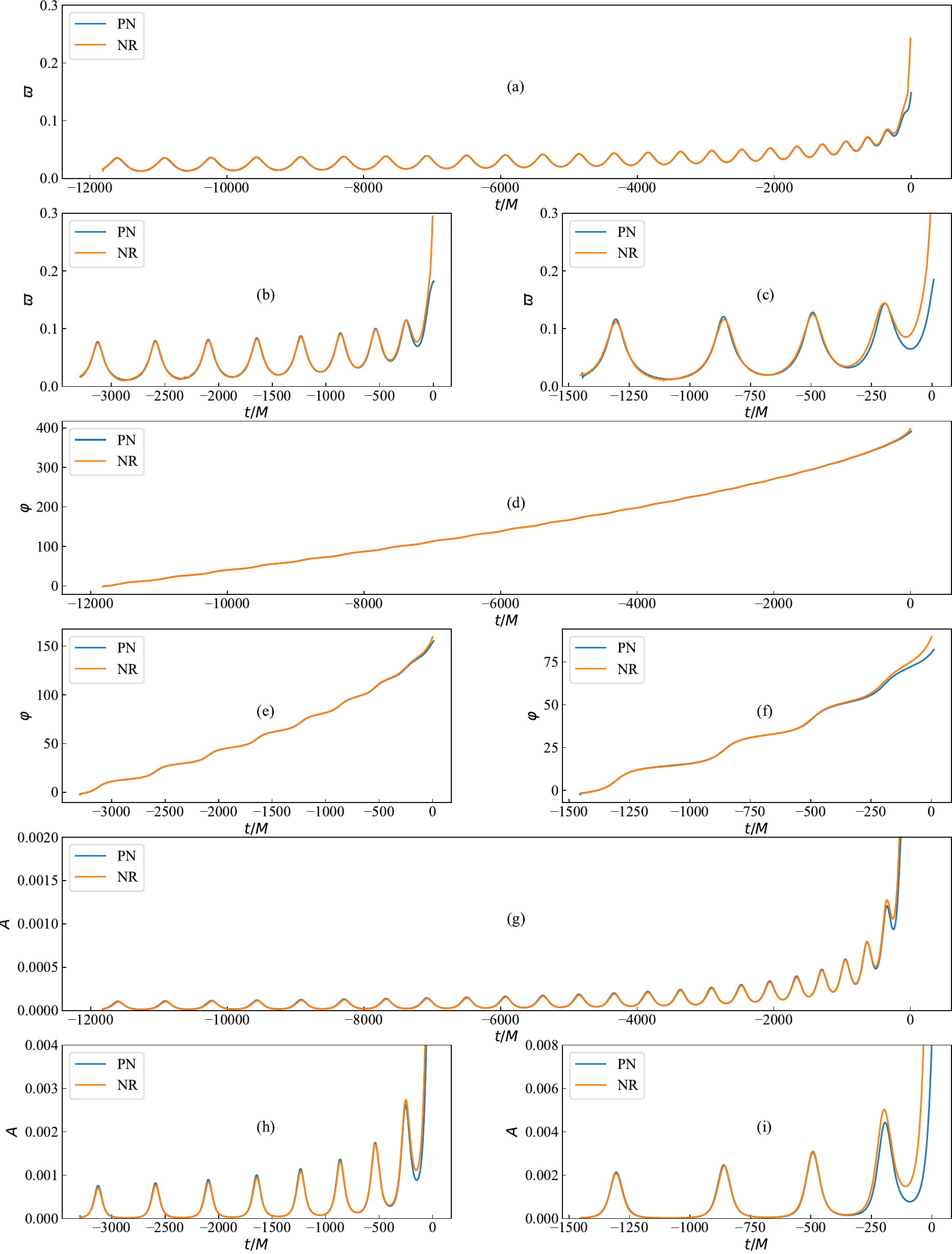}
\caption{\label{FIG:8}Frequency, phase, and leading Newtonian order amplitude of three sets of nonspinning waveforms RIT:eBBH:1330 (panel (a), (d), (g)), RIT:eBBH:1331 (panel (b), (e), (h)), RIT:eBBH:1332 (panel (c), (f), (i)), along with their PN fitting.}
\end {figure*}

The direct PN fitting results shown in FIGs. \ref{FIG:8} and \ref{FIG:9} do not provide a clear quantitative measure of the error between PN fits and NR waveforms. To address this, we introduce new metrics to describe these differences more precisely. We define the frequency relative error as: \begin{equation}\label{eq:69}
\frac{|\delta \varpi|}{\varpi_{\mathrm{NR}}}=|\frac{{\varpi_{\mathrm{PN}}}-{\varpi_{\mathrm{NR}}}}{\varpi_{\mathrm{NR}}}| \times 100\%,
\end{equation} 
the phase absolute error as: \begin{equation}\label{eq:70}
|\delta \varphi| = |\varphi_{\mathrm{PN}}-\varphi_{\mathrm{NR}}|,
\end{equation}
and the amplitude relative error as: \begin{equation}\label{eq:71}
\frac{|\delta A|}{A_{\mathrm{NR}}}=|\frac{{A_{\mathrm{PN}}}-{A_{\mathrm{NR}}}}{A_{\mathrm{NR}}}| \times 100\%,
\end{equation}
These metrics provide a quantitative description of the differences between PN fitting and NR waveforms.

\begin{figure*}[htbp!]
\centering
\includegraphics[width=15cm,height=21cm]{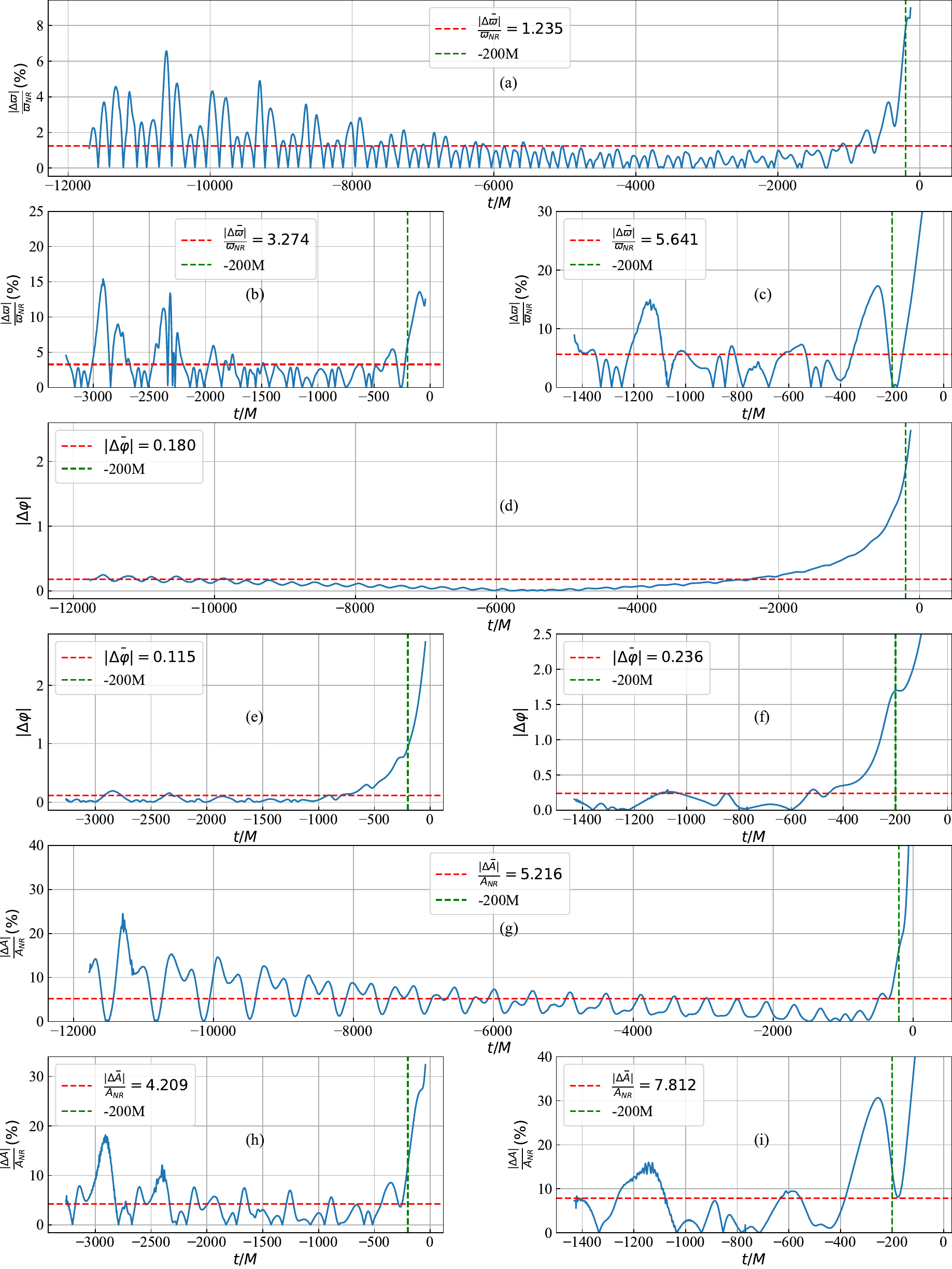}
\caption{\label{FIG:10}Frequency relative error $\frac{|\delta \varpi|}{\varpi_{\mathrm{NR}}}$, phase absolute error $|\delta \varphi|$, and amplitude relative error $\frac{|\delta A|}{A_{\mathrm{NR}}}$ of three sets of nonspinning waveforms RIT:eBBH:1330 (panel (a), (d), (g)), RIT:eBBH:1331 (panel (b), (e), (h)), RIT:eBBH:1332 RIT:eBBH:1332 (panel (c), (f), (i)). The horizontal red line represents their average value, and the vertical green line represents time $-200M$.}
\end {figure*}

\begin{figure*}[htbp!]
\centering
\includegraphics[width=15cm,height=21cm]{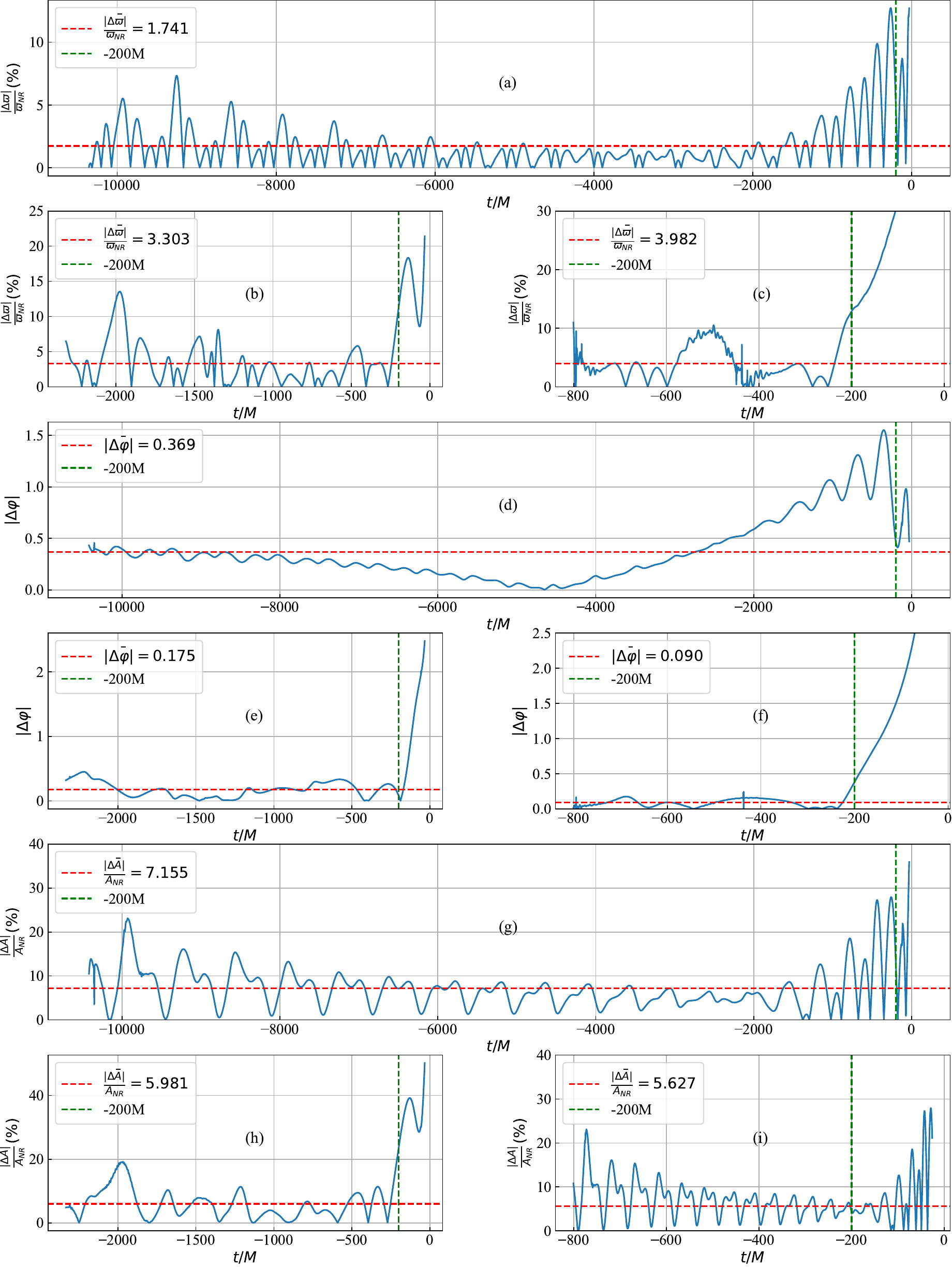}
\caption{\label{FIG:11}Frequency relative error $\frac{|\delta \varpi|}{\varpi_{\mathrm{NR}}}$, phase absolute error $|\delta \varphi|$, and amplitude relative error $\frac{|\delta A|}{A_{\mathrm{NR}}}$ of three sets of spin-aligned waveforms RIT:eBBH:1899 (panel (a), (d), (g)), RIT:eBBH:1900 (panel (b), (e), (h)), RIT:eBBH:1901 (panel (c), (f), (i)). The horizontal red line represents their average value, and the vertical green line represents time $-200M$.}
\end {figure*}

In FIG. \ref{FIG:10}, we present the frequency relative error $\frac{|\delta \varpi|}{\varpi_{\mathrm{NR}}}$, phase absolute error $|\delta \varphi|$, and amplitude relative error $\frac{|\delta A|}{A_{\mathrm{NR}}}$ for three sets of nonspinning waveforms: RIT:eBBH:1330 (panel (a), (d), (g)), RIT:eBBH:1331 (panel (b), (e), (h)), and RIT:eBBH:1332 (panel (c), (f), (i)). FIG. \ref{FIG:11} displays the corresponding frequency relative error $\frac{|\delta \varpi|}{\varpi_{\mathrm{NR}}}$, phase absolute error $|\delta \varphi|$, and amplitude relative error $\frac{|\delta A|}{A_{\mathrm{NR}}}$ for three sets of spin-aligned waveforms: RIT:eBBH:1899 (panel (a), (d), (g)), RIT:eBBH:1900 (panel (b), (e), (h)), and RIT:eBBH:1901 (panel (c), (f), (i)).

From FIGs. \ref{FIG:10} and \ref{FIG:11}, we observe that these errors oscillate over time, which contrasts with the behavior seen in circular orbits. In circular orbits, the frequency, phase, and amplitude increase monotonically over time, leading to a monotonic difference between PN and NR. However, in eccentric orbits, the frequency, phase, and amplitude oscillate, which causes the differences between PN and NR to oscillate as well. This oscillation spans the entire time range, as evidenced by the patterns in FIGs. \ref{FIG:10} and \ref{FIG:11}. Moreover, when viewed over shorter time intervals, these oscillations exhibit no clear regularity, sometimes the error is large, and other times it is nearly zero, making it difficult to accurately describe the error between PN and NR at any specific frequency or time. As eccentricity increases, the errors in all panels for frequency, phase, and amplitude exhibit an upward trend, with oscillations becoming more pronounced. In panel (a) of FIG. \ref{FIG:10}, the relative frequency error fluctuates between 0\% and 6\%, increasing near the merger but remaining stable until approximately $200M$ before the merger. In panel (b), the error ranges from 0\% to 15\%, with local increases potentially due to waveform artifacts that introduce fluctuations (these artifacts arise from taking the second derivative of $h$ to obtain $\Psi_4$). Panel (c) shows errors between 0\% and 20\%. In panel (d), the absolute phase error remains around 0 to 0.3 before the merger, while panel (e) shows an error around 0.4. Panel (f) exhibits a phase error under 0.5. The relative amplitude error in panel (g) stays below 25\%, while in panels (h) and (i), it fluctuates within 20\% and 35\%, respectively. FIG. \ref{FIG:11} shows similar trends to FIG. \ref{FIG:10}, but with noticeably larger frequency, phase, and amplitude errors in each panel. This discrepancy between FIG. \ref{FIG:10} and FIG. \ref{FIG:11} may result from lingering inconsistencies between local PN and NR frameworks, or it could indicate that the spin-aligned PN waveform lacks the necessary precision. 

From the preceding analysis, it is clear that the oscillatory errors in frequency, phase, and amplitude, as depicted in FIGs. \ref{FIG:10} and \ref{FIG:11}, cannot be accurately represented by singular values at specific moments in time. A measurement at any given moment could coincide with the trough of an oscillation, yielding a lower error, or it could align with the peak, resulting in a higher error. In such cases, drawing definitive conclusions is difficult. To address this, we calculate average errors over the entire time span, excluding the region where PN fails, specifically, from $200M$ before the merger. The average errors in frequency, phase, and amplitude are denoted as $\bar{\frac{|\delta \varpi|}{\varpi_{\mathrm{NR}}}}$, $\bar{|\delta \varphi|}$, and $\bar{\frac{|\delta A|}{A_{\mathrm{NR}}}}$. In FIGs. \ref{FIG:10} and \ref{FIG:11}, we mark the $-200M$ point, corresponding to a frequency near 0.1. FIGs. \ref{FIG:10} and \ref{FIG:11} show the average errors for frequency, phase, and amplitude for the six waveforms, providing a qualitative measure of their magnitudes. The exact average error values are annotated in these figures, offering a comprehensive and quantitative depiction of error across the datasets by averaging over the oscillations. These averages, positioned at the midpoint of the oscillations, serve as a balanced metric for assessing error. From FIGs. \ref{FIG:10} and \ref{FIG:11}, we observe that the average error values generally increase with eccentricity, though occasional decreases may occur, potentially due to factors such as waveform length influencing the fitting process and some other numerical errors. It is important to note that phase represents the integral of frequency, meaning that phase errors accumulate over time. Consequently, the phase difference is affected by the length of the waveform. However, in the context of eccentricity-induced oscillations, relying solely on phase error as a precise metric is challenging due to the complexity of these oscillations. The three average errors we employ here should be viewed as reference values rather than definitive measures, as the errors at specific times can significantly exceed the average error magnitude.

\begin{figure*}[htbp!]
\centering
\includegraphics[width=15cm,height=11.25cm]{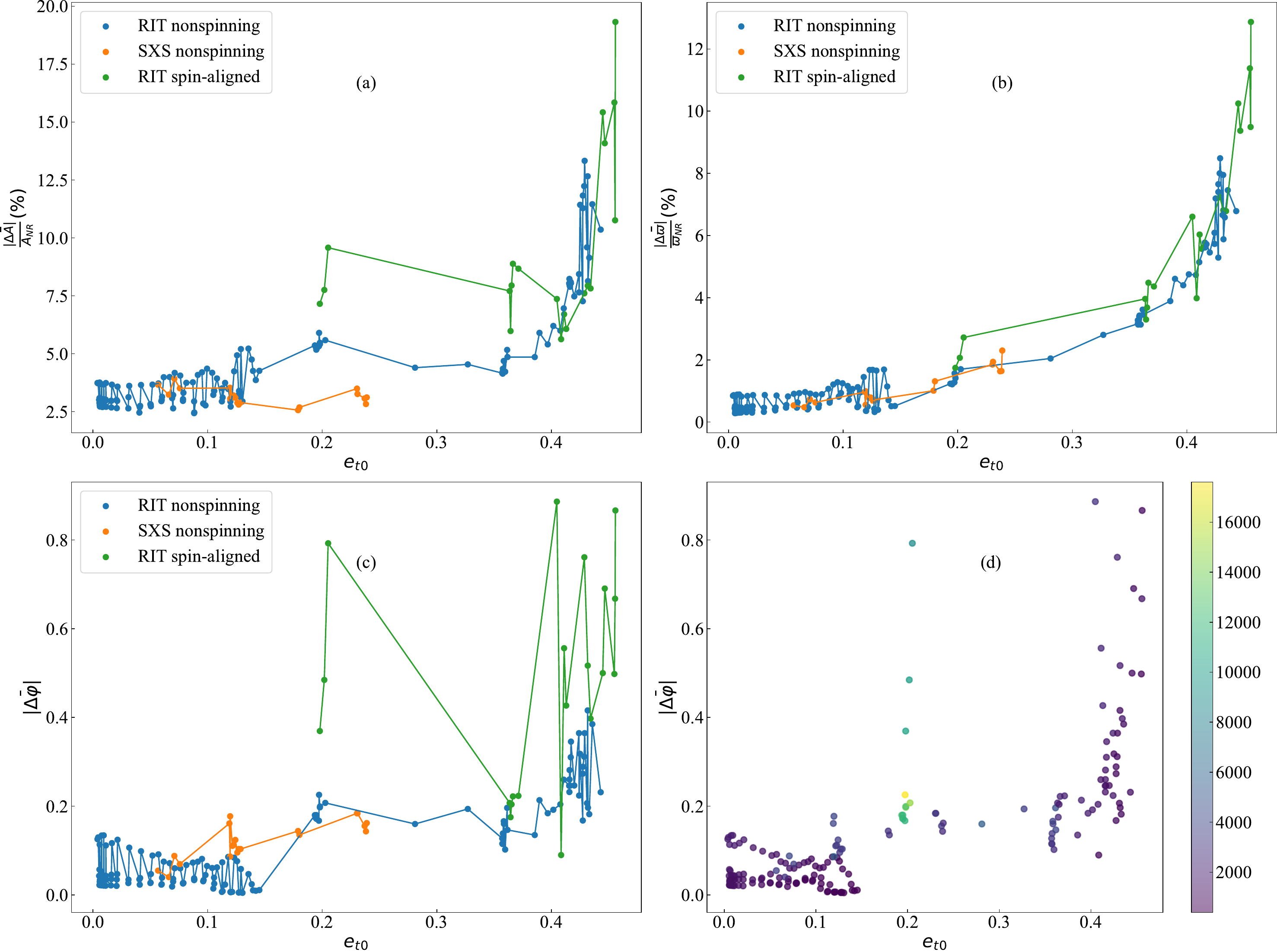}
\caption{\label{FIG:12}Correlation between the average amplitude relative error $\bar{\frac{|\delta A|}{A_{\mathrm{NR}}}}$ (a), the average frequency relative error $\bar{\frac{|\delta \varpi|}{\varpi_{\mathrm{NR}}}}$ ((b)), and the average phase absolute error $\bar{|\delta \varphi|}$ (c) resulting from the PN fitting of the (2,2) mode and the initial eccentricity parameters ${e_t}_0$. Panel (d) illustrates the relationship between phase error, eccentricity, and waveform duration, with the color bar denoting the length of waveforms.}
\end {figure*}

As highlighted in Sec. \ref{sec:II:B}, we have performed PN fitting on a comprehensive set of 180 eccentric NR waveforms. Given the extensive nature of these results, presenting detailed figures for each case would be impractical. Therefore, in FIG. \ref{FIG:12}, we provide the average errors in frequency, phase, and amplitude across these 180 waveforms as quantitative metrics to assess their accuracy. We also distinguish between the errors for RIT and SXS waveforms, as well as between nonspinning and spin-aligned RIT waveforms, and waveform length. 

In panel (a) of FIG. \ref{FIG:12}, we show the relationship between the average relative amplitude error, $\bar{\frac{|\delta A|}{A_{\mathrm{NR}}}}$, and eccentricity ${e_t}_0$. The error increases gradually with rising eccentricity. For eccentricities between 0 and 0.2, the maximum error reaches 6\%. In the eccentricity range of 0.2 to 0.4, differences emerge between nonspinning and spin-aligned waveforms: for nonspinning, $\bar{\frac{|\delta A|}{A_{\mathrm{NR}}}}$ ranges from 5\% to 7.5\%, while for spin-aligned waveforms, it ranges from 6\% to 10\%. At higher eccentricities, the amplitude error surpasses 10\%. 

Panel (b) of FIG. \ref{FIG:12} presents the evolution of the average relative frequency error, $\bar{\frac{|\delta \varpi|}{\varpi_{\mathrm{NR}}}}$, as a function of eccentricity. Similar to the amplitude error, the frequency error increases with eccentricity. In the 0 to 0.2 range, $\bar{\frac{|\delta \varpi|}{\varpi_{\mathrm{NR}}}}$ ranges between 0\% and 2\%. For eccentricity between 0.2 and 0.4, the nonspinning case shows frequency errors between 2\% and 6\%, while the spin-aligned case exhibits errors between 2\% and 8\%. For eccentricities beyond 0.4, the frequency error exceeds 8\%. Overall, the frequency error is generally smaller than the amplitude error. 

In panel (c) of FIG. \ref{FIG:12}, we examine the average phase error, $\bar{|\delta \varphi|}$, as a function of initial eccentricity ${e_t}_0$. For nonspinning cases with eccentricities between 0 and 0.2, the phase error ranges from 0 to 0.2. As eccentricity increases to between 0.2 and 0.4, $\bar{|\delta \varphi|}$ ranges from 0.1 to 0.3, and for eccentricities above 0.4, it spans from 0.3 to 0.5. In the spin-aligned case, the phase error is larger than in the nonspinning case. For eccentricities between 0.2 and 0.45, $\bar{|\delta \varphi|}$ fluctuates between 0.2 and 0.9, reflecting lower accuracy for spin-aligned waveforms compared to nonspinning ones. 

Panel (d) of FIG. \ref{FIG:12} explores the relationship between phase error, eccentricity, and waveform duration, with the color bar indicating waveform length. Shorter waveforms tend to exhibit lower phase errors, clustering towards the lower end of panel (d). Additionally, higher eccentricities correspond to larger phase errors for shorter waveforms, highlighting the combined influence of waveform length and eccentricity on phase accuracy. 

\subsection{Higher modes}\label{sec:III:D}
In the previous section, we concentrated on the dominant (2,2) mode. In this section, we shift our focus to the PN fitting results for higher-order modes, including the (2,1), (3,3), (3,2), (4,4), (4,3), and (5,5) modes. It should be noted that the RIT catalog does not include the (5,5) mode, with the (4,4) mode being the highest one available. Unlike the (2,2) mode, higher-order modes do not undergo iterative fitting. Instead, we directly utilize the fitting outcome of the (2,2) mode as the initial parameter to generate waveforms for these higher-order modes. This approach is not intended to suggest that higher-order modes are unsuitable for fitting but rather serves to cross-check the consistency of the parameters derived from fitting the (2,2) mode. 

Due to symmetry, not all higher-order harmonic modes exist in BBH systems. For instance, modes with odd values of $m$ in Eq. (\ref{eq:19}) are absent in equal-mass systems. In the RIT catalog, many of the expected higher-order modes are missing, and some of the available higher-order modes display unphysical behavior, which may result from errors in the extraction or expansion of $\Psi_4$. 

For brevity, we do not display all the fitting results for higher-order modes here. Instead, we present a selection of the most representative waveforms. If needed, any waveform can be reproduced using the fitting parameters provided in Appendix \ref{App:B}. 

FIG. \ref{FIG:13} presents the PN fitting results for the frequency (panels (a)–(f)), phase (panels (g)–(l)), and amplitude (panels (m)–(r)) of the higher-order modes, including the (2,1), (3,3), (3,2), (4,4), (4,3), and (5,5) modes, for waveform SXS:BBH:1364. This particular waveform was chosen for its mass ratio of $q=1/2$ and the presence of well-behaved higher-order modes. For amplitude, we consider only the leading PN moment, following the approach used for the dominant (2,2) mode. This choice is motivated by its demonstrated accuracy, which will be discussed later. From FIG. \ref{FIG:13}, it is clear that the PN fitting for the frequency and phase of the higher-order modes is of comparable quality to that of the dominant (2,2) mode. However, the amplitude fitting is less accurate and shows distinct behavior compared to the dominant mode. 

\begin{figure*}[htbp!]
\centering
\includegraphics[width=15cm,height=15cm]{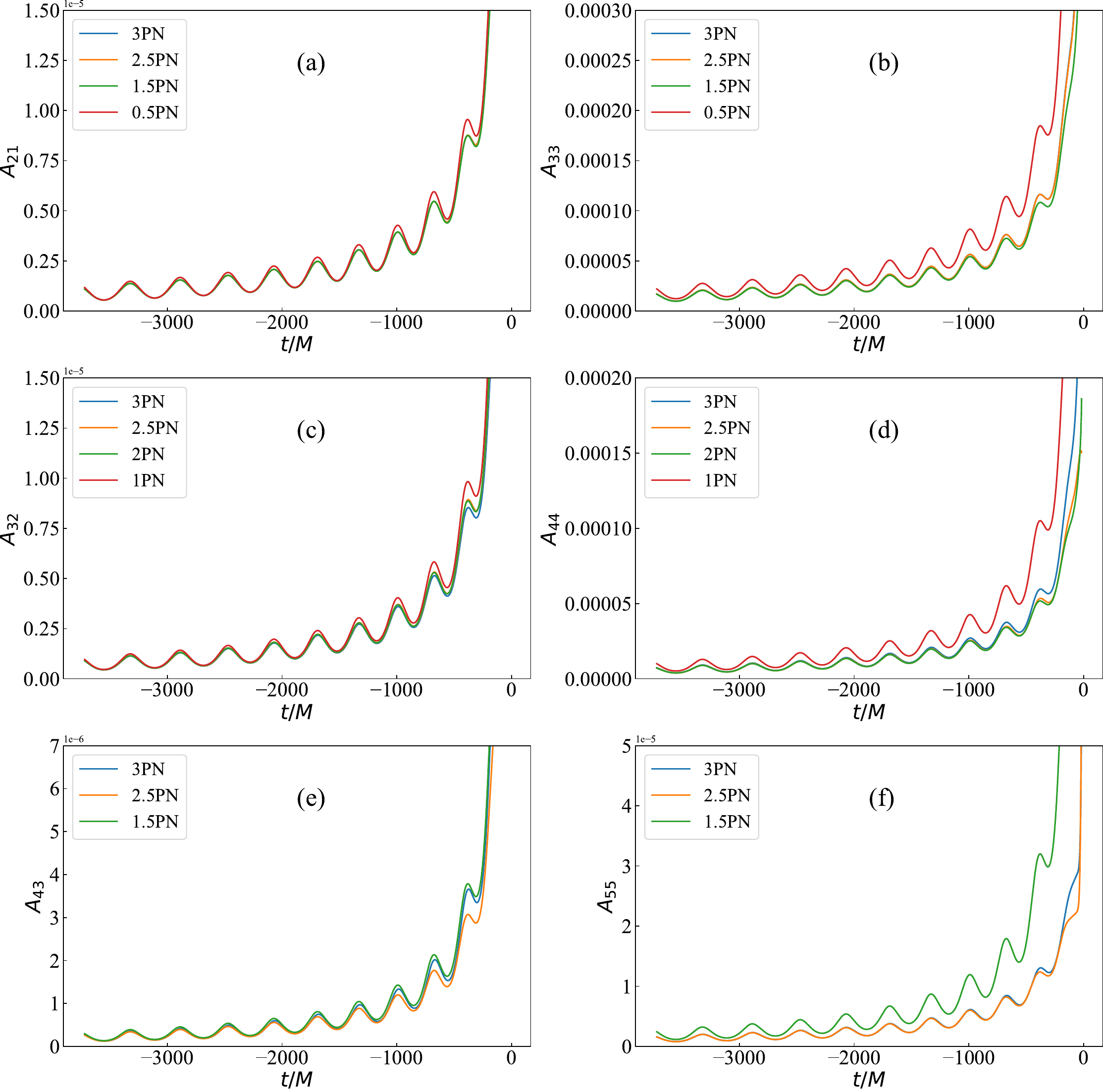}
\caption{\label{FIG:14}Amplitudes of various PN order moments for the higher-order modes (2,1) (a), (3,3) (b), (3,2) (c), (4,4) (d), (4,3) (e), and (5,5) (f) of the waveform SXS:BBH:1364.}
\end {figure*}

In FIG. \ref{FIG:14}, we illustrate the amplitudes of various PN order moments for the higher-order modes (2,1) (panel (a)), (3,3) (panel (b)), (3,2) (panel (c)), (4,4) (panel (d)), (4,3) (panel (e)), and (5,5) (panel (f)) of the waveform SXS:BBH:1364. The amplitudes of these modes arise from different PN orders and exhibit significant variation in magnitude. Notably, $A_{21}$, $A_{32}$, and $A_{43}$ share similar behavior, while $A_{33}$, $A_{44}$, and $A_{55}$ display analogous patterns. Despite these variations, a common feature is that the leading-order moment produces the largest amplitude, closely approximating the values seen in NR simulations. 

\begin{figure*}[htbp!]
\centering
\includegraphics[width=15cm,height=21cm]{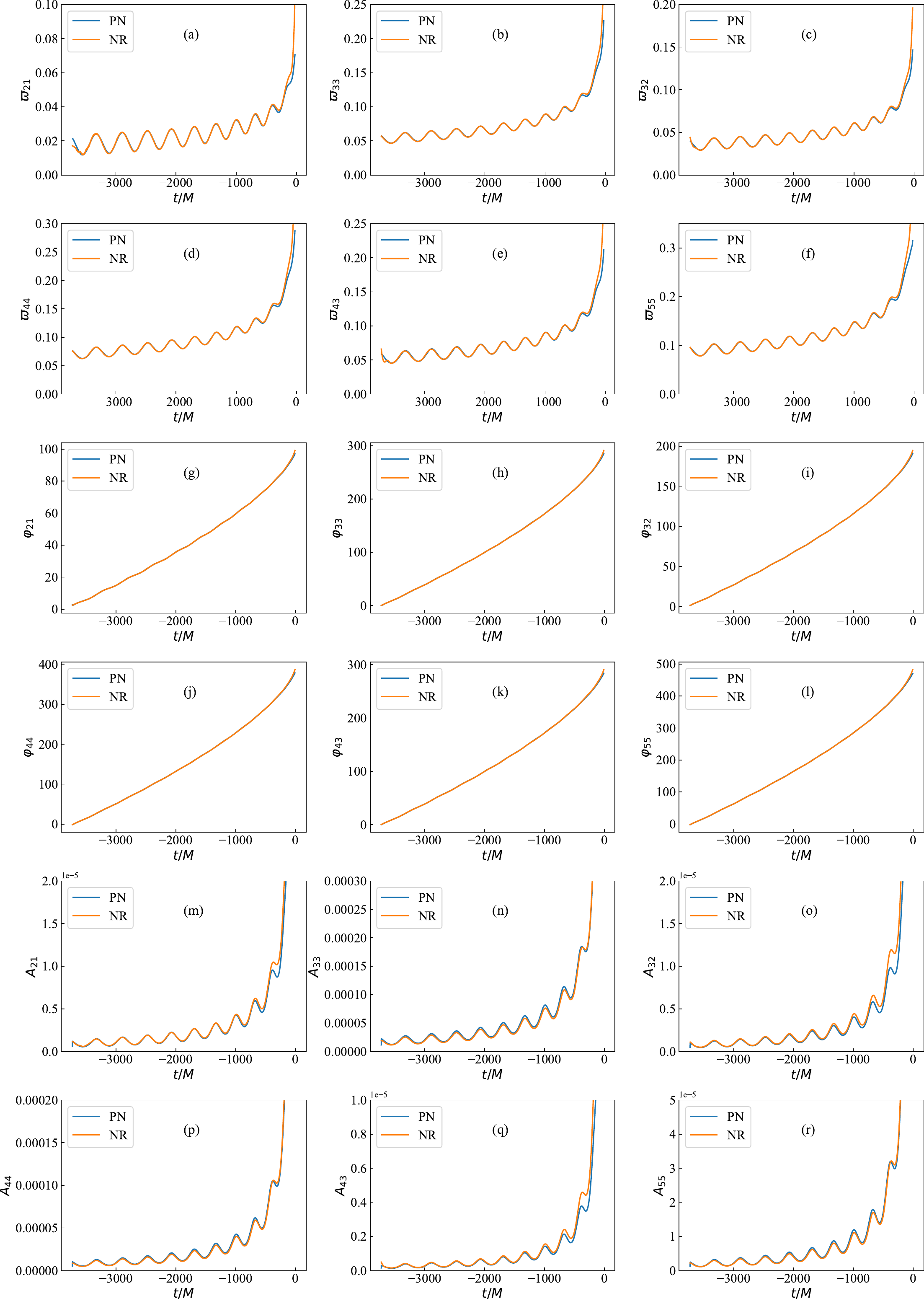}
\caption{\label{FIG:13}PN fitting outcomes for the frequency ((a), (b), (c), (d), (e), (f)), phase ((g), (h), (i), (j), (k), (l)), and amplitude ((m), (n), (o), (p), (q), (r)) of the higher-order modes including (2,1), (3,3), (3,2), (4,4), (4,3), and (5,5) modes of the waveform SXS:BBH:1364.}
\end {figure*}

\begin{figure*}[htbp!]
\centering
\includegraphics[width=15cm,height=21cm]{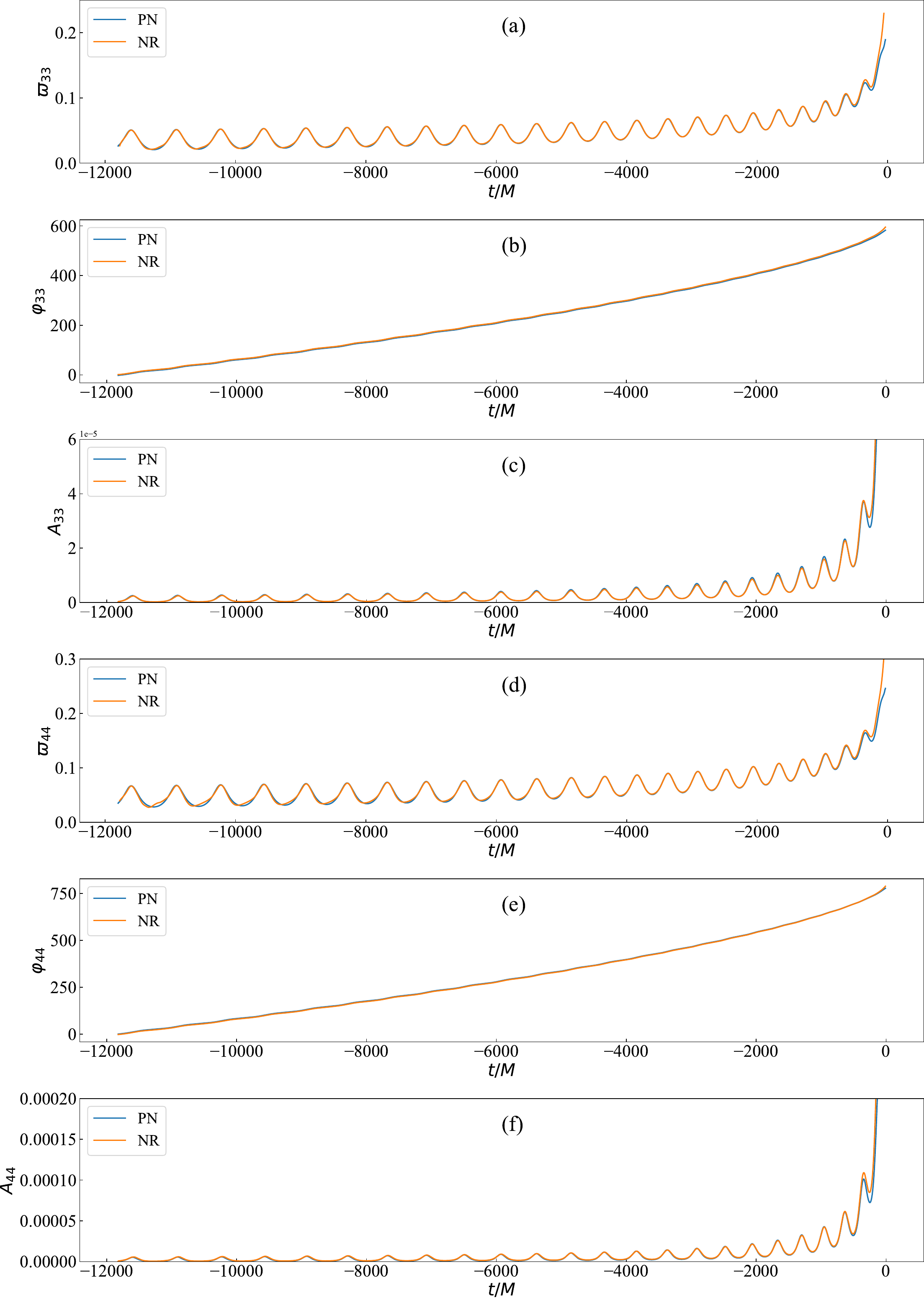}
\caption{\label{FIG:15}PN fitting results of frequency ((a), (d)), phase ((b), (e)) and amplitude ((c), (f)) of the (3,3) and (4,4) modes of the long nonspinning  waveform RIT:eBBH:1330.}
\end {figure*}

In contrast to the comprehensive coverage of higher-order modes in the SXS catalog, the RIT catalog provides incomplete data for these modes. Apart from the (3,3) and (4,4) modes, most higher-order modes contain disordered data that render them unsuitable for comparison. Even some instances of the (3,3) and (4,4) modes show irregularities. As a result, our comparisons are restricted to a limited number of well-behaved higher-order modes. For these modes, we present the frequency ((a), (d)), phase ((b), (e)), and amplitude ((c), (f)) PN fitting results of the (3,3) and (4,4) modes of the long waveform RIT:eBBH:1330 in FIG. \ref{FIG:15}. The fitting results for frequency and phase are comparable to those of the (2,2) mode shown in FIG. \ref{FIG:8}, while the amplitude fitting shows slightly worse performance. 

\begin{figure*}[htbp!]
\centering
\includegraphics[width=15cm,height=10.5cm]{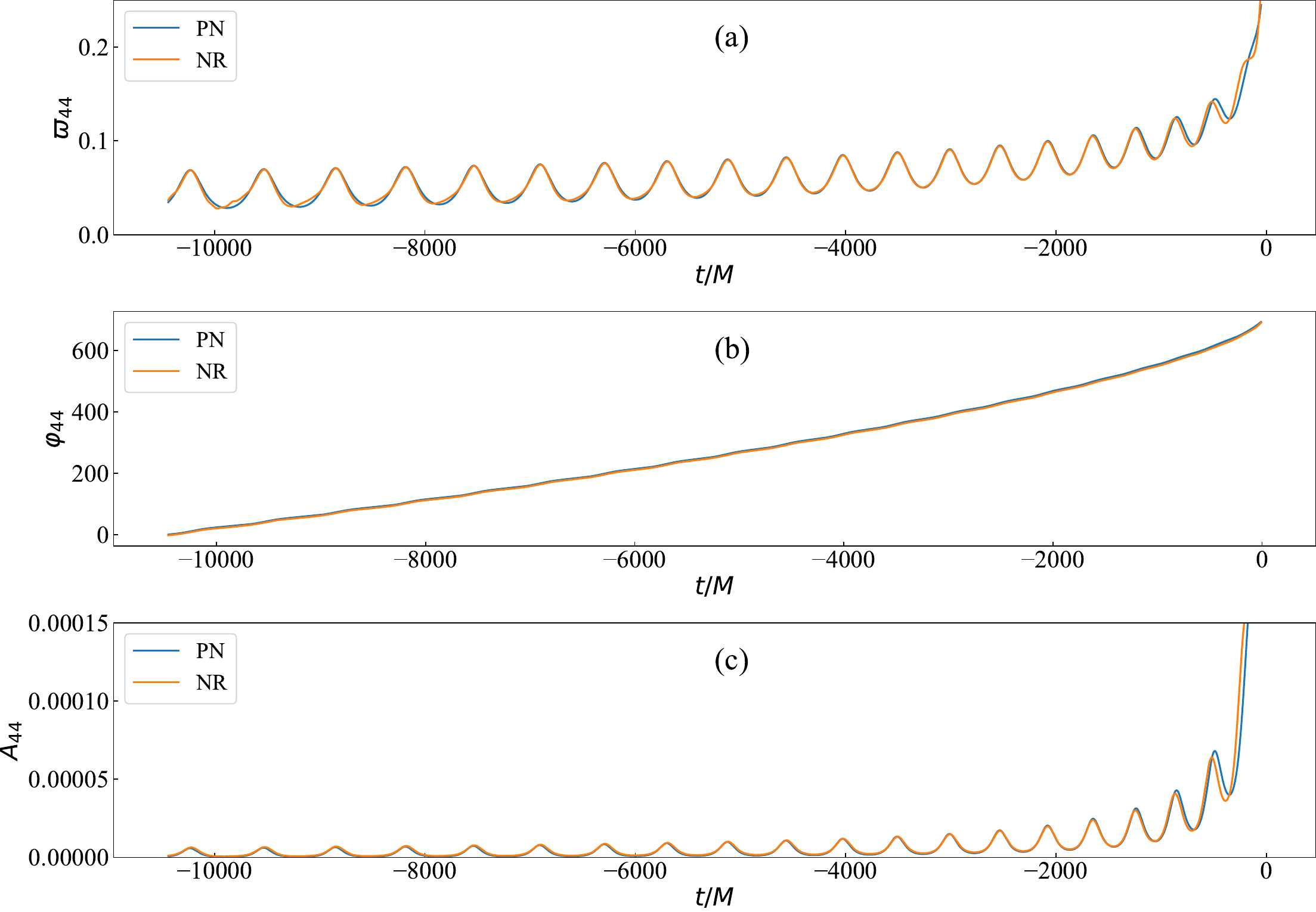}
\caption{\label{FIG:16}PN fitting results of the frequency (a), phase (b), and amplitude (c) of the (4,4) modes of long spin-aligned waveform RIT:eBBH:1899.}
\end {figure*}

In the spin-aligned case, the RIT catalog includes even fewer higher-order modes, with only the (4,4) mode available. To maintain consistency with the discussion of the (2,2) mode, we present in FIG. \ref{FIG:16} the PN fitting results for the frequency (a), phase (b), and amplitude (c) of the (4,4) mode from waveform RIT:eBBH:1899. The results indicate that the frequency and phase fitting for the (4,4) mode in the spin-aligned case are similar to those of the (2,2) mode in FIG. \ref{FIG:9}, albeit with slightly poorer amplitude fitting. Additionally, the overall quality of the (4,4) mode's fitting in the spin-aligned case is slightly degraded compared to the nonspinning case.

\begin{figure*}[htbp!]
\centering
\includegraphics[width=15cm,height=21cm]{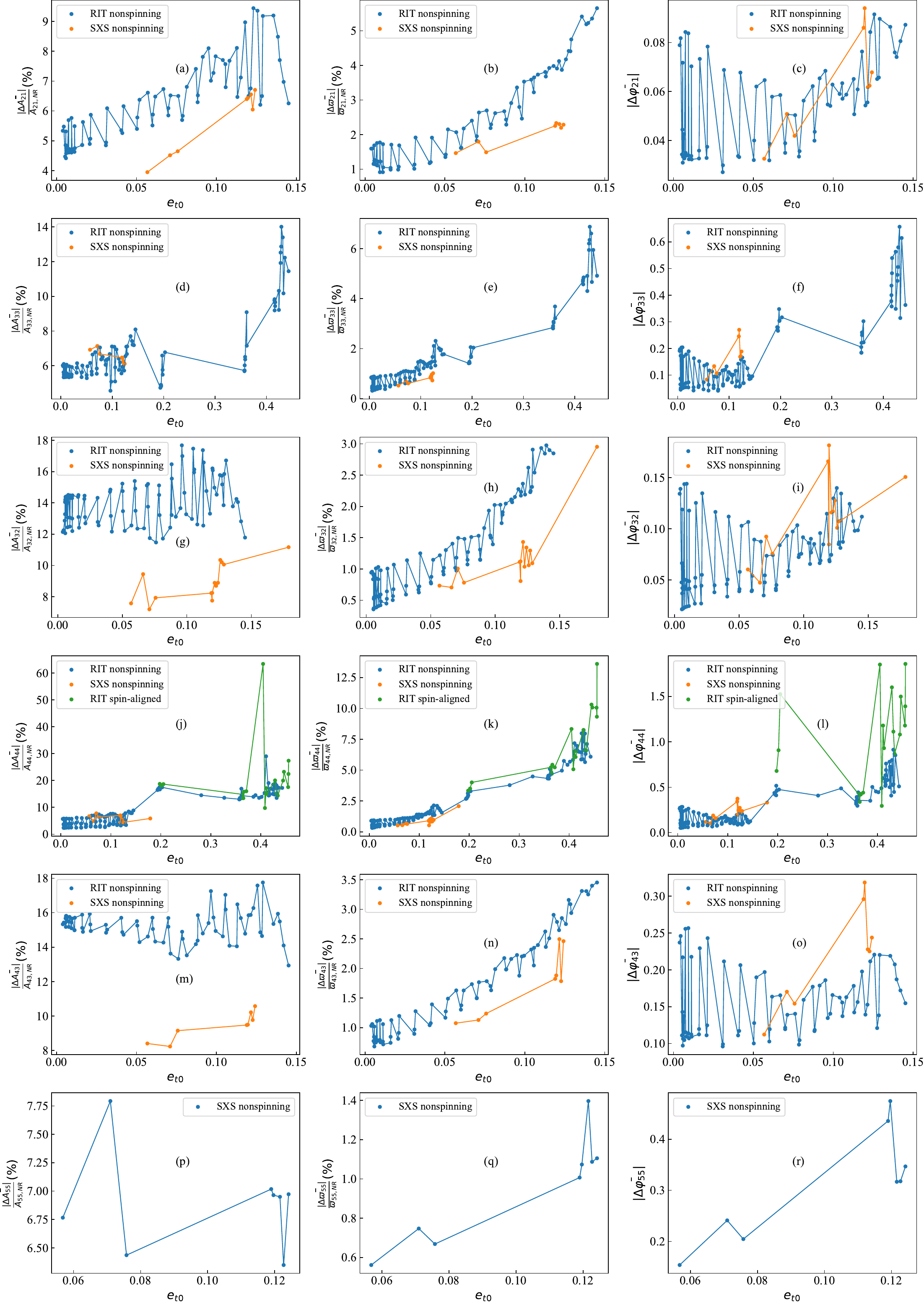}
\caption{\label{FIG:17}Correlation between the average amplitude relative error $\bar{\frac{|\delta A_{\ell m}|}{A_{\ell m,\mathrm{NR}}}}$ ((a), (d), (g), (j), (m), (p)), the average frequency relative error $\bar{\frac{|\delta \varpi_{\ell m}|}{\varpi_{\ell m,\mathrm{NR}}}}$ ((b), (e), (h), (k), (n), (q)), and the average phase absolute error $\bar{|\delta \varphi_{\ell m}|}$ ((c), (f), (i), (l), (o), (r)) resulting from the PN fitting of the high-order mode and the initial eccentricity parameters ${e_t}_0$.}
\end {figure*}

FIGs. \ref{FIG:13}, \ref{FIG:15}, and \ref{FIG:16} offer illustrative examples of the PN fitting results for higher-order modes, providing an intuitive understanding of the outcomes. Given the large number of waveforms analyzed, it is impractical to display all the fitting results individually. However, higher-order modes for any waveform can be generated using the parameters in Appendix \ref{App:B} in conjunction with the PN equations. As with the (2,2) mode, we quantify the discrepancies between the PN fitting waveforms and the NR waveforms using the average amplitude relative error $\bar{\frac{|\delta A_{\ell m}|}{A_{\ell m,\mathrm{NR}}}}$, the average frequency relative error $\bar{\frac{|\delta \varpi_{\ell m}|}{\varpi_{\ell m,\mathrm{NR}}}}$, and the average phase absolute error $\bar{|\delta \varphi_{\ell m}|}$ for each higher-order mode, where $\ell m$ represents the harmonic mode. FIG. \ref{FIG:17} shows the relationship between the average amplitude relative error ((a), (d), (g), (j), (m), (p)), the average frequency relative error ((b), (e), (h), (k), (n), (q)), and the average phase absolute error ((c), (f), (i), (l), (o), (r)) as a function of the initial eccentricity parameter ${e_t}_0$. 

For the average amplitude relative error $\bar{\frac{|\delta A_{\ell m}|}{A_{\ell m,\mathrm{NR}}}}$, identifying a clear trend with eccentricity is difficult due to the narrow eccentricity range in most cases. Nevertheless, $A_{21}$, $A_{33}$, and $A_{44}$ show an increasing trend with eccentricity, whereas the behavior of $A_{32}$, $A_{43}$, and $A_{55}$ is less discernible. The amplitude $A_{21}$ has the smallest error and $A_{44}$ the largest. In general, the errors in RIT waveforms exceed those in SXS waveforms, and errors in the spin-aligned scenario are larger than in the nonspinning case. Specific error values are presented in FIG. \ref{FIG:17}. 

For the average frequency error $\bar{\frac{|\delta \varpi_{\ell m}|}{\varpi_{\ell m,\mathrm{NR}}}}$, all higher-order modes show an increasing trend with eccentricity, with errors significantly smaller than those for amplitude. These results suggest that the PN fitting method captures frequency more accurately than amplitude. For eccentricities between 0 and 0.2, the errors remain below 4\%, while for eccentricities from 0.2 to 0.4, they stay below 8\%. The errors in SXS waveforms are lower than those in RIT waveforms. 

The average phase absolute error, $\bar{|\delta \varphi_{\ell m}|}$, follows a pattern similar to that of the (2,2) mode, depending on both the waveform duration and the initial eccentricity. For clarity, we focus on the relationship between phase error and eccentricity. Across all waveform lengths, the phase error remains below 0.3 for eccentricities in the range of 0 to 0.2, demonstrating high precision for higher-order modes. For eccentricities between 0.2 and 0.4, the phase error stays below 1 for nonspinning waveforms and under 2 for spin-aligned waveforms, indicating a decline in phase accuracy at higher eccentricities. 

In summary, the frequency and phase behaviors of the higher-order modes are similar to those of the (2,2) mode, showing comparable average errors. This consistency demonstrates that the parameters obtained from fitting the (2,2) mode are valid for higher-order modes. While the amplitudes of higher-order modes are best estimated by their leading-order moment, the errors tend to be larger than for the (2,2) mode. Additionally, the errors in higher-order modes for RIT waveforms generally exceed those in SXS waveforms, and spin-aligned scenarios exhibit larger errors compared to nonspinning cases. 

\subsection{Small mass ratio fitting problems and implications}\label{sec:III:E}
In Sec. \ref{sec:III:A}, we highlighted the issue of increasing fitting residuals with the mass ratio. This trend may arise from the growing error in NR simulations of eccentric orbit BBH systems as the mass ratio decreases. The RIT simulations include a lower mass ratio of 1/7, which we provides lengthy waveforms suitable for fitting but discards excessively short waveforms here. For this study, we focus exclusively on waveforms with a mass ratio of 1/4, as previously mentioned, and omit those with mass ratios of 1/5, 1/6, and 1/7. This section clarifies the rationale behind excluding these specific mass ratios. 

\begin{figure*}[htbp!]
\centering
\includegraphics[width=15cm,height=10.5cm]{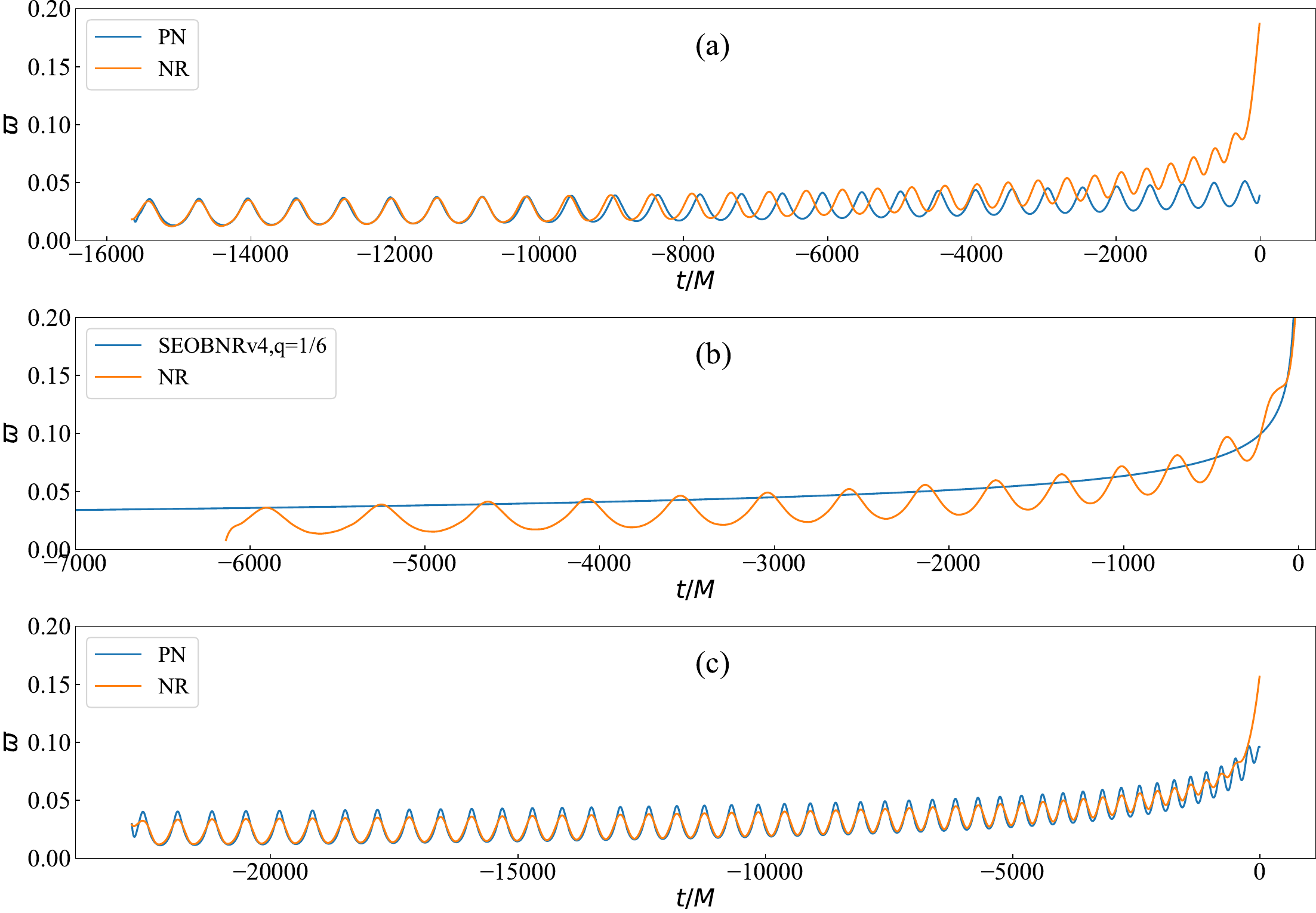}
\caption{\label{FIG:18}Frequencies of the (2,2) modes for the longest waveforms RIT:eBBH:1514 (a) and RIT:eBBH:1560 (c) with mass ratios of 1/5 and 1/7 from the RIT catalog alongside their corresponding PN fitting outcomes. Panel (b) displays the frequencies of waveform RIT:eBBH:1537 and the circular orbit waveform with the equivalent mass ratio $q=1/6$, sourced from SEOBNRv4 of Pycbc.}
\end {figure*}

FIG. \ref{FIG:18} displays the frequencies of the (2,2) modes for the long waveforms RIT:eBBH:1514 (panel (a)) and RIT:eBBH:1560 (panel (c)), with mass ratios of 1/5 and 1/7 from the RIT catalog, along with their corresponding PN fitting outcomes. Panel (b) shows the frequency of waveform RIT:eBBH:1537 and the circular orbit waveform with an equivalent mass ratio $q=1/6$, sourced from SEOBNRv4 of PyCBC. Panel (b) serves a similar purpose as FIG. \ref{FIG:5}. 

Waveforms RIT:eBBH:1514, RIT:eBBH:1537, and RIT:eBBH:1560 share identical initial parameters with other long waveforms in the RIT catalog, specifically, no spin, an initial coordinate separation of $24.6M$, and a similar initial eccentricity of approximately 0.19. The only difference is the mass ratio. According to PN and NR theories, under similar initial conditions, a smaller mass ratio should result in a longer merger time and thus a longer waveform. However, waveforms RIT:eBBH:1514 and RIT:eBBH:1537 exhibit shorter waveform durations compared to RIT:eBBH:1491, which has a larger mass ratio of $q=1/4$. This discrepancy suggests potential issues with the NR simulations of RIT:eBBH:1514 and RIT:eBBH:1537. 

In contrast to waveform RIT:eBBH:1282, panel (a) of FIG. \ref{FIG:18} shows the PN fitting results for waveform RIT:eBBH:1514 over the interval $[-15552M,-10000M]$. Due to the infeasibility of fitting the entire waveform, only the initial segment was considered. While the fitting of this segment appears satisfactory, significant deviations are evident between the PN waveform and the NR waveform in subsequent segments, indicating possible issues with the NR simulation. 

The discrepancy in the (2,2) mode frequency of waveform RIT:eBBH:1537, shown in panel (b) of FIG. \ref{FIG:18}, is notably more pronounced. Its significantly shorter waveform duration compared to that in panel (a) highlights this issue. Additionally, plotting a circular orbit waveform with the same mass ratio of $q=1/6$ reveals a marked deviation from the average orbital frequency of the eccentric waveform. This deviation contradicts our expectations based on FIG. \ref{FIG:5}, where the circular orbit waveform frequency should approximately align with the midpoint of the eccentric waveform frequency deviation. This pronounced discrepancy strongly suggests problems within the NR simulation of RIT:eBBH:1537.

Regarding waveform RIT:eBBH:1560, shown in panel (c) of FIG. \ref{FIG:18}, a complete waveform fit is possible. However, the fitting results reveal a discrepancy: the initial eccentricity inferred from radiation reaction differs from that derived via quasi-Keplerian parameterization. Partial fits lead to larger deviations, which are not shown here. This inconsistency causes a divergence between the PN fitting and the NR waveform, suggesting potential issues in the NR simulation of waveform RIT:eBBH:1560. 

In conclusion, NR simulations of BBH in eccentric orbits, particularly for small mass ratios, may encounter challenges, resulting in inaccuracies in the waveforms. This underscores the need for careful selection of waveforms with smaller mass ratios when constructing templates and highlights the importance of exploring waveforms from diverse dynamical perspectives for cross-validation. The findings from NR simulations in small mass ratio cases emphasize the critical need for developing more accurate and refined NR algorithms to improve gravitational wave parameter extraction and reduce systematic errors.

\section{Conclusion and Outlook}\label{sec:IV}
In this study, we present a comprehensive comparison between post-Newtonian and numerically relativistic waveforms in eccentric orbits, encompassing both nonspinning and spin-aligned configurations. The comparison delves into the frequency, amplitude, and phase characteristics of various harmonic modes, including the (2,2), (2,1), (3,3), (3,2), (4,4), (4,3), and (5,5) modes. The nonspinning and spin-aligned eccentric PN waveforms utilized in our analysis are founded on the 3PN quasi-Keplerian parameterization, incorporating 3PN radiative reaction incorporating instantaneous and hereditary contributions. Furthermore, we utilize the gravitational wave amplitude from higher-order moments at 3PN, which exceed the Newtonian quadrupole moment. The NR waveforms considered in our study comprise a diverse array of eccentric orbit waveforms sourced from the RIT and SXS catalogs. These waveforms span mass ratios ranging from 1/4 to 1, eccentricities from 0 to 0.45, and waveform durations extending beyond $17000M$, across both nonspinning and spin-aligned configurations. We present in detail the fitting parameters of the eccentricity orbit waveforms for the SXS and RIT catalogs used in all studies in Appendix \ref{App:B}.

In our investigation, we opted to focus on the (2,2) mode as the target for fitting and proceeded to compare the frequencies of the quadrupole moments and higher-order moments of $\Psi_4^{22}$ and $h^{22}$. To ensure a comprehensive analysis, we utilized the fitting parameters of the $\Psi_4^{22}$ of higher-order moment as the reference standard. Throughout our analysis, we compared the amplitudes of the quadrupole and higher-order moments of $\Psi_4^{22}$ and $h^{22}$. Notably, we found that the amplitude derived from the quadrupole moment of $\Psi_4^{22}$ demonstrated superior accuracy, while the amplitudes associated with other PN order moments showed relatively diminished values in comparison.

Throughout our study, we conducted fitting exercises on a total of 180 sets of eccentricity waveforms, revealing a trend where fitting residuals increased in correspondence with rising eccentricity. Notably, these residuals proved to be independent of waveform length, catalog source, and spin but exhibited a correlation with the mass ratio. Specifically, we noted that smaller mass ratios correlated with larger fitting residuals, potentially indicating reduced accuracy in PN or NR involving small mass ratios.

Moreover, our analysis involved a comparison of the initial eccentricity obtained through PN fitting with that derived from the 3PN quasi-Keplerian parameterization and the initial eccentricity provided by the RIT and SXS catalogs. Notably, we observed that the initial eccentricity from RIT closely approximated ${e_t}_0$, followed by the quasi-Keplerian parameterization, while SXS exhibited the least alignment.

Further investigations delved into comparing the frequency, phase, and amplitude characteristics of the (2,2) modes. Noteworthy findings included a strong consistency between PN and NR in the inspiral segment, particularly evident in extended waveforms. Nevertheless, discrepancies became apparent as the nonspinning PN model started to diverge roughly at $200M$ before the merger, while the spin-aligned PN waveform exhibited failures before the $200M$. Furthermore, discrepancies between PN and NR waveforms magnified with increasing initial eccentricity, notably apparent at the periastron point and merger phase, where strong-field effects came into play.

Given the oscillatory nature of frequency, phase, and amplitude in eccentric waveforms, the discrepancies between PN and NR waveforms also oscillate, rendering it challenging to pinpoint a specific moment for error assessment. Consequently, in our study, we utilized the average errors $\bar{\frac{|\delta \varpi|}{\varpi_{\mathrm{NR}}}}$, $\bar{|\delta \varphi|}$, and $\bar{\frac{|\delta A|}{A_{\mathrm{NR}}}}$ up to $200M$ pre-merger to encapsulate the overall waveform behavior.
Our analysis revealed that the average errors in frequency, amplitude, and phase of the (2,2) modes amplified alongside increasing eccentricity levels. For eccentricities ranging from 0 to 0.2, we observed frequency errors below 3\%, phase errors below 0.2, and amplitude errors under 6\%. In cases where eccentricities spanned 0.2 to 0.4, nonspinning scenarios exhibited frequency errors between 2\% and 6\%, phase errors ranging from 0.1 to 0.3, and amplitude errors of 5\% to 7.5\%. Conversely, in spin-aligned cases within the same eccentricity range, frequency errors were between 2\% and 8\%, phase errors ranged from 0.3 to 0.5, and amplitude errors fell within 6\% to 10\%. As eccentricity surpassed 0.4, errors escalated. Furthermore, phase errors demonstrated an increase with elongating waveform lengths.

Subsequently, we conducted a comparative analysis of the amplitudes, frequencies, and phases of the higher-order modes, including the (2,1), (3,3), (3,2), (4,4), (4,3), and (5,5) modes. Our observations indicated that the frequency and phase characteristics mirrored those of the 22 mode, whereas the amplitude behavior diverged. Notably, the leading-order moment's amplitude consistently provided the most precise results across these modes.
Utilizing the average error metric, we quantified the disparities between the PN and NR waveforms for the higher-order modes. Our findings revealed that the frequency and amplitude errors exhibited similarities to those of the (2,2) mode, although the amplitude errors were comparatively larger. Each mode displayed distinct error patterns, highlighting the distinctive behaviors inherent to these higher-order modes.

In the last part, we performed fittings on three distinctive eccentric waveforms from RIT featuring lower mass ratios, uncovering varied challenges during the fitting procedures. We attribute these discrepancies to issues within NR simulations when dealing with low mass ratio scenarios, underscoring the imperative need for the advancement of a more precise NR simulation framework to address these specific challenges.

Through a comprehensive comparison of PN and NR waveforms in eccentric orbits, we have delved deeply into the existing challenges within both methodologies. Our analysis underscores that there is still room for refinement in PN waveforms, particularly in enhancing accuracy across higher PN orders. Notably, for spin-aligned configurations, PN exhibits persistent shortcomings, experiencing failures preceding mergers and in proximity to periastron passages.

NR simulations, on the other hand, necessitate enhancements, particularly in handling small mass ratios and waveform extraction, with a specific focus on high-order mode extractions. To advance the construction of more precise gravitational wave templates for eccentric orbits, addressing these issues is paramount. By progressively resolving these challenges, we can systematically diminish systematic errors in parameter estimation, attain more precise gravitational wave parameters, and significantly advance gravitational wave detection efforts.

\begin{acknowledgments}
The authors are very grateful to RIT collaboration and SXS collaboration for the numerical simulation of eccentric BBH mergers, and thanks to Yan-Fang Huang, Zhou-Jian Cao, Duan-Yuan Gao, Lin Zhou, Yuan-Yuan Zuo, Jun-Yi Shen and Shi-Yan Tian for their helpful discussions. The computation is partially completed in the HPC Platform of Huazhong University of Science and Technology. The language was polished by ChatGPT during the revision of the draft. This work is supported by the National Key R\&D Program of China (2021YFA0718504). Z. L. is supported by China Scholarship Council, No.\,202306340128.
\end{acknowledgments}

\bibliographystyle{apsrev4-2}

\bibliography{ref}
\appendix

\section{eccentric PN expression}\label{App:A}
\subsection{Nonspinning waveforms}
In this section, we present the comprehensive formulation of the nonspinning BBH 3PN conservative dynamics. This description encompasses the computation of variables such as $r$, $\dot{r}$, $\phi$, $\dot{\phi}$, $l$, and $n$. To streamline the presentation, we provide expressions for $r$ and $\dot{\phi}$ herein, with $\dot{r}$ and $\phi$ derivable through differentiation and integration. Regarding the nonspinning component, we direct the reader to Ref. \cite{Hinder:2008kv}, where these quantities are articulated in terms of the expansion parameter $x$ and temporal eccentricity $e_t$.

\begin{widetext}
The orbital radius $r^\mathrm{NS}$ of the nonspinning eccentric BBH can be expressed as
\begin{equation}\label{eq:A1}
r^\mathrm{NS}=  r_{\mathrm{Newt}}^\mathrm{NS} x^{-1}+r_{1 \mathrm{PN}}^\mathrm{NS}+r_{2 \mathrm{PN}}^\mathrm{NS} x
+r_{3 \mathrm{PN}}^\mathrm{NS} x^2 
+\mathcal{O}\left(x^3\right),
\end{equation}
where the coefficients can be expressed as
\begin{equation}\label{eq:A2}
r_{\mathrm{Newt}}^\mathrm{NS}=1-e_t \cos (u),
\end{equation}

\begin{equation}\label{eq:A3}
r_{1 \mathrm{PN}}^\mathrm{NS}=\frac{2(e_t \cos (u)-1)}{{e_t}^2-1}+\frac{1}{6}(2(\eta-9)+{e_t}(7 \eta-6) \cos (u)),    \end{equation}

\begin{equation}\label{eq:A4}
\begin{aligned}
r_{2 \mathrm{PN}}^\mathrm{NS}= & \frac{1}{\left(1-{e_t}^2\right)^2}\left[\frac{1}{72}\left(8 \eta^2+30 \eta+72\right) {e_t}^4+\frac{1}{72}\left(-16 \eta^2-876 \eta+756\right) {e_t}^2+\frac{1}{72}\left(8 \eta^2+198 \eta+360\right)\right. \\
& +\left(\frac{1}{72}\left(-35 \eta^2+231 \eta-72\right) {e_t}^5+\frac{1}{72}\left(70 \eta^2-150 \eta-468\right) {e_t}^3+\frac{1}{72}\left(-35 \eta^2+567 \eta-648\right) {e_t}\right) \cos (u) \\
& \left.+\sqrt{1-{e_t}^2}\left(\frac{1}{72}(360-144 \eta) {e_t}^2+\frac{1}{72}(144 \eta-360)+\left(\frac{1}{72}(180-72 \eta) {e_t}^3+\frac{1}{72}(72 \eta-180) {e_t}\right) \cos (u)\right)\right],
\end{aligned}    
\end{equation}

\begin{equation}\label{eq:A5}
\begin{aligned}
r_{3 \mathrm{PN}}^\mathrm{NS}= & \frac{1}{181440\left(1-{e_t}^2\right)^{7 / 2}}\left[\left(-665280 \eta^2+1753920 \eta-1814400\right) {e_t}^6\right. \\
& +\left(725760 \eta^2-77490 \pi^2 \eta+5523840 \eta-3628800\right) {e_t}^4+\left(544320 \eta^2+154980 \pi^2 \eta-14132160 \eta+7257600\right) {e_t}^2 \\
& -604800 \eta^2+6854400 \eta+\left(\left(302400 \eta^2-1254960 \eta+453600\right) {e_t}^7+\left(-1542240 \eta^2-38745 \pi^2 \eta\right.\right. \\
& +6980400 \eta-453600) {e_t}^5+\left(2177280 \eta^2+77490 \pi^2 \eta-12373200 \eta+4989600\right) {e_t}^3 \\
& \left.+\left(-937440 \eta^2-38745 \pi^2 \eta+6647760 \eta-4989600\right) {e_t}\right) \cos (u) \\
& +\sqrt{1-{e_t}^2}\left(\left(-4480 \eta^3-25200 \eta^2+22680 \eta-120960\right) {e_t}^6+\left(13440 \eta^3+4404960 \eta^2+116235 \pi^2 \eta\right.\right. \\
& -12718296 \eta+5261760) {e_t}^4+\left(-13440 \eta^3+2242800 \eta^2+348705 \pi^2 \eta-19225080 \eta+16148160\right) {e_t}^2 \\
& +4480 \eta^3+45360 \eta^2-8600904 \eta+\left(\left(-6860 \eta^3+550620 \eta^2-986580 \eta+120960\right) {e_t}^7\right. \\
& +\left(20580 \eta^3-2458260 \eta^2+3458700 \eta-2358720\right) {e_t}^5+\left(-20580 \eta^3-3539340 \eta^2-116235 \pi^2 \eta\right. \\
& \left.+20173860 \eta-16148160) {e_t}^3+\left(6860 \eta^3-1220940 \eta^2-464940 \pi^2 \eta+17875620 \eta-4717440\right) {e_t}\right) \cos (u) \\
& \left.\left.+116235 \eta \pi^2+1814400\right)-77490 \eta \pi^2-1814400\right] .
\end{aligned}    
\end{equation}

The relative angular velocity $\dot{\phi} ^\mathrm{NS}$ of the nonspinning eccentric BBH can be expressed as

\begin{equation}\label{eq:A6}
\dot{\phi} ^\mathrm{NS}=  \dot{\phi}_{ \mathrm{Newt}}^\mathrm{NS} x^{3 / 2}+\dot{\phi}_{1 \mathrm{PN}}^\mathrm{NS} x^{5 / 2} +\dot{\phi}_{2 \mathrm{PN}}^\mathrm{NS} x^{7 / 2} 
+\dot{\phi}_{3 \mathrm{PN}}^\mathrm{NS} x^{9 / 2}+\mathcal{O}\left(x^{11 / 2}\right),
\end{equation}
where the coefficients can be expressed as

\begin{equation}\label{eq:A7}
\dot{\phi}_{ \mathrm{Newt}}^\mathrm{NS}=\frac{\sqrt{1-{e_t}^2}}{({e_t} \cos (u)-1)^2},
\end{equation}

\begin{equation}\label{eq:A8}
\dot{\phi}_{1 \mathrm{PN}}^\mathrm{NS}=-\frac{{e_t}(\eta-4)({e_t}-\cos (u))}{\sqrt{1-{e_t}^2}({e_t} \cos (u)-1)^3},    
\end{equation}

\begin{equation}\label{eq:A9}
\begin{aligned}
\dot{\phi}_{2 \mathrm{PN}}^\mathrm{NS}= & \frac{1}{12\left(1-{e_t}^2\right)^{3 / 2}({e_t} \cos (u)-1)^5}\left[\left(-12 \eta^2-18 \eta\right) {e_t}^6+\left(20 \eta^2-26 \eta-60\right) {e_t}^4+\left(-2 \eta^2+50 \eta+75\right) {e_t}^2\right. \\
& +\left[\left(-14 \eta^2+8 \eta-147\right) {e_t}^5+\left(8 \eta^2+22 \eta+42\right) {e_t}^3\right] \cos ^3(u)+\left[\left(17 \eta^2-17 \eta+48\right) {e_t}^6+\left(-4 \eta^2-38 \eta \right. \right. \\ & \left. \left.
+153\right) {e_t}^4 +\left(5 \eta^2-35 \eta+114\right) {e_t}^2\right] \cos ^2(u)-36 \eta+\left[\left(-\eta^2+97 \eta+12\right) {e_t}^5+\left(-16 \eta^2-74 \eta-81\right) {e_t}^3\right. \\
& \left.+\left(-\eta^2+67 \eta-246\right) {e_t}\right] \cos (u)+\sqrt{1-{e_t}^2}\left[{e_t}^3(36 \eta-90) \cos ^3(u)+\left((180-72 \eta) {e_t}^4 \right. \right. \\ & \left. \left. +(90-36 \eta) {e_t}^2\right) \cos ^2(u) \left.+\left((144 \eta-360) {e_t}^3+(90-36 \eta) {e_t}\right) \cos (u)+{e_t}^2(180-72 \eta)+36 \eta-90\right]+90\right],
\end{aligned}  
\end{equation}

\begin{equation}\label{eq:A10}
\begin{aligned}
\dot{\phi}_{3 \mathrm{PN}}^\mathrm{NS}=&\frac{1}{13440\left(1-{e_t}^2\right)^{5 / 2}({e_t} \cos (u)-1)^7}\left[\left(10080 \eta^3+40320 \eta^2-15120 \eta\right) {e_t}^{10}+\left(-52640 \eta^3-13440 \eta^2+483280 \eta\right) {e_t}^8\right.\\
&+\left(84000 \eta^3-190400 \eta^2-17220 \pi^2 \eta-50048 \eta-241920\right) {e_t}^6+\left(-52640 \eta^3+516880 \eta^2+68880 \pi^2 \eta\right.\\
&-1916048 \eta+262080) {e_t}^4+\left(4480 \eta^3-412160 \eta^2-30135 \pi^2 \eta+553008 \eta+342720\right) {e_t}^2\\
&+\left(\left(13440 \eta^3+94640 \eta^2-113680 \eta-221760\right) {e_t}^9+\left(-11200 \eta^3-112000 \eta^2+12915 \pi^2 \eta+692928 \eta\right.\right.\\
&\left.-194880) {e_t}^7+\left(4480 \eta^3+8960 \eta^2-43050 \pi^2 \eta+1127280 \eta-147840\right) {e_t}^5\right) \cos ^5(u)+\left(\left(-16240 \eta^3+12880 \eta^2\right.\right.\\
&+18480 \eta) {e_t}^{10}+\left(16240 \eta^3-91840 \eta^2+17220 \pi^2 \eta-652192 \eta+100800\right) {e_t}^8+\left(-55440 \eta^3+34160 \eta^2\right.\\
&\left.\left.-30135 \pi^2 \eta-2185040 \eta+2493120\right) {e_t}^6+\left(21840 \eta^3+86800 \eta^2+163590 \pi^2 \eta-5713888 \eta+228480\right) {e_t}^4\right)\\
&\times \cos ^4(u)+\left(\left(560 \eta^3-137200 \eta^2+388640 \eta+241920\right) {e_t}^9+\left(30800 \eta^3-264880 \eta^2-68880 \pi^2 \eta+624128 \eta\right.\right.\\
&+766080) {e_t}^7+\left(66640 \eta^3+612080 \eta^2-8610 \pi^2 \eta+6666080 \eta-6652800\right) {e_t}^5+\left(-30800 \eta^3-294000 \eta^2\right.\\
&\left.\left.-223860 \pi^2 \eta+9386432 \eta\right) {e_t}^3\right) \cos ^3(u)+67200 \eta^2+\left(\left(4480 \eta^3-20160 \eta^2+16800 \eta\right) {e_t}^{10}\right.\\
&+\left(3920 \eta^3+475440 \eta^2-17220 \pi^2 \eta+831952 \eta-725760\right) {e_t}^8+\left(-75600 \eta^3+96880 \eta^2+154980 \pi^2 \eta\right.\\
&-3249488 \eta-685440) {e_t}^6+\left(5040 \eta^3-659120 \eta^2+25830 \pi^2 \eta-7356624 \eta+6948480\right) {e_t}^4\\
&\left.+\left(-5040 \eta^3+190960 \eta^2+137760 \pi^2 \eta-7307920 \eta+107520\right) {e_t}^2\right) \cos ^2(u)-761600 \eta \\
& +\left(\left(-2240 \eta^3-168000 \eta^2-424480 \eta\right) {e_t}^9+\left(28560 \eta^3+242480 \eta^2+34440 \pi^2 \eta-1340224 \eta+725760\right) {e_t}^7\right. \\
& +\left(-33040 \eta^3-754880 \eta^2-172200 \pi^2 \eta+5458480 \eta-221760\right) {e_t}^5+\left(40880 \eta^3+738640 \eta^2+30135 \pi^2 \eta\right. \\
& \left.+1554048 \eta-2936640) {e_t}^3+\left(-560 \eta^3-100240 \eta^2-43050 \pi^2 \eta+3284816 \eta-389760\right) {e_t}\right) \cos (u) \\
& +\sqrt{1-{e_t}^2}\left(\left(\left(-127680 \eta^2+544320 \eta-739200\right) {e_t}^7+\left(-53760 \eta^2-8610 \pi^2 \eta+674240 \eta-67200\right) {e_t}^5\right) \cos ^5(u)\right. \\
& +\left(\left(161280 \eta^2-477120 \eta+537600\right) {e_t}^8+\left(477120 \eta^2+17220 \pi^2 \eta-2894080 \eta+2217600\right) {e_t}^6\right. \\
& \left.+\left(268800 \eta^2+25830 \pi^2 \eta-2721600 \eta+1276800\right) {e_t}^4\right) \cos ^4(u)+\left(\left(-524160 \eta^2+1122240 \eta-940800\right) {e_t}^7\right. \\
& +\left(-873600 \eta^2-68880 \pi^2 \eta+7705600 \eta-3897600\right) {e_t}^5+\left(-416640 \eta^2-17220 \pi^2 \eta\right. \\
& \left.+3357760 \eta-3225600) {e_t}^3\right) \cos ^3(u)+\left(\left(604800 \eta^2-504000 \eta-403200\right) {e_t}^6+\left(1034880 \eta^2+103320 \pi^2 \eta\right.\right. \\
& \left.-11195520 \eta+5779200) {e_t}^4+\left(174720 \eta^2-17220 \pi^2 \eta-486080 \eta+2688000\right) {e_t}^2\right) \cos ^2(u) \\
& +\left(\left(-282240 \eta^2-450240 \eta+1478400\right) {e_t}^5+\left(-719040 \eta^2-68880 \pi^2 \eta+8128960 \eta-5040000\right) {e_t}^3\right. \\
& \left.+\left(94080 \eta^2+25830 \pi^2 \eta-1585920 \eta-470400\right) {e_t}\right) \cos (u)-67200 \eta^2+761600 \eta \\
& +{e_t}^4\left(40320 \eta^2+309120 \eta-672000\right)+{e_t}^2\left(208320 \eta^2+17220 \pi^2 \eta-2289280 \eta+1680000\right) \\
& \left.\left.-8610 \eta \pi^2-201600\right)+8610 \eta \pi^2+201600\right].
\end{aligned}
\end{equation}
The mean anomaly $l^\mathrm{NS}$ of the nonspinning eccentric BBH can be expressed as
\begin{equation}\label{eq:A11}
l^\mathrm{NS}=u-e_t \sin u+l_{2 \mathrm{PN}}^\mathrm{NS} x^2+l_{3 \mathrm{PN}}^\mathrm{NS} x^3+\mathcal{O}\left(x^4\right),
\end{equation}
where the coefficients can be expressed as
\begin{equation}\label{eq:A12}
l_{2 \mathrm{PN}}^\mathrm{NS}=\frac{1}{8 \sqrt{1-{e_t}^2}(1-{e_t} \cos (u))}\left[24(2 \eta-5)\tan ^{-1}\left(\frac{\sin (u) \beta_\phi}{1-\cos (u) \beta_\phi}\right)({e_t} \cos (u)-1)-{e_t} \sqrt{1-{e_t}^2}(\eta-15) \eta \sin (u)\right]
\end{equation}

\begin{equation}\label{eq:A13}
\begin{aligned}
l_{3 \mathrm{PN}}^\mathrm{NS}= & \frac{1}{6720\left(1-{e_t}^2\right)^{3 / 2}(1-{e_t} \cos (u))^3}\left[35 \left(96\left(11 \eta^2-29 \eta+30\right) {e_t}^2+960 \eta^2\right.\right. \\
& \left.-2\eta\left(-13184+123 \pi^2\right)+8640\right)\tan ^{-1}\left(\frac{\sin (u) \beta_\phi}{1-\cos (u) \beta_\phi}\right)({e_t} \cos (u)-1)^3+3360(24(2 \eta-5) \tan ^{-1}\left(\frac{\sin (u) \beta_\phi}{1-\cos (u) \beta_\phi}\right) \\
& \left.+12 {e_t}(2 \eta-5) \cos (u)(-2 \tan ^{-1}\left(\frac{\sin (u) \beta_\phi}{1-\cos (u) \beta_\phi}\right))+{e_t} \sqrt{1-{e_t}^2}(\eta-15) \eta \sin (u)\right)({e_t} \cos (u)-1)^2 \\
& +{e_t} \sqrt{1-{e_t}^2}\left(140\left(13 {e_t}^4-11 {e_t}^2-2\right) \eta^3-140\left(73 {e_t}^4-325 {e_t}^2+444\right) \eta^2+\left(3220 {e_t}^4-148960 {e_t}^2-4305 \pi^2\right.\right. \\
& +143868) \eta+{e_t}^2\left(1820\left({e_t}^2-1\right) \eta^3-140\left(83 {e_t}^2+109\right) \eta^2-\left(1120 {e_t}^2+4305 \pi^2+752\right) \eta+67200\right) \cos ^2(u) \\
& \left.\left.-2 {e_t}\left(1960\left({e_t}^2-1\right) \eta^3+6720\left({e_t}^2-5\right) \eta^2+\left(-71820 {e_t}^2-4305 \pi^2+69948\right) \eta+67200\right) \cos (u)+67200\right) \sin (u)\right],
\end{aligned} 
\end{equation}
where $\beta_\phi$ is given by
\begin{equation}\label{eq:A14}
\beta_\phi=\frac{1-\sqrt{1-e_\phi^2}}{e_\phi},
\end{equation}
where phase eccentricity $e_\phi$ is given by
\begin{equation}\label{eq:A15}
e_\phi=e_t+e_{\phi 1 \mathrm{PN}} x+e_{\phi 2 \mathrm{PN}} x^2+e_{\phi 3 \mathrm{PN}} x^3+\mathcal{O}\left(x^4\right),
\end{equation}
where the coefficients can be expressed as
\begin{equation}\label{eq:A16}
e_{\phi 1 \mathrm{PN}}=-e_t(\eta-4),
\end{equation}
\begin{equation}\label{eq:A17}
e_{\phi 2 \mathrm{PN}}=  \frac{{e_t}}{96\left({e_t}^2-1\right)}\left[\left(41 \eta^2-659 \eta+1152\right) {e_t}^2+4 \eta^2\right. 
 \left.+68 \eta+\sqrt{1-{e_t}^2}(288 \eta-720)-1248\right],
\end{equation}

\begin{equation}\label{eq:A18}
\begin{aligned}
e_{\phi 3 \mathrm{PN}}= & -\frac{{e_t}}{26880\left(1-{e_t}^2\right)^{5 / 2}}\left[\left(13440 \eta^2+483840 \eta-940800\right) {e_t}^4+\left(255360 \eta^2+17220 \pi^2 \eta-2880640 \eta \right.\right. \\ & \left. +2688000\right) {e_t}^2 -268800 \eta^2+2396800 \eta+\sqrt{1-{e_t}^2}\left(\left(1050 \eta^3-134050 \eta^2+786310 \eta-860160\right) {e_t}^4\right. \\
& +\left(-18900 \eta^3+553980 \eta^2+4305 \pi^2 \eta-1246368 \eta+2042880\right) {e_t}^2+276640 \eta^2+2674480 \eta-17220 \eta \pi^2 \\
& \left.-1451520)-17220 \eta \pi^2-1747200\right].
\end{aligned}
\end{equation}

The mean motion $n^\mathrm{NS} $ of the nonspinning eccentric BBH can be expressed as
\begin{equation}\label{eq:A19}
\dot{l}^\mathrm{NS}  =n^\mathrm{NS} 
=x^{3 / 2}+n_{1 \mathrm{PN}}^\mathrm{NS} x^{5 / 2}+n_{2 \mathrm{PN}} ^\mathrm{NS}x^{7 / 2}+n_{3 \mathrm{PN}}^\mathrm{NS} x^{9 / 2}
+\mathcal{O}\left(x^{11 / 2}\right),
\end{equation}
where the coefficients can be expressed as
\begin{equation}\label{eq:A20}
n_{1 \mathrm{PN}}^\mathrm{NS}=\frac{3}{{e_t}^2-1},
\end{equation}

\begin{equation}\label{eq:A21}
n_{2 \mathrm{PN}} ^\mathrm{NS}=\frac{(26 \eta-51) {e_t}^2+28 \eta-18}{4\left({e_t}^2-1\right)^2},
\end{equation}

\begin{equation}
\begin{aligned}
n_{3 \mathrm{PN}}^\mathrm{NS}= & \frac{-1}{128\left(1-{e_t}^2\right)^{7 / 2}}\left[(1536 \eta-3840) {e_t}^4\right. 
 +(1920-768 \eta) {e_t}^2-768 \eta 
 +\sqrt{1-{e_t}^2}\left(\left(1040 \eta^2-1760 \eta+2496\right) {e_t}^4\right. \\
& +\left(5120 \eta^2+123 \pi^2 \eta-17856 \eta+8544\right) {e_t}^2  \left.\left.+896 \eta^2-14624 \eta+492 \eta \pi^2-192\right)+1920\right]
\end{aligned}
\end{equation}

The radiative dynamics at the 3PN level that we employ is sourced from Ref. \cite{Arun:2009mc,Arun:2007rg,Arun:2007sg}, encompassing both instantaneous and hereditary components expressed in terms of PN expansion parameters $x$ and temporal eccentricity $e_t$.
The PN expansion parameters $\dot{x}^\mathrm{NS}$ for the nonspinning eccentric BBH system under the ADM coordinate system can be decomposed into instantaneous and hereditary components as
\begin{equation}\label{eq:A23}
\dot{x}^\mathrm{NS}= \dot{x}_{\mathrm{inst}}^\mathrm{NS}+\dot{x}_{\mathrm{hered}}^\mathrm{NS},
\end{equation}
where instantaneous component is given by
\begin{equation}\label{eq:A24}
\dot{x}_{\mathrm{inst}}^\mathrm{NS}=  \frac{2 c^3 \eta}{3 G M} x^5 \left(\dot{x}_{\mathrm{Newt}}^\mathrm{NS}+\dot{x}_{1 \mathrm{PN}}^\mathrm{NS} x+\dot{x}_{2 \mathrm{PN}}^\mathrm{NS} x^2 \right. 
\left. +\dot{x}_{3 \mathrm{PN}}^\mathrm{NS} x^3\right),
\end{equation}
where the coefficients can be expressed as
\begin{equation}\label{eq:A25}
\dot{x}_{\mathrm{Newt}}^\mathrm{NS}=\frac{1}{\left(1-e_t^2\right)^{7 / 2}}\left(\frac{96}{5}+\frac{292}{5} e_t^2+\frac{37}{5} e_t^4\right),
\end{equation}

\begin{equation}\label{eq:A26}
\dot{x}_{1 \mathrm{PN}}^\mathrm{NS}=\frac{1}{\left(1-e_t^2\right)^{9 / 2}}\left\{-\frac{1486}{35}-\frac{264}{5} \eta+e_t^2\left(\frac{2193}{7}-570 \eta\right)+e_t^4\left(\frac{12217}{20}-\frac{5061}{10} \eta\right)+e_t^6\left(\frac{11717}{280}-\frac{148}{5} \eta\right)\right\},
\end{equation}

\begin{equation}\label{eq:A27}
\begin{aligned}
\dot{x}_{2 \mathrm{PN}}^\mathrm{NS}= & \frac{1}{\left(1-e_t^2\right)^{11 / 2}}\left\{-\frac{11257}{945}+\frac{15677}{105} \eta+\frac{944}{15} \eta^2+e_t^2\left(-\frac{2960801}{945}-\frac{2781}{5} \eta+\frac{182387}{90} \eta^2\right)+e_t^4\left(-\frac{68647}{1260}\right.\right. \\
& \left.-\frac{1150631}{140} \eta+\frac{396443}{72} \eta^2\right)+e_t^6\left(\frac{925073}{336}-\frac{199939}{48} \eta+\frac{192943}{90} \eta^2\right)+e_t^8\left(\frac{391457}{3360}-\frac{6037}{56} \eta+\frac{2923}{45} \eta^2\right) \\
& \left.+\sqrt{1-e_t^2}\left[\left(48-\frac{96}{5} \eta\right)+e_t^2\left(2134-\frac{4268}{5} \eta\right)+e_t^4\left(2193-\frac{4386}{5} \eta\right)+e_t^6\left(\frac{175}{2}-35 \eta\right)\right]\right\},
\end{aligned}
\end{equation}

\begin{equation}\label{eq:A28}
\begin{aligned}
\dot{x}_{3 \mathrm{PN}}^\mathrm{NS}= & \frac{1}{\left(1-e_t^2\right)^{13 / 2}}\left\{\frac{614389219}{148500}+\left[-\frac{57265081}{11340}+\frac{369}{2} \pi^2\right] \eta-\frac{16073}{140} \eta^2-\frac{1121}{27} \eta^3+e_t^2\left(\frac{19769277811}{693000}\right.\right. \\
& \left.+\left[\frac{66358561}{3240}+\frac{42571}{80} \pi^2\right] \eta-\frac{3161701}{840} \eta^2-\frac{1287385}{324} \eta^3\right)+e_t^4\left(-\frac{3983966927}{8316000}+\left[\frac{6451690597}{90720}\right.\right. \\
& \left.\left.-\frac{12403}{64} \pi^2\right] \eta+\frac{34877019}{1120} \eta^2-\frac{33769597}{1296} \eta^3\right)+e_t^6\left(-\frac{4548320963}{5544000}+\left[-\frac{59823689}{4032}-\frac{242563}{640} \pi^2\right] \eta\right. \\
& \left.+\frac{411401857}{6720} \eta^2-\frac{3200965}{108} \eta^3\right)+e_t^8\left(\frac{19593451667}{2464000}+\left[-\frac{6614711}{480}-\frac{12177}{640} \pi^2\right] \eta+\frac{92762}{7} \eta^2-\frac{982645}{162} \eta^3\right) \\
& +e_t^{10}\left(\frac{33332681}{197120}-\frac{1874543}{10080} \eta+\frac{109733}{840} \eta^2-\frac{8288}{81} \eta^3\right)+\sqrt{1-e_t^2}\left[-\frac{1425319}{1125}+\left[\frac{9874}{105}-\frac{41}{10} \pi^2\right] \eta+\frac{632}{5} \eta^2\right. \\
& +e_t^2\left(\frac{933454}{375}+\left[-\frac{2257181}{63}+\frac{45961}{240} \pi^2\right] \eta+\frac{125278}{15} \eta^2\right)+e_t^4\left(\frac{840635951}{21000}+\left[-\frac{4927789}{60}+\frac{6191}{32} \pi^2\right] \eta\right. \\
& \left.\left.+\frac{317273}{15} \eta^2\right)+e_t^6\left(\frac{702667207}{31500}+\left[-\frac{6830419}{252}+\frac{287}{960} \pi^2\right] \eta+\frac{232177}{30} \eta^2\right)+e_t^8\left(\frac{56403}{112}-\frac{427733}{840} \eta+\frac{4739}{30} \eta^2\right)\right] \\
& \left.+\left(\frac{54784}{175}+\frac{465664}{105} e_t^2+\frac{4426376}{525} e_t^4+\frac{1498856}{525} e_t^6+\frac{31779}{350} e_t^8\right) \ln \left[\frac{x}{x_0} \frac{1+\sqrt{1-e_t^2}}{2\left(1-e_t^2\right)}\right]\right\},
\end{aligned}
\end{equation}
where the constant $x_0$ serves the purpose of removing the logarithmic term related to gauge dependence. This constant is present in the hereditary term as well, and their contributions can offset each other. Consequently, in Eq. (\ref{eq:A28}), the choice of $x_0$ does not impact the ultimate computational outcome.

And hereditary component is given by
\begin{equation}\label{eq:A29}
\dot{x}_{\mathrm{hered}}^\mathrm{NS}= \frac{64c^3 \eta}{5 G M} x^4 \left(\dot{x}_{1.5\mathrm{PN}}^\mathrm{NS} x^{3/2}+\dot{x}_{2.5 \mathrm{PN}}^\mathrm{NS} x^{5/2} \right. 
\left. +\dot{x}_{3 \mathrm{PN}}^\mathrm{NS} x^3 \right),
\end{equation}
where the coefficients can be expressed as
\begin{equation}\label{eq:A30}
\dot{x}_{1.5\mathrm{PN}}^\mathrm{NS} = 4 \pi  \varphi\left(e_t\right),
\end{equation}

\begin{equation}\label{eq:A31}
\dot{x}_{2.5 \mathrm{PN}}^\mathrm{NS} = \pi \left[-\frac{4159}{672} \psi_x \left(e_t\right)-\frac{189}{8} \eta \zeta_x \left(e_t\right)\right],
\end{equation}

\begin{equation}\label{eq:A32}
\dot{x}_{3 \mathrm{PN}}^\mathrm{NS} = -\frac{116761}{3675} \kappa\left(e_t\right)+\left[\frac{16}{3} \pi^2-\frac{1712}{105} C-\frac{1712}{105} \ln \left(\frac{6x}{x_0}\right)\right] F\left(e_t\right),
\end{equation}
where $C$ represents Euler's constant, with a value of 0.577. And $\varphi\left(e_t\right)$, $\psi_x \left(e_t\right)$, $\zeta_x \left(e_t\right)$, $\kappa\left(e_t\right)$ and $F(e_t)$ represent some special functions, among which only $F(e_t)$ has an analytical form, which can be expressed as
\begin{equation}\label{eq:A33}
F(e_t)=\frac{1+\frac{85}{6} {e_t}^2+\frac{5171}{192} {e_t}^4+\frac{1751}{192} {e_t}^6+\frac{297}{1024} {e_t}^8}{\left(1-{e_t}^2\right)^{13 / 2}}.
\end{equation}
While other special functions can be represented in series form, the analytical formulas are not provided in Ref. \cite{Arun:2009mc,Arun:2007rg,Arun:2007sg}; only expansions for small eccentricities are offered. Nonetheless, these expansions are unsuitable for the waveform under consideration as they are designed for eccentricities below 0.1 and may provide inaccurate outcomes when applied beyond this threshold. Nevertheless, by interpolating the data from the table in Appendix B of Ref. \cite{Arun:2009mc}, accurate results can still be obtained.

In principle, the calculation results should be converted into the harmonic coordinate system. However, as $x$ is a gauge invariant quantity, such a transformation is deemed unnecessary.

The temporal eccentricity $\dot{e_t}^\mathrm{NS}$ for the nonspinning eccentric BBH system under the ADM coordinate system can be decomposed into instantaneous and hereditary components as
\begin{equation}\label{eq:A34}
\dot{e_t}^\mathrm{NS}= \dot{e_t}_{\mathrm{inst}}^\mathrm{NS}+\dot{e_t}_{\mathrm{hered}}^\mathrm{NS},
\end{equation}
where instantaneous component is given by
\begin{equation}\label{eq:A35}
\dot{e_t}_{\mathrm{inst}}^\mathrm{NS}=  -\frac{c^3 \eta}{G M} e_t x^4 \left(\dot{e_t}_{\mathrm{Newt}}^\mathrm{NS}+\dot{e_t}_{1 \mathrm{PN}}^\mathrm{NS} x+\dot{e_t}_{2 \mathrm{PN}}^\mathrm{NS} x^2 \right. 
 \left. +\dot{e_t}_{3 \mathrm{PN}}^\mathrm{NS} x^3 \right),
\end{equation}
where the coefficients can be expressed as
\begin{equation}
\dot{e_t}_{\mathrm{Newt}}^\mathrm{NS}=\frac{1}{\left(1-e_t^2\right)^{5 / 2}}\left\{\frac{304}{15}+\frac{121 e_t^2}{15}\right\}
\end{equation}

\begin{equation}
\dot{e_t}_{1 \mathrm{PN}}^\mathrm{NS}=\frac{1}{\left(1-e_t^2\right)^{7 / 2}}\left\{-\frac{939}{35}-\frac{4084}{45} \eta+e_t^2\left(\frac{29917}{105}-\frac{7753}{30} \eta\right)+e_t^4\left(\frac{13929}{280}-\frac{1664}{45} \eta\right)\right\},
\end{equation}

\begin{equation}
\begin{aligned}
\dot{e_t}_{2 \mathrm{PN}}^\mathrm{NS}= & \frac{1}{\left(1-e_t^2\right)^{9 / 2}}\left\{-\frac{961973}{1890}+\frac{70967}{210} \eta+\frac{752}{5} \eta^2+e_t^2\left(-\frac{3180307}{2520}-\frac{1541059}{840} \eta+\frac{64433}{40} \eta^2\right)\right. \\
& +e_t^4\left(\frac{23222071}{15120}-\frac{13402843}{5040} \eta+\frac{127411}{90} \eta^2\right)+e_t^6\left(\frac{420727}{3360}-\frac{362071}{2520} \eta+\frac{821}{9} \eta^2\right) \\
& \left.+\sqrt{1-e_t^2}\left[\frac{1336}{3}-\frac{2672}{15} \eta+e_t^2\left(\frac{2321}{2}-\frac{2321}{5} \eta\right)+e_t^4\left(\frac{565}{6}-\frac{113}{3} \eta\right)\right]\right\},
\end{aligned}
\end{equation}

\begin{equation}
\begin{aligned}
\dot{e_t}_{3 \mathrm{PN}}^\mathrm{NS}= & \frac{1}{\left(1-e_t^2\right)^{11 / 2}}\left\{\frac{54177075619}{6237000}+\left[\frac{7198067}{22680}+\frac{1283}{10} \pi^2\right] \eta-\frac{3000281}{2520} \eta^2-\frac{61001}{486} \eta^3+e_t^2\left(\frac{6346360709}{891000}\right.\right. \\
& \left.+\left[\frac{9569213}{360}+\frac{54001}{960} \pi^2\right] \eta+\frac{12478601}{15120} \eta^2-\frac{86910509}{19440} \eta^3\right)+e_t^4\left(-\frac{126288160777}{16632000}+\left[\frac{418129451}{181440}\right.\right. \\
& \left.\left.-\frac{254903}{1920} \pi^2\right] \eta+\frac{478808759}{20160} \eta^2-\frac{2223241}{180} \eta^3\right)+e_t^6\left(\frac{5845342193}{1232000}+\left[-\frac{98425673}{10080}-\frac{6519}{640} \pi^2\right] \eta\right. \\
& \left.+\frac{6538757}{630} \eta^2-\frac{11792069}{2430} \eta^3\right)+e_t^8\left(\frac{302322169}{1774080}-\frac{1921387}{10080} \eta+\frac{41179}{216} \eta^2-\frac{193396}{1215} \eta^3\right) \\
& +\sqrt{1-e_t^2}\left[-\frac{22713049}{15750}+\left[-\frac{5526991}{945}+\frac{8323}{180} \pi^2\right] \eta+\frac{54332}{45} \eta^2+e_t^2\left(\frac{89395687}{7875}+\left[-\frac{38295557}{1260}\right.\right.\right. \\
& \left.\left.+\frac{94177}{960} \pi^2\right] \eta+\frac{681989}{90} \eta^2\right)+e_t^4\left(\frac{5321445613}{378000}+\left[-\frac{26478311}{1512}+\frac{2501}{2880} \pi^2\right] \eta+\frac{225106}{45} \eta^2\right) \\
& \left.+e_t^6\left(\frac{186961}{336}-\frac{289691}{504} \eta+\frac{3197}{18} \eta^2\right)\right]+\frac{730168}{23625} \frac{1}{1+\sqrt{1-e_t^2}} \\
& \left.+\frac{304}{15}\left(\frac{82283}{1995}+\frac{297674}{1995} e_t^2+\frac{1147147}{15960} e_t^4+\frac{61311}{21280} e_t^6\right) \ln \left[\frac{x}{x_0} \frac{1+\sqrt{1-e_t^2}}{2\left(1-e_t^2\right)}\right]\right\} .
\end{aligned}
\end{equation}
And hereditary component is given by
\begin{equation}\label{eq:A40}
\dot{e_t}_{\mathrm{hered}}^\mathrm{NS}=  \frac{32c^3 \eta}{5 G M} e_t x^4 \left(\dot{e_t}_{1.5\mathrm{PN}}^\mathrm{NS} x^{3/2}+\dot{e_t}_{2.5 \mathrm{PN}}^\mathrm{NS} x^{5/2} \right. 
\left. +\dot{e_t}_{3 \mathrm{PN}}^\mathrm{NS} x^3 \right),
\end{equation}
where the coefficients can be expressed as
\begin{equation}\label{eq:A41}
\dot{e_t}_{1.5\mathrm{PN}}^\mathrm{NS} = -\frac{985}{48} \pi \varphi_e\left(e_t\right),
\end{equation}

\begin{equation}\label{eq:A42}
\dot{e_t}_{2.5 \mathrm{PN}}^\mathrm{NS} = \pi \left[\frac{55691}{1344} \psi_e\left(e_t\right)+\frac{19067}{126} \eta \zeta_e\left(e_t\right)\right],
\end{equation}

\begin{equation}\label{eq:A43}
\dot{e_t}_{3 \mathrm{PN}}^\mathrm{NS} =
\left(\frac{89789209}{352800}-\frac{87419}{630} \ln 2+\frac{78003}{560} \ln 3\right)
 \times \kappa_e\left(e_t\right)
 -\frac{769}{96}\left[\frac{16}{3} \pi^2-\frac{1712}{105} C-\frac{1712}{105} \ln  
\left(\frac{6x}{x_0}\right)\right] F_e\left(e_t\right),
\end{equation}
where $\varphi_e\left(e_t\right)$, $\psi_e \left(e_t\right)$, $\zeta_e \left(e_t\right)$, $\kappa_e \left(e_t\right)$ and $F_e \left(e_t\right)$ represent some special functions, among which only $F_e (e_t)$ has an analytical form, which can be expressed as

\begin{equation}
F_e(e_t)=\frac{1+\frac{2782}{769} {e_t}^2+\frac{10721}{6152} {e_t}^4+\frac{1719}{24608} {e_t}^6}{\left(1-{e_t}^2\right)^{11 / 2}}.
\end{equation}
Similarly, accurate results for the other four special functions can be obtained by interpolating the data presented in the table in Appendix B of Ref. \cite{Arun:2009mc}. Moreover, as previously mentioned, it is imperative to convert $e_t$ from ADM coordinates to harmonic coordinates. The transformation formula is as follows \cite{Arun:2009mc}:

\begin{equation}
\frac{e_t^{\mathrm{MH}}}{e_t^{\mathrm{ADM}}}=  1+\frac{x^2}{1-e_t^2}\left(-\frac{1}{4}-\frac{17}{4} \eta \right) 
 +\frac{x^3}{\left(1-e_t^2\right)^2}\left(-\frac{1}{2}+\left[-\frac{16739}{1680}+\frac{21}{16} \pi^2\right] \eta+\frac{83}{24} \eta^2+e_t^2\left(-\frac{1}{2}-\frac{249}{16} \eta+\frac{241}{24} \eta^2\right)\right],
\end{equation}
where the superscript MH denotes modified harmonic coordinates, with the right-hand side of the equation represented in ADM coordinates.

The (2,2) mode of the gravitational waveform $h^{22, \mathrm{NS}}$ for the nonspinning eccentric BBH system, as provided in Ref. \cite{Mishra:2015bqa,Boetzel:2019nfw,Ebersold:2019kdc}, can be expressed as
\begin{equation}\label{eq:A46}
h^{22, \mathrm{NS}}=\frac{4 G M \eta }{c^2 R} \sqrt{\frac{\pi}{5}} e^{-2 i \phi} H^{22, \mathrm{NS}},
\end{equation}
where amplitude $H^{22, \mathrm{NS}}$ can be expressed as the sum of different PN orders,
\begin{equation}\label{eq:A47}
H^{22, \mathrm{NS}}=H_{\text {Newt }}^{22, \mathrm{NS}}+H_{1 \mathrm{PN}}^{22, \mathrm{NS}}+H_{2 \mathrm{PN}}^{22, \mathrm{NS}}+H_{2.5 \mathrm{PN}}^{22, \mathrm{NS}}+H_{3 \mathrm{PN}}^{22, \mathrm{NS}},
\end{equation}
where the individual PN terms read as
\begin{equation}\label{eq:A48}
H_{\text {Newt }}^{22, \mathrm{NS}}=\frac{G M}{r}+r^2 \dot{\phi}^2+2 i r \dot{r} \dot{\phi}-\dot{r}^2,
\end{equation}
\begin{equation}\label{eq:A49}
\begin{aligned}
H_{1 \mathrm{PN}}^{22, \mathrm{NS}}= & \frac{1}{c^2}\left[\frac{G^2 M^2}{r^2}\left(-5+\frac{\eta}{2}\right)+\frac{G M \dot{r}^2}{r}\left(-\frac{15}{14}-\frac{16 \eta}{7}\right)+\left(-\frac{9}{14}+\frac{27 \eta}{14}\right) \dot{r}^4+r\left(\frac{9 i}{7}-\frac{27 i \eta}{7}\right) \dot{r}^3 \dot{\phi}\right. \\
& \left.+G M r\left(\frac{11}{42}+\frac{26 \eta}{7}\right) \dot{\phi}^2+r^4\left(\frac{9}{14}-\frac{27 \eta}{14}\right) \dot{\phi}^4+\dot{r}\left(G M\left(\frac{25 i}{21}+\frac{45 i \eta}{7}\right) \dot{\phi}+r^3\left(\frac{9 i}{7}-\frac{27 i \eta}{7}\right) \dot{\phi}^3\right)\right]
\end{aligned}
\end{equation}
\begin{equation}\label{eq:A50}
\begin{aligned}
H_{2 \mathrm{PN}}^{22, \mathrm{NS}}= & \frac{1}{c^4}\left[\frac{G^3 M^3}{r^3}\left(\frac{757}{63}+\frac{181 \eta}{36}+\frac{79 \eta^2}{126}\right)+\left(-\frac{83}{168}+\frac{589 \eta}{168}-\frac{1111 \eta^2}{168}\right) \dot{r}^6+r\left(\frac{83 i}{84}-\frac{589 i \eta}{84}+\frac{1111 i \eta^2}{84}\right) \dot{r}^5 \dot{\phi}\right. \\
& +G^2 M^2\left(-\frac{11891}{1512}-\frac{5225 \eta}{216}+\frac{13133 \eta^2}{1512}\right) \dot{\phi}^2+G M r^3\left(\frac{835}{252}+\frac{19 \eta}{252}-\frac{2995 \eta^2}{252}\right) \dot{\phi}^4 \\
& +r^6\left(\frac{83}{168}-\frac{589 \eta}{168}+\frac{1111 \eta^2}{168}\right) \dot{\phi}^6+\dot{r}^4\left(\frac{G M}{r}\left(-\frac{557}{168}+\frac{83 \eta}{21}+\frac{214 \eta^2}{21}\right)+r^2\left(-\frac{83}{168}+\frac{589 \eta}{168}-\frac{1111 \eta^2}{168}\right) \dot{\phi}^2\right) \\
& +\dot{r}^3\left(G M\left(\frac{863 i}{126}-\frac{731 i \eta}{63}-\frac{211 i \eta^2}{9}\right) \dot{\phi}+r^3\left(\frac{83 i}{42}-\frac{589 i \eta}{42}+\frac{1111 i \eta^2}{42}\right) \dot{\phi}^3\right) \\
& +\dot{r}^2\left(\frac{G^2 M^2}{r^2}\left(\frac{619}{252}-\frac{2789 \eta}{252}-\frac{467 \eta^2}{126}\right)+G M r\left(\frac{11}{28}-\frac{169 \eta}{14}-\frac{58 \eta^2}{21}\right) \dot{\phi}^2+r^4\left(\frac{83}{168}-\frac{589 \eta}{168}+\frac{1111 \eta^2}{168}\right) \dot{\phi}^4\right) \\
& +\dot{r}\left(\frac{G^2 M^2}{r}\left(-\frac{773 i}{189}-\frac{3767 i \eta}{189}+\frac{2852 i \eta^2}{189}\right) \dot{\phi}+G M r^2\left(\frac{433 i}{84}+\frac{103 i \eta}{12}-\frac{1703 i \eta^2}{84}\right) \dot{\phi}^3\right. \\
& \left.\left.+r^5\left(\frac{83 i}{84}-\frac{589 i \eta}{84}+\frac{1111 i \eta^2}{84}\right) \dot{\phi}^5\right)\right],
\end{aligned}
\end{equation}
\begin{equation}\label{eq:A51}
H_{2.5 \mathrm{PN}}^{22, \mathrm{NS}}=\frac{1}{c^5}\left[-\frac{122 G^2 M^2 \eta \dot{r}^3}{35 r^2}-\frac{468 i G^3 M^3 \eta \dot{\phi}}{35 r^2}+\frac{184 i G^2 M^2 \eta \dot{r}^2 \dot{\phi}}{35 r}-\frac{316}{35} i G^2 M^2 r \eta \dot{\phi}^3+\dot{r}\left(\frac{2 G^3 M^3 \eta}{105 r^3}-\frac{121}{5} G^2 M^2 \eta \dot{\phi}^2\right)\right],
\end{equation}
\begin{equation}\label{eq:A52}
\begin{aligned}
H_{3 \mathrm{PN}}^{22, \mathrm{NS}}= & \frac{1}{c^6}\left[\frac{G^4 M^4}{r^4}\left(-\frac{512714}{51975}+\left(-\frac{1375951}{13860}+\frac{41 \pi^2}{16}\right) \eta+\frac{1615 \eta^2}{616}+\frac{2963 \eta^3}{4158}\right)\right. \\
& +\left(-\frac{507}{1232}+\frac{6101 \eta}{1232}-\frac{12525 \eta^2}{616}+\frac{34525 \eta^3}{1232}\right) \dot{r}^8+r\left(\frac{507 i}{616}-\frac{6101 i \eta}{616}+\frac{12525 i \eta^2}{308}-\frac{34525 i \eta^3}{616}\right) \dot{r}^7 \dot{\phi} \\
& +\frac{G^3 m^3}{r}\left(\frac{42188851}{415800}+\left(\frac{190703}{3465}-\frac{123 \pi^2}{64}\right) \eta-\frac{18415 \eta^2}{308}+\frac{281473 \eta^3}{16632}\right) \dot{\phi}^2 \\
& +G^2 M^2 r^2\left(\frac{328813}{55440}-\frac{374651 \eta}{33264}+\frac{249035 \eta^2}{4158}-\frac{1340869 \eta^3}{33264}\right) \dot{\phi}^4 \\
& +G M r^5\left(\frac{12203}{2772}-\frac{36427 \eta}{2772}-\frac{13667 \eta^2}{1386}+\frac{49729 \eta^3}{924}\right) \dot{\phi}^6+r^8\left(\frac{507}{1232}-\frac{6101 \eta}{1232}+\frac{12525 \eta^2}{616}-\frac{34525 \eta^3}{1232}\right) \dot{\phi}^8 \\
& +\dot{r}^4\left(\frac{G^2 M^2}{r^2}\left(-\frac{92567}{13860}+\frac{7751 \eta}{396}+\frac{400943 \eta^2}{11088}+\frac{120695 \eta^3}{3696}\right)\right. \\
& \left.+G M r\left(-\frac{42811}{11088}+\frac{6749 \eta}{1386}+\frac{19321 \eta^2}{693}-\frac{58855 \eta^3}{1386}\right) \dot{\phi}^2\right)+\dot{r}^6\left(\frac{G M}{r}\left(-\frac{5581}{1232}+\frac{4694 \eta}{231}-\frac{3365 \eta^2}{462}-\frac{1850 \eta^3}{33}\right)\right. \\
& \left.+r^2\left(-\frac{507}{616}+\frac{6101 \eta}{616}-\frac{12525 \eta^2}{308}+\frac{34525 \eta^3}{616}\right) \dot{\phi}^2\right)+\dot{r}^5\left(G M\left(\frac{17233 i}{1848}-\frac{31532 i \eta}{693}+\frac{65575 i \eta^2}{2772}+\frac{85145 i \eta^3}{693}\right) \dot{\phi}\right. \\
& \left.+r^3\left(\frac{1521 i}{616}-\frac{18303 i \eta}{616}+\frac{37575 i \eta^2}{308}-\frac{103575 i \eta^3}{616}\right) \dot{\phi}^3\right)\\ & +\dot{r}^3\left(\frac{G^2 M^2}{r}\left(\frac{39052 i}{3465}-\frac{154114 i \eta}{2079}-\frac{246065 i \eta^2}{4158}-\frac{365725 i \eta^3}{4158}\right) \dot{\phi}\right. \\
& \left.+G M r^2\left(\frac{13867 i}{792}-\frac{191995 i \eta}{2772}-\frac{8741 i \eta^2}{5544}+\frac{52700 i \eta^3}{231}\right) \dot{\phi}^3+r^5\left(\frac{1521 i}{616}-\frac{18303 i \eta}{616}+\frac{37575 i \eta^2}{308}-\frac{103575 i \eta^3}{616}\right) \dot{\phi}^5\right) \\
& +\dot{r}^2\left(\frac{G^3 M^3}{r^3}\left(\frac{913799}{29700}+\left(\frac{174679}{2310}+\frac{123 \pi^2}{32}\right) \eta-\frac{158215 \eta^2}{2772}-\frac{12731 \eta^3}{4158}\right)\right. \\
& +G^2 M^2\left(\frac{20191}{18480}-\frac{3879065 \eta}{33264}-\frac{411899 \eta^2}{8316}-\frac{522547 \eta^3}{33264}\right) \dot{\phi}^2 \\
& \left.+G M r^3\left(\frac{381}{77}-\frac{101237 \eta}{2772}+\frac{247505 \eta^2}{5544}+\frac{394771 \eta^3}{5544}\right) \dot{\phi}^4+r^6\left(\frac{507}{616}-\frac{6101 \eta}{616}+\frac{12525 \eta^2}{308}-\frac{34525 \eta^3}{616}\right) \dot{\phi}^6\right) \\
& +\dot{r}\left(\frac{G^3 M^3}{r^2}\left(-\frac{68735 i}{378}+\left(-\frac{57788 i}{315}+\frac{123 i \pi^2}{32}\right) \eta-\frac{701 i \eta^2}{27}+\frac{11365 i \eta^3}{378}\right) \dot{\phi}\right. \\
& +G^2 M^2 r\left(\frac{91229 i}{13860}+\frac{97861 i \eta}{4158}+\frac{919811 i \eta^2}{8316}-\frac{556601 i \eta^3}{8316}\right) \dot{\phi}^3 \\ & +G M r^4\left(\frac{6299 i}{792}-\frac{68279 i \eta}{5544}-\frac{147673 i \eta^2}{2772}+\frac{541693 i \eta^3}{5544}\right) \dot{\phi}^5 \left.\left.+r^7\left(\frac{507 i}{616}-\frac{6101 i \eta}{616}+\frac{12525 i \eta^2}{308}-\frac{34525 i \eta^3}{616}\right) \dot{\phi}^7\right)\right] .
\end{aligned}
\end{equation}

The high-order harmonic mode for nonspinning BBH can be expressed as
\begin{equation}\label{eq:A53}
h^{\ell m, \mathrm{NS}}=\frac{4 G M \eta}{c^4 R} \sqrt{\frac{\pi}{5}} e^{-i m \phi} H^{\ell m, \mathrm{NS}}.
\end{equation}
For the high-order modes (2,1), (3,3), (3,2), (4,4), (4,3), and (5,5), we simplify by presenting only the amplitude of leading PN order waveform. In regard to the phase, it is important to emphasize that distinct high-order modes require different values of $m$ in Eq. (\ref{eq:19}) and Eq. (\ref{eq:A53}).

The amplitudes of the high-order modes (2,1), (3,3), (3,2), (4,4), (4,3), and (5,5) at the leading PN order for nonspinning BBH can be expressed as
\begin{equation}\label{eq:A54}
H_{\text {0.5PN }}^{21, \mathrm{NS}} = \frac{1}{c}\left(\frac{2}{3} i G M \dot{\phi} \sqrt{1-4 \eta}\right),
\end{equation}

\begin{equation}\label{eq:A55}
H_{\text {0.5PN }}^{33, \mathrm{NS}}= \frac{1}{c} \sqrt{1 - 4 \eta} \left(
-\frac{i}{2} \sqrt{\frac{35}{6}} G M \dot{\phi} 
- i \sqrt{\frac{5}{42}} \dot{\phi}^3 r^3 
+ \frac{\sqrt{\frac{10}{21}} G M \dot{r}}{r} \right.
 \left. + \sqrt{\frac{15}{14}} \dot{\phi}^2 r^2 \dot{r}
+ i \sqrt{\frac{15}{14}} \dot{\phi} r \dot{r}^2 
- \sqrt{\frac{5}{42}} \dot{r}^3  \right),
\end{equation}
\begin{equation}\label{eq:A56}
H_{\text {1PN }}^{32, \mathrm{NS}} = \frac{1}{c^2} \sqrt{\frac{5}{7}}\left(\frac{2}{3} G M \dot{\phi}^2 r 
+ \frac{i}{6}  G M \dot{\phi} \dot{r} 
- 2 G M \dot{\phi}^2 r \eta 
- \frac{i}{2} G M \dot{\phi} \dot{r} \eta \right),
\end{equation}
\begin{equation}\label{eq:A57}
\begin{aligned}
H_{\text {1PN }}^{44, \mathrm{NS}} = & \frac{1}{c^2} \left(-\frac{\sqrt{35} G^2 M^2}{36 r^2} 
- \frac{17}{12} \sqrt{\frac{5}{7}} G M \dot{\phi}^2 r 
- \frac{1}{6} \sqrt{\frac{5}{7}} \dot{\phi}^4 r^4 
- \frac{3i}{2} \sqrt{\frac{5}{7}} G M \dot{\phi} \dot{r} 
- \frac{2i}{3} \sqrt{\frac{5}{7}} \dot{\phi}^3 r^3 \dot{r} 
 + \frac{\sqrt{\frac{5}{7}} G M \dot{r}^2}{2 r} \right.\\
&+ \sqrt{\frac{5}{7}} \dot{\phi}^2 r^2 \dot{r}^2  
+ \frac{2i}{3} \sqrt{\frac{5}{7}} \dot{\phi} r \dot{r}^3 
- \frac{1}{6} \sqrt{\frac{5}{7}} \dot{r}^4 
 + \frac{\sqrt{35} G^2 M^2 \eta}{12 r^2} 
+ \frac{17}{4} \sqrt{\frac{5}{7}} G M \dot{\phi}^2 r \eta 
+ \frac{1}{2} \sqrt{\frac{5}{7}} \dot{\phi}^4 r^4 \eta \\
& \left.+ \frac{9i}{2} \sqrt{\frac{5}{7}} G M \dot{\phi} \dot{r} \eta 
 + 2i \sqrt{\frac{5}{7}} \dot{\phi}^3 r^3 \dot{r} \eta 
- \frac{3 \sqrt{\frac{5}{7}} G M \dot{r}^2 \eta}{2 r} 
- 3 \sqrt{\frac{5}{7}} \dot{\phi}^2 r^2 \dot{r}^2 \eta 
- 2i \sqrt{\frac{5}{7}} \dot{\phi} r \dot{r}^3 \eta 
+ \frac{1}{2} \sqrt{\frac{5}{7}} \dot{r}^4 \eta \right),
\end{aligned}
\end{equation}
\begin{equation}\label{eq:A58}
\begin{aligned}
H_{\text {1.5PN }}^{43, \mathrm{NS}} = & \frac{1}{c^3} \sqrt{1 - 4 \eta} \left(-\frac{i \sqrt{\frac{2}{35}} G^2 M^2 \dot{\phi} }{3 r} 
- \frac{23 i G M \dot{\phi}^3 r^2 }{6 \sqrt{70}} 
+ \frac{\sqrt{\frac{5}{14}} G M \dot{\phi}^2 r \dot{r} }{3} 
 + \frac{i G M \dot{\phi} \dot{r}^2 }{3 \sqrt{70}}\right.\\
&\left. + \frac{2 i \sqrt{\frac{2}{35}} G^2 M^2 \dot{\phi} \eta}{3 r} 
+ \frac{23 i G M \dot{\phi}^3 r^2 \eta}{3 \sqrt{70}} 
 - \frac{\sqrt{\frac{10}{7}} G M \dot{\phi}^2 r \dot{r} \eta}{3} 
- \frac{i \sqrt{\frac{2}{35}} G M \dot{\phi} \dot{r}^2 \eta}{3}\right),
\end{aligned}
\end{equation}
\begin{equation}\label{eq:A59}
\begin{aligned}
H_{\text {1.5PN }}^{55, \mathrm{NS}} =& \frac{1}{c^3} \sqrt{1 - 4 \eta} \left( \frac{43 i G^2 M^2 \dot{\phi} }{12 \sqrt{66} \, r} 
+ \frac{13 i \sqrt{\frac{11}{6}} G M \dot{\phi}^3 r^2 }{16} 
+ \frac{i \dot{\phi}^5 r^5 }{2 \sqrt{66}} 
- \frac{41 G^2 M^2 \dot{r}}{24 \sqrt{66} \, r^2} 
- \frac{13 G M \dot{\phi}^2 r \dot{r}}{\sqrt{66}} 
 - \frac{5 \dot{\phi}^4 r^4 \dot{r}}{2 \sqrt{66}} \right.\\ 
&- \frac{i \sqrt{\frac{33}{2}} G M \dot{\phi} \dot{r}^2}{4} 
- \frac{5 i \dot{\phi}^3 r^3 \dot{r}^2}{\sqrt{66}} 
+ \frac{\sqrt{\frac{2}{33}} G M \dot{r}^3 }{r} 
+ \frac{5 \dot{\phi}^2 r^2 \dot{r}^3 }{\sqrt{66}} 
 + \frac{5 i \dot{\phi} r \dot{r}^4}{2 \sqrt{66}} 
- \frac{\dot{r}^5}{2 \sqrt{66}} 
- \frac{43 i G^2 M^2 \dot{\phi} \eta}{6 \sqrt{66} \, r}\\ 
&- \frac{13 i \sqrt{\frac{11}{6}} G M \dot{\phi}^3 r^2 \eta}{8} 
 - \frac{i \dot{\phi}^5 r^5 \eta}{\sqrt{66}} 
+ \frac{41 G^2 M^2 \dot{r} \eta}{12 \sqrt{66} \, r^2} 
+ \frac{13 \sqrt{\frac{2}{33}} G M \dot{\phi}^2 r \dot{r} \eta}{\eta} 
+ \frac{5 \dot{\phi}^4 r^4 \dot{r} \eta}{\sqrt{66}} \\
&\left. + \frac{i \sqrt{\frac{33}{2}} G M \dot{\phi} \dot{r}^2 \eta}{2} 
+ \frac{5 i \sqrt{\frac{2}{33}} \dot{\phi}^3 r^3 \dot{r}^2 \eta}{\eta} 
- \frac{2 \sqrt{\frac{2}{33}} G M \dot{r}^3 \eta}{r} 
- \frac{5 \sqrt{\frac{2}{33}} \dot{\phi}^2 r^2 \dot{r}^3 \eta}{\eta} 
 - \frac{5 i \dot{\phi} r \dot{r}^4 \eta}{\sqrt{66}} 
+ \frac{\dot{r}^5 \eta}{\sqrt{66}}\right).
\end{aligned}
\end{equation}

\subsection{Spin-aligned waveforms}
Building upon the preceding nonspinning setup, this section delves into the 3PN dynamics of spin-aligned BBH in eccentric orbits. We commence by introducing the conservative dynamics at the 3PN level, as detailed in Ref. \cite{Henry:2023tka}. The orbital radius $r^\mathrm{SP}$ of the spin-aligned eccentric BBH can be expressed as
\begin{equation}\label{eq:A60}
r^{\mathrm{SP}}=r^{\mathrm{NS}} + r^{\mathrm{SO}} + r^{\mathrm{SS}},
\end{equation}
where the nonspinning component $r^{\mathrm{NS}}$ has been detailed in the preceding section, and the spin-orbit coupling component $r^{\mathrm{SO}}$ can be expressed as
\begin{equation}\label{eq:A61}
r^{\mathrm{SO}}=r_{1.5 \mathrm{PN}}^{\mathrm{SO}} + r_{2.5 \mathrm{PN}}^{\mathrm{SO}},
\end{equation}

\begin{equation}\label{eq:A62}
r_{1.5 \mathrm{PN}}^{\mathrm{SO}} =-\frac{2}{3x} \left(\frac{x}{1 - e_t^2}\right)^{3/2} \left[\delta \chi_A \left(1 + 3 e_t^2 \right) + \chi_S - 3 e_t^2 (1 - \eta) \chi_S + \eta \chi_S - 4 e_t \delta \chi_A \cos{u} + 2 e_t (2 - \eta) \chi_S \cos{u} \right],
\end{equation}
\begin{equation}\label{eq:A63}
\begin{aligned}
r_{2.5 \mathrm{PN}}^{\mathrm{SO}} = & \frac{x^{3/2} \epsilon^5}{3 (1 - e_t^2)^3} 
\Bigg[2 (1 - e_t^2)^2 \Big(2 \delta (-3 + \eta) \chi_A - (6 - 8 \eta + \eta^2) \chi_S \Big) (2 + e_t \cos{u}) \\
&+ \sqrt{1 - e_t^2} \Big(
\delta \Big( e_t^2 (84 - 33 \eta) + e_t^4 (12 - 8 \eta) - 4 (-6 + \eta) \Big) \chi_A \\ 
&+ \Big( 24 - 28 \eta + 4 e_t^4 (3 - 5 \eta + 2 \eta^2) + e_t^2 (84 - 87 \eta + 10 \eta^2) \Big) \chi_S \\
&+ e_t \Big( \delta \Big( -60 + 17 \eta + 4 e_t^2 (-15 + 7 \eta) \Big) \chi_A 
- \Big( 60 - 77 \eta + 4 \eta^2 + 2 e_t^2 (30 - 29 \eta + 7 \eta^2) \Big) \chi_S \Big) \cos{u}\Big)\Bigg].
\end{aligned}
\end{equation}
The spin-spin coupling component $r^{\mathrm{SS}}$ can be expressed as
\begin{equation}\label{eq:A64}
r^{\mathrm{SS}}=r_{2 \mathrm{PN}}^{\mathrm{SS}} + r_{3 \mathrm{PN}}^{\mathrm{SS}},
\end{equation}
\begin{equation}\label{eq:A65}
r_{2 \mathrm{PN}}^{\mathrm{SS}} = \frac{x}{2 (-1 + e_t^2)^2} \left[ \kappa_S - 2 \kappa_S \eta + \chi_A^2 - 4 \eta \chi_A^2 + \chi_S^2 + \delta (\kappa_A + 2 \chi_A \chi_S) \right] \left(1 + e_t^2 - 2 e_t \cos{u}\right)
\end{equation}

\begin{equation}\label{eq:A66}
\begin{aligned}
r_{3 \mathrm{PN}}^{\mathrm{SS}} = & \frac{x^2}{18 (1 - e_t^2)^3}  \bigg\{99 \kappa_S + 306 e_t^2 \kappa_S + 45 e_t^4 \kappa_S - 252 \kappa_S \eta - 750 e_t^2 \kappa_S \eta - 105 e_t^4 \kappa_S \eta + 18 \kappa_S \eta^2 + 168 e_t^2 \kappa_S \eta^2  \\ & + 48 e_t^4 \kappa_S \eta^2 
- 3 \delta \kappa_A \left(-33 + 5 e_t^4 (-3 + \eta) + 18 \eta + 2 e_t^2 (-51 + 23 \eta)\right) + 139 \chi_A^2 + 474 e_t^2 \chi_A^2 + 45 e_t^4 \chi_A^2 \\ & - 574 \eta \chi_A^2 - 2007 e_t^2 \eta \chi_A^2 - 213 e_t^4 \eta \chi_A^2 
+ 36 \eta^2 \chi_A^2 + 336 e_t^2 \eta^2 \chi_A^2 + 96 e_t^4 \eta^2 \chi_A^2 + 2 \delta \left(139 + e_t^2 (474 - 309 \eta) \right. \\ & \left. + e_t^4 (45 - 33 \eta) - 122 \eta\right) \chi_A \chi_S + 139 \chi_S^2 
+ 474 e_t^2 \chi_S^2 + 45 e_t^4 \chi_S^2 - 226 \eta \chi_S^2 - 507 e_t^2 \eta \chi_S^2 - 33 e_t^4 \eta \chi_S^2 \\ & + 40 \eta^2 \chi_S^2 + 120 e_t^2 \eta^2 \chi_S^2 
+ e_t \left[-3 \kappa_S \left(72 - 178 \eta + 20 \eta^2 + e_t^2 \left(78 - 191 \eta + 58 \eta^2\right)\right) \right. \\ & \left. + \delta \left(3 \kappa_A \left(-72 + 34 \eta + e_t^2 \left(-78 + 35 \eta\right)\right) + 4 \left(-176 + 157 \eta + 3 e_t^2 \left(-51 + 25 \eta\right)\right) \chi_A \chi_S\right] \right. \\ 
& \left. - 2 \left[\left(176 - 737 \eta + 60 \eta^2 + 3 e_t^2 \left(51 - 220 \eta + 58 \eta^2\right)\right) \chi_A^2 + \left(176 - 281 \eta + 80 \eta^2 - 51 e_t^2 \left(-3 + 2 \eta\right)\right) \chi_S^2\right] \right] \cos u \\ 
&+ 3 (1 - e_t^2)^{3/2} \left[\delta \kappa_A (-14 + 5 \eta) + \kappa_S (-14 + 33 \eta - 6 \eta^2) + 4 \delta (-11 + 9 \eta) \chi_A \chi_S - 2 \left[\left(11 - 46 \eta + 6 \eta^2\right) \chi_A^2 \right. \right. \\ & \left. \left. + \left(11 - 16 \eta + 4 \eta^2\right) \chi_S^2\right]\right] (2 + e_t \cos u)\bigg\},
\end{aligned}
\end{equation}
where we introduce some new quantities 
\begin{equation}\label{eq:A67}
\chi_S \equiv \frac{1}{2}\left(\chi_1+\chi_2\right),
\end{equation}
\begin{equation}\label{eq:A68}
\chi_A \equiv \frac{1}{2}\left(\chi_1-\chi_2\right),
\end{equation}

\begin{equation}\label{eq:A69}
\kappa_S \equiv \frac{1}{2}\left((\kappa_1 - 1) {\chi_1}^2 + (\kappa_2 - 1) {\chi_2}^2 \right),
\end{equation}

\begin{equation}\label{eq:A70}
\kappa_A \equiv \frac{1}{2}\left((\kappa_1 - 1) {\chi_1}^2 - (\kappa_2 - 1) {\chi_2}^2 \right),
\end{equation}
where $\kappa_1$ and $\kappa_2$ are the spin quadrupole constants, which equal 1 for black holes.

The relative angular velocity $\dot{\phi} ^\mathrm{NS}$ of the spin-aligned eccentric BBH can be expressed as
\begin{equation}\label{eq:A71}
\dot{\phi}^{\mathrm{SP}}=\dot{\phi}^{\mathrm{NS}} + \dot{\phi}^{\mathrm{SO}} + \dot{\phi}^{\mathrm{SS}},
\end{equation}
where the spin-orbit coupling component $\dot{\phi}^{\mathrm{SO}}$ can be expressed as
\begin{equation}\label{eq:A72}
\dot{\phi}^{\mathrm{SO}}=\dot{\phi}_{1.5 \mathrm{PN}}^{\mathrm{SO}} + \dot{\phi}_{2.5 \mathrm{PN}}^{\mathrm{SO}},
\end{equation}
\begin{equation}\label{eq:A73}
\dot{\phi}_{1.5 \mathrm{PN}}^{\mathrm{SO}} =  \frac{2 e_t x^3 \left(\delta \chi_A + \chi_S\right) \left(e_t - \cos u\right)}{\left(1 - e_t^2\right) \left(1 - e_t \cos u\right)^3},
\end{equation}
\begin{equation}\label{eq:A74}
\begin{aligned}
\dot{\phi}_{2.5 \mathrm{PN}}^{\mathrm{SO}} = &\frac{x^4}{6 (1 - e_t^2)^2 (-1 + e_t \cos u)^5} \Bigg[ -\delta \left( e_t^4 (36 - 54 \eta) - 24 (-3 + \eta) + 6 e_t^6 (-2 + \eta) + e_t^2 (56 + 31 \eta) \right) \chi_A \\
& +\left( -12 (6 - 8 \eta + \eta^2) + 6 e_t^6 (2 - 3 \eta + 2 \eta^2) - 6 e_t^4 (6 - 27 \eta + 8 \eta^2) + e_t^2 (-56 - 57 \eta + 50 \eta^2) \right) \chi_S \\
&-e_t \left[\delta \left( -224 + 44 \eta + 6 e_t^4 (-6 + 11 \eta) + e_t^2 (-196 + 13 \eta) \right) \chi_A + \left( -4 (56 - 60 \eta + \eta^2) \right. \right. \\ 
& \left. \left. - 6 e_t^4 (6 - 23 \eta + 4 \eta^2) + e_t^2 (-196 + 171 \eta + 34 \eta^2) \right) \chi_S \right] \cos u  + e_t^2 \left[\delta \left( 8 (-29 + 5 \eta) + 6 e_t^4 (-6 + 5 \eta) \right. \right. \\ & \left. \left. + e_t^2 (-188 + 53 \eta) \right) \chi_A + \left( -4 (58 - 81 \eta + 2 \eta^2) - 6 e_t^4 (6 - 5 \eta + 2 \eta^2) + e_t^2 (-188 + 195 \eta + 26 \eta^2) \right) \chi_S \right] \cos^2 u \\ &  -e_t^3 \left[\delta \left( -68 + 14 \eta + 3 e_t^2 (-28 + 9 \eta) \right) \chi_A + \left( -68 + 162 \eta - 4 \eta^2 + 3 e_t^2 (-28 + 7 \eta + 2 \eta^2) \right) \chi_S \right] \cos^3 u \\ &  -12 \sqrt{1 - e_t^2} \left[ 2 \delta (-3 + \eta) \chi_A - (6 - 8 \eta + \eta^2) \chi_S \right] (1 - e_t \cos u)^2 (1 - 2 e_t^2 + e_t \cos u) \Bigg].
\end{aligned}
\end{equation}
And the spin-spin coupling component $\dot{\phi}^{\mathrm{SS}}$ can be expressed as
\begin{equation}\label{eq:A75}
\dot{\phi}^{\mathrm{SS}}=\dot{\phi}_{2 \mathrm{PN}}^{\mathrm{SS}} + \dot{\phi}_{3 \mathrm{PN}}^{\mathrm{SS}},
\end{equation}
\begin{equation}\label{eq:A76}
\dot{\phi}_{2 \mathrm{PN}}^{\mathrm{SS}} = \frac{e_t x^{7/2} \left( \kappa_S - 2 \kappa_S \eta + \chi_A^2 - 4 \eta \chi_A^2 + \chi_S^2 + \delta (\kappa_A + 2 \chi_A \chi_S) \right) (e_t - \cos u)}{(1 - e_t^2)^{3/2} (-1 + e_t \cos u)^3},
\end{equation}
\begin{equation*}
\begin{aligned} 
\dot{\phi}_{3 \mathrm{PN}}^{\mathrm{SS}} = &\frac{x^4}{12 (1 - e_t^2)^3 (1 - e_t \cos u)^5} \left( 6 \sqrt{x} (-1 + e_t^2) \Big[ \delta \kappa_A (-14 + 5 \eta) + \kappa_S (-14 + 33 \eta - 6 \eta^2) + 4 \delta (-11 + 9 \eta) \chi_A \chi_S \right. \\ & - 2 \left( (11 - 46 \eta + 6 \eta^2) \chi_A^2 + (11 - 16 \eta + 4 \eta^2) \chi_S^2 \right) \Big] (1 - e_t \cos u)^2 (1 - 2 e_t^2 + e_t \cos u) \\ &
- 24 \sqrt{x - e_t^2 x} (1 - e_t \cos u)^2 \Bigg[ -(1 + 4 e_t^2 + 3 e_t^4) (-1 + 4 \eta) \chi_A^2 + 2 \delta (1 + e_t^4 (3 - 6 \eta) - e_t^2 (-4 + \eta) + \eta) \chi_A \chi_S \\ & 
+ \left( (1 + \eta)^2 + 3 e_t^4 (1 - 4 \eta + 3 \eta^2) - 2 e_t^2 (-2 + \eta + 3 \eta^2) \right) \chi_S^2 \\ & 
+ e_t (-1 + 3 e_t^2) \Big[ (1 + 3 e_t^2) (-1 + 4 \eta) \chi_A^2 + 2 \delta (-1 + 3 e_t^2 (-1 + \eta)) \chi_A \chi_S - (1 + 3 e_t^2 (-1 + \eta) - \eta) (-1 + \eta) \chi_S^2 \Big] \cos u \Bigg] \\ & + \sqrt{x - e_t^2 x} \Bigg[ -84 \kappa_S - 114 e_t^2 \kappa_S - 72 e_t^4 \kappa_S + 24 e_t^6 \kappa_S + 198 \kappa_S \eta + 311 e_t^2 \kappa_S \eta + 160 e_t^4 \kappa_S \eta - 48 e_t^6 \kappa_S \eta - 36 \kappa_S \eta^2 \\ & + 38 e_t^2 \kappa_S \eta^2 - 128 e_t^4 \kappa_S \eta^2 + 24 e_t^6 \kappa_S \eta^2 + \delta \kappa_A (-84 + 24 e_t^6 + 30 \eta + 8 e_t^4 (-9 + 2 \eta) + e_t^2 (-114 + 83 \eta)) - 108 \chi_A^2 \\ & - 138 e_t^2 \chi_A^2 + 24 e_t^6 \chi_A^2 + 456 \eta \chi_A^2 + 512 e_t^2 \eta \chi_A^2 + 112 e_t^4 \eta \chi_A^2 - 120 e_t^6 \eta \chi_A^2 - 72 \eta^2 \chi_A^2 + 76 e_t^2 \eta^2 \chi_A^2 - 256 e_t^4 \eta^2 \chi_A^2 \\ & + 48 e_t^6 \eta^2 \chi_A^2 + 4 \delta \Big[ -54 + e_t^6 (12 - 24 \eta) + e_t^2 (-69 + \eta) + 66 \eta + 56 e_t^4 \eta \Big] \chi_A \chi_S - 108 \chi_S^2 - 138 e_t^2 \chi_S^2 + 24 e_t^6 \chi_S^2 \\ & + 240 \eta \chi_S^2 + 44 e_t^2 \eta \chi_S^2 + 112 e_t^4 \eta \chi_S^2 - 72 e_t^6 \eta \chi_S^2  - 24 \eta^2 \chi_S^2 - 96 e_t^2 \eta^2 \chi_S^2 + 72 e_t^4 \eta^2 \chi_S^2 \\ & - e_t \Big[ -\kappa_S (300 - 740 \eta + 64 \eta^2 + 2 e_t^4 (36 - 97 \eta + 74 \eta^2) + e_t^2 (366 - 929 \eta + 94 \eta^2)) + \delta \Big[ \kappa_A \left( 20 (-15 + 7 \eta) \right. \\ & \left. + e_t^4 (-72 + 50 \eta) + e_t^2 (-366 + 197 \eta) \right) + 4 (-258 + e_t^4 (144 - 167 \eta) + 211 \eta + e_t^2 (-219 + 253 \eta)) \chi_A \chi_S \Big] \\ & - 2 \Big[ (258 - 1057 \eta + 64 \eta^2 + e_t^2 (219 - 910 \eta + 94 \eta^2) + e_t^4 (-144 + 527 \eta + 148 \eta^2)) \chi_A^2 \\ & + (258 - 397 \eta + 36 \eta^2 + e_t^4 (-144 + 383 \eta - 252 \eta^2) + e_t^2 (219 - 472 \eta + 288 \eta^2)) \chi_S^2 \Big] \cos u \\ &
+ e_t^2 \Big[ -\kappa_S (348 - 868 \eta + 56 \eta^2 + 2 e_t^4 (36 - 89 \eta + 46 \eta^2) + e_t^2 (318 - 817 \eta + 158 \eta^2)) \\ & + \delta \Big[ \kappa_A (e_t^4 (-72 + 34 \eta) + 4 (-87 + 43 \eta) + e_t^2 (-318 + 181 \eta)) + 4 (-354 + e_t^4 (216 - 259 \eta) + 323 \eta
\end{aligned}
\end{equation*}
\begin{equation}\label{eq:A77}
\begin{aligned}
&+ e_t^2 (-195 + 233 \eta)) \chi_A \chi_S \Big] - 2 \Big[ (354 - 1457 \eta + 56 \eta^2 + e_t^4 (-216 + 835 \eta + 92 \eta^2) + e_t^2 (195 - 818 \eta + 158 \eta^2)) \chi_A^2 \\ & + (354 - 605 \eta + 108 \eta^2 + e_t^4 (-216 + 547 \eta - 324 \eta^2) + e_t^2 (195 - 428 \eta + 288 \eta^2)) \chi_S^2 \Big] \cos u^2 \\ & + e_t^3 \Big[ \kappa_S (108 - 278 \eta + 4 \eta^2 + e_t^2 (138 - 343 \eta + 98 \eta^2)) + \delta \Big[ \kappa_A (108 + e_t^2 (138 - 67 \eta) - 62 \eta) + 4 (138 + e_t^2 (81 - 53 \eta) \\ & + 108 e_t^4 (-1 + \eta) - 154 \eta) \chi_A \chi_S \Big] + 2 \Big[ (138 - 568 \eta + 4 \eta^2 + 108 e_t^4 (-1 + 4 \eta) + e_t^2 (81 - 344 \eta + 98 \eta^2)) \chi_A^2 \\ &\left. + (138 - 108 e_t^4 (-1 + \eta)^2 - 292 \eta + 84 \eta^2 + e_t^2 (81 - 86 \eta + 48 \eta^2)) \chi_S^2 \Big] \cos u^3 \Bigg] \right).
\end{aligned}
\end{equation}

The mean anomaly $l^\mathrm{SP}$ of the spin-aligned eccentric BBH can be expressed as
\begin{equation}\label{eq:A78}
l^{\mathrm{SP}}=l^{\mathrm{NS}} + l^{\mathrm{SO}} + l^{\mathrm{SS}},
\end{equation}
where the spin-orbit coupling component $l^{\mathrm{SO}}$ can be expressed as
\begin{equation}\label{eq:A79}
\begin{aligned}
l^{\mathrm{SO}}=&l_{2.5 \mathrm{PN}}^{\mathrm{SO}} \\
=& \frac{x^{5/2}}{2 (1 - e_t^2)} \Bigg[ 8 u \delta (-3 + \eta) \chi_A - 4 u (6 - 8 \eta + \eta^2) \chi_S \\ & - 8 \Big( 2 \delta (-3 + \eta) \chi_A - (6 - 8 \eta + \eta^2) \chi_S \Big)  \left(-\tan ^{-1}\left(\frac{\sin (u) \beta_\phi}{1-\cos (u) \beta_\phi}\right)+u\right) \\ &+ e_t \Big( \delta (4 + \eta) \chi_A + (4 - 3 \eta - 2 \eta^2) \chi_S \Big)  \sin \left( 2  \left(-\tan ^{-1}\left(\frac{\sin (u) \beta_\phi}{1-\cos (u) \beta_\phi}\right)+u\right) \right) \Bigg],
\end{aligned}
\end{equation}
and the spin-spin coupling component $l^{\mathrm{SS}}$ can be expressed as
\begin{equation}\label{eq:A80}
\begin{aligned}
l^{\mathrm{SS}}=&l_{3 \mathrm{PN}}^{\mathrm{SS}}\\
=&\frac{x^3}{4 (1 - e_t^2)^{3/2}} \Bigg[ 2 u \Big( -14 \kappa_S + 33 \kappa_S \eta - 6 \kappa_S \eta^2 + \delta \kappa_A (-14 + 5 \eta) - 22 \chi_A^2 + 92 \eta \chi_A^2 - 12 \eta^2 \chi_A^2 \\ & + 4 \delta (-11 + 9 \eta) \chi_A \chi_S - 22 \chi_S^2 + 32 \eta \chi_S^2 - 8 \eta^2 \chi_S^2 \Big) + 4 \Big( \delta \kappa_A (14 - 5 \eta) + \kappa_S (14 - 33 \eta + 6 \eta^2) \\ & + 22 \chi_A^2 - 92 \eta \chi_A^2 + 12 \eta^2 \chi_A^2 + 4 \delta (11 - 9 \eta) \chi_A \chi_S + 22 \chi_S^2 - 32 \eta \chi_S^2 + 8 \eta^2 \chi_S^2 \Big)  \left(-\tan ^{-1}\left(\frac{\sin (u) \beta_\phi}{1-\cos (u) \beta_\phi}\right)+u\right) \\ & + e_t \Big( \delta \kappa_A (8 - 3 \eta) + \kappa_S (8 - 19 \eta + 2 \eta^2) + 8 \chi_A^2 - 31 \eta \chi_A^2 + 4 \eta^2 \chi_A^2 + 2 \delta (8 - 7 \eta) \chi_A \chi_S + 8 \chi_S^2 \\ & - 15 \eta \chi_S^2 \Big)  \sin \left( 2  \left(-\tan ^{-1}\left(\frac{\sin (u) \beta_\phi}{1-\cos (u) \beta_\phi}\right)+u\right) \right) \Bigg],
\end{aligned}
\end{equation}
where the expression for $\beta_{\phi}$ is provided in Eq. (\ref{eq:A14}). Furthermore, in Eqs. (\ref{eq:A79}) and (\ref{eq:A80}), we have refined the equations based on the guidelines outlined in Ref. \cite{Hinder:2008kv}, effectively mitigating the issue of local divergence during waveform computation.

The mean motion $n^{\mathrm{SP}}$ of the spin-aligned eccentric BBH can be expressed as
\begin{equation}\label{eq:A81}
n^{\mathrm{SP}}=n^{\mathrm{NS}} + n^{\mathrm{SO}} + n^{\mathrm{SS}},
\end{equation}
where the spin-orbit coupling component $n^{\mathrm{SO}}$ can be expressed as
\begin{equation}\label{eq:82}
n^{\mathrm{SO}}=n_{1.5 \mathrm{PN}}^{\mathrm{SO}} + n_{2.5 \mathrm{PN}}^{\mathrm{SO}},
\end{equation}
\begin{equation}\label{eq:83}
n_{1.5 \mathrm{PN}}^{\mathrm{SO}} = \frac{2 x^3   \left( 2 \delta \chi_A - (-2 + \eta) \chi_S \right)}{(1 - e_t^2)^{3/2}},
\end{equation}
\begin{equation}\label{eq:84}
n_{2.5 \mathrm{PN}}^{\mathrm{SO}} = - \frac{x^4  \left( \delta \left( -20 + 17 \eta + 4 e_t^2 (-15 + 7 \eta) \right) \chi_A - \left( 20 - 57 \eta + 4 \eta^2 + 2 e_t^2 (30 - 29 \eta + 7 \eta^2) \right) \chi_S \right)}{2  (1 - e_t^2)^{5/2}}.
\end{equation}
And the spin-spin coupling component $n^{\mathrm{SS}}$ can be expressed as
\begin{equation}\label{eq:85}
n^{\mathrm{SS}}=n_{2 \mathrm{PN}}^{\mathrm{SS}} + n_{3 \mathrm{PN}}^{\mathrm{SS}},
\end{equation}
\begin{equation}\label{eq:A86}
n_{2 \mathrm{PN}}^{\mathrm{SS}} = - \frac{3 x^{7/2}  \left( \kappa_S - 2 \kappa_S \eta + \chi_A^2 - 4 \eta \chi_A^2 + \chi_S^2 + \delta (\kappa_A + 2 \chi_A \chi_S) \right)}{2 (1 - e_t^2)^2},
\end{equation}
\begin{equation}\label{eq:A87}
\begin{aligned}
n_{3 \mathrm{PN}}^{\mathrm{SS}} =& \frac{x^{9/2}}{4 (1 - e_t^2)^3} \left( \kappa_S \left( 42 - 118 \eta + 20 \eta^2 + e_t^2 (78 - 191 \eta + 58 \eta^2) \right) + \delta \left( \kappa_A \left( 42 + e_t^2 (78 - 35 \eta) - 34 \eta \right) \right.\right. \\ & \left. \left.+ 4 (17 + e_t^2 (51 - 25 \eta) - 39 \eta) \chi_A \chi_S \right) + 2 \left( (17 - 79 \eta + 20 \eta^2 + e_t^2 (51 - 220 \eta + 58 \eta^2)) \chi_A^2 \right. \right. \\ & \left. \left. + (17 + e_t^2 (51 - 34 \eta) - 67 \eta + 20 \eta^2) \chi_S^2 \right) \right).
\end{aligned}
\end{equation}

The radiative dynamics up to the 3PN order under the harmonic coordinate system for spin-aligned BBH utilized in this study are derived from Ref. \cite{Henry:2023tka}, covering instantaneous and hereditary contributions described in relation to PN expansion parameters $x$ and temporal eccentricity $e_t$. These components can be represented as the combination of the nonspinning portion, spin-orbit coupling, and spin-spin coupling, given by
\begin{equation}\label{eq:A88}
\dot{x}^{\mathrm{SP}}=\dot{x}^{\mathrm{NS}} + \dot{x}^{\mathrm{SO}} + \dot{x}^{\mathrm{SS}},
\end{equation}
and 
\begin{equation}\label{eq:A89}
\dot{e_t}^{\mathrm{SP}}=\dot{e_t}^{\mathrm{NS}} + \dot{e_t}^{\mathrm{SO}} + \dot{e_t}^{\mathrm{SS}}.
\end{equation}
They can be further partitioned into instantaneous and hereditary constituents as follows:
\begin{equation}\label{eq:A90}
\dot{x}^{\mathrm{SO}}= \dot{x}_{\mathrm{inst}}^\mathrm{SO}+\dot{x}_{\mathrm{hered}}^\mathrm{SO},
\end{equation}

\begin{equation}\label{eq:A91}
\dot{x}^{\mathrm{SS}}= \dot{x}_{\mathrm{inst}}^\mathrm{SS},
\end{equation}

\begin{equation}\label{eq:A92}
\dot{e_t}^{\mathrm{SO}}= \dot{e_t}_{\mathrm{inst}}^\mathrm{SO}+\dot{e_t}_{\mathrm{hered}}^\mathrm{SO},
\end{equation}

\begin{equation}\label{eq:A93}
\dot{e_t}^{\mathrm{SS}}= \dot{e_t}_{\mathrm{inst}}^\mathrm{SS}.
\end{equation}
It is worth mentioning that the hereditary term of spin-spin coupling comes from beyond the 3PN order, so it is absent here. The specific expressions of the PN order of the above equations are
\begin{equation}\label{eq:A94}
\dot{x}_{\mathrm{inst}}^\mathrm{SO}= \dot{x}_{1.5 \mathrm{PN}}^\mathrm{SO}+\dot{x}_{2.5 \mathrm{PN}}^\mathrm{SO},
\end{equation}
\begin{equation}\label{eq:A95}
\begin{aligned}
\dot{x}_{1.5 \mathrm{PN}}^\mathrm{SO} = &\frac{x^{13/2}  \eta}{45 (-1 + {e_t}^2)^5 M} \left( (5424 + 27608 {e_t}^2 + 16694 {e_t}^4 + 585 e^6)  \delta  \chi_A + (5424 + {e_t}^4 (16694 - 4072 \eta) \right. \\ & \left. - 3648 \eta + 9 {e_t}^6 (65 + 8 \eta) - 8 {e_t}^2 (-3451 + 1670 \eta))  \chi_S \right),
\end{aligned}
\end{equation}
\begin{equation}\label{eq:A96}
\begin{aligned}
\dot{x}_{2.5 \mathrm{PN}}^\mathrm{SO} = & \frac{x^{15/2} \eta}{10080 (1 - {e_t}^2)^6 M} \left( \delta \left( 45 {e_t}^8 (-17967 + 14560 \eta) + 128 (-31319 + 48678 \eta) + 224 {e_t}^4 (-505492 + 493737 \eta) \right. \right. \\ & \left. + 128 {e_t}^2 (-334209 + 496090 \eta) + 16 {e_t}^6 (-2581907 + 2083557 \eta) \right) \chi_A  + \left( 45 {e_t}^8 (-17967 + 12924 \eta + 1792 \eta^2) \right. \\ & \left. \left. - 128 (31319 - 91900 \eta + 26544 \eta^2) - 448 {e_t}^4 (252746 - 403471 \eta + 98766 \eta^2) \right. \right. \\ & \left. \left. - 128 {e_t}^2 (334209 - 828514 \eta + 266042 \eta^2) - 16 {e_t}^6 (2581907 - 2894533 \eta \right. \right. \\ & \left. \left. + 469140 \eta^2) \right) \chi_S + 896 {e_t}^2 \sqrt{1 - {e_t}^2} (4376 + 4458 {e_t}^2 + 91 {e_t}^4) \left( 2 \delta (-3 + \eta) \chi_A - (6 - 8 \eta + \eta^2) \chi_S \right) \right),
\end{aligned}
\end{equation}

\begin{equation}\label{eq:A97}
\dot{x}_{\mathrm{hered}}^\mathrm{SO}= \dot{x}_{3 \mathrm{PN}}^\mathrm{SO},
\end{equation}
\begin{equation}\label{eq:A98}
\begin{aligned}
\dot{x}_{3 \mathrm{PN}}^\mathrm{SO} = & -\frac{\pi  x^8  \eta}{103680 (1 - {e_t}^2)^{13/2} M} \left( (49766400 + 528887808 {e_t}^2 + 814424832 {e_t}^4 + 213166272 {e_t}^6 + 3911917 {e_t}^8)  \delta  \chi_A \right. \\ & \left. + \left( -221184 (-225 + 148 \eta) - 18432 {e_t}^2 (-28694 + 13789 \eta) - 2304 {e_t}^4 (-353483 + 101614 \eta) \right. \right. \\ & \left. \left. - 96 {e_t}^6 (-2220482 + 214925 \eta) + {e_t}^8 (3911917 + 1055680 \eta) \right)  \chi_S \right),
\end{aligned}
\end{equation}

\begin{equation}\label{eq:A99}
\dot{x}_{\mathrm{inst}}^\mathrm{SS}= \dot{x}_{2 \mathrm{PN}}^\mathrm{SS}+\dot{x}_{3 \mathrm{PN}}^\mathrm{SS},
\end{equation}
\begin{equation}\label{eq:A100}
\begin{aligned}
\dot{x}_{2 \mathrm{PN}}^\mathrm{SS} = & \frac{x^7 \eta}{60 (1 - {e_t}^2)^{11/2}  M} \left( 4 (960 + 5384 {e_t}^2 + 3736 {e_t}^4 + 177 {e_t}^6)  \delta \kappa_A - 4 (960 + 5384 {e_t}^2 + 3736 {e_t}^4 + 177 {e_t}^6) \kappa_S (-1 + 2 \eta) \right. \\ & \left. + 3888 \chi_A^2 + 21992 {e_t}^2 \chi_A^2 + 15358 {e_t}^4 \chi_A^2 + 735 {e_t}^6 \chi_A^2 - 15360 \eta \chi_A^2 - 86144 {e_t}^2 \eta \chi_A^2 - 59776 {e_t}^4 \eta \chi_A^2 - 2832 {e_t}^6 \eta \chi_A^2 \right. \\ & \left. + 2 (3888 + 21992 {e_t}^2 + 15358 {e_t}^4 + 735 {e_t}^6) \delta \chi_A \chi_S + 3888 \chi_S^2 + 21992 {e_t}^2 \chi_S^2 + 15358 {e_t}^4 \chi_S^2 + 735 {e_t}^6 \chi_S^2 - 192 \eta \chi_S^2 \right. \\ & \left. - 1824 {e_t}^2 \eta \chi_S^2 - 1656 {e_t}^4 \eta \chi_S^2 - 108 {e_t}^6 \eta \chi_S^2 \right),
\end{aligned}
\end{equation}
\begin{equation}\label{eq:A101}
\begin{aligned}
\dot{x}_{3 \mathrm{PN}}^\mathrm{SS} = & \frac{x^8 \eta}{30240 M} \bigg( \frac{1}{(1 - {e_t}^2)^{13/2}} \left( 6 \kappa_S \left( 192 (8963 - 31632 \eta + 14784 \eta^2) + 224 {e_t}^2 (70203 - 254204 \eta + 160324 \eta^2) \right. \right. \\ & + 3 {e_t}^8 (164778 - 456935 \eta + 251104 \eta^2) + 28 {e_t}^6 (624771 - 1759522 \eta + 937088 \eta^2) + 8 {e_t}^4 (5381787 \\ & \left. \left. \left. - 16375838 \eta + 9140320 \eta^2) \right) + 32150592 \chi_A^2 + 236283488 {e_t}^2 \chi_A^2 + 439738616 {e_t}^4 \chi_A^2 + 141676164 {e_t}^6 \chi_A^2 \right. \right.  \\ & + 3374811 {e_t}^8 \chi_A^2 - 139984704 \eta \chi_A^2 - 1054958240 {e_t}^2  \eta \chi_A^2 - 1959961928 {e_t}^4 \eta \chi_A^2 - 633764796 {e_t}^6 \eta  \chi_A^2 \\ & - 15271560 {e_t}^8 \eta \chi_A^2 + 34062336 \eta^2 \chi_A^2 + 430950912 {e_t}^2 \eta^2 \chi_A^2 + 877470720 {e_t}^4 \eta^2 \chi_A^2 + 314861568 {e_t}^6 \eta^2 \chi_A^2 \\ & + 9039744 {e_t}^8 \eta^2 \chi_A^2 + 32150592 \chi_S^2 + 236283488 {e_t}^2 \chi_S^2 + 439738616 {e_t}^4 \chi_S^2 + 141676164 {e_t}^6 \chi_S^2 \\ & + 3374811 {e_t}^8 \chi_S^2 - 67507776 \eta \chi_S^2 - 389245472 {e_t}^2 \eta \chi_S^2 - 507132584 {e_t}^4 \eta \chi_S^2 - 121365132 {e_t}^6 \eta \chi_S^2 \\ & - 2752848 {e_t}^8 \eta \chi_S^2 + 19272960 \eta^2 \chi_S^2 + 71582336 {e_t}^2 \eta^2 \chi_S^2 + 57059744 {e_t}^4 \eta^2 \chi_S^2 + 9706032 {e_t}^6 \eta^2 \chi_S^2 \\ & \left. + 344736 {e_t}^8 \eta^2 \chi_S^2 \right) - 2 \delta \left( 3 \kappa_A \left( 192 (-8963 + 13706 \eta) + 224 {e_t}^2 (-70203 + 113798 \eta) + {e_t}^8 (-494334 \right. \right.  \\ & \left. \left. \left. + 382137 \eta) + 28 {e_t}^6 (-624771 + 509980 \eta) + 8 {e_t}^4 (-5381787 + 5612264 \eta) \right) + \left( 576 (-55817 + 68481 \eta) \right. \right. \right.  \\ & + 9 {e_t}^8 (-374979 + 251398 \eta) + 224 {e_t}^2 (-1054837 + 1113995 \eta) + 84 {e_t}^6 (-1686621 + 1121579 \eta) \\ & \left. \left. \left. + 8 {e_t}^4 (-54967327 + 44258753 \eta) \right) \chi_A \chi_S \right) - \frac{672 (-96 + 4484 {e_t}^2 + 4530 {e_t}^4 + 7 {e_t}^6)}{(1 - {e_t}^2)^6} \left( \delta \kappa_A (-14 + 5 \eta) \right.\right. \\ &  \left. + \kappa_S (-14 + 33 \eta - 6 \eta^2) + 4 \delta (-11 + 9 \eta) \chi_A \chi_S - 2 \left( (11 - 46 \eta + 6 \eta^2) \chi_A^2 + (11 - 16 \eta + 4 \eta^2) \chi_S^2 \right) \right) \bigg) ,
\end{aligned}
\end{equation}

\begin{equation}\label{eq:A102}
\dot{e_t}_{\mathrm{inst}}^\mathrm{SO}= \dot{e_t}_{1.5 \mathrm{PN}}^\mathrm{SO}+\dot{e_t}_{2.5 \mathrm{PN}}^\mathrm{SO},
\end{equation}

\begin{equation}\label{eq:A103}
\begin{aligned}
\dot{e_t}_{1.5 \mathrm{PN}}^\mathrm{SO} = & \frac{{e_t} x^{11/2}  \eta}{90 (1 - {e_t}^2)^4 M} \left( (19688 + 28256 {e_t}^2 + 2367 {e_t}^4) \delta \chi_A \right. \\ & \left. + \left( 19688 + {e_t}^2 (28256 - 7972 \eta) - 13312 \eta + 3 {e_t}^4 (789 + 92 \eta) \right) \chi_S \right),
\end{aligned}
\end{equation}

\begin{equation}\label{eq:A104}
\begin{aligned}
\dot{e_t}_{2.5 \mathrm{PN}}^\mathrm{SO} = & \frac{{e_t} x^{13/2} \eta}{60480 M} \bigg( - \frac{2688 (2960 + 6927 {e_t}^2 + 313 {e_t}^4) \left( 2 \delta (-3 + \eta) \chi_A - (6 - 8 \eta + \eta^2) \chi_S \right)}{(1 - {e_t}^2)^{9/2}} \\ & \left. + \frac{1}{(-1 + {e_t}^2)^5} \left( \delta \left( 9 {e_t}^6 (-1037433 + 955808 \eta) + 64 (-324747 + 1197434 \eta) + 56 {e_t}^4 (-4389075 \right. \right. \right. \\ & \left. \left. \left. + 3676346 \eta) + 48 {e_t}^2 (-6885449 + 7330106 \eta) \right) \chi_A + \left( 3 {e_t}^6 (-3112299 + 2662908 \eta + 336896 \eta^2) \right. \right. \right.\\ & - 64 (324747 - 1788860 \eta + 740824 \eta^2) - 48 {e_t}^2 (6885449 - 12914742 \eta + 3551324 \eta^2) \\ &  \left. \left. - 8 {e_t}^4 (30723525 - 36750368 \eta + 6313216 \eta^2) \right) \chi_S \right) \bigg) ,
\end{aligned}
\end{equation}

\begin{equation}\label{eq:A105}
\dot{e_t}_{\mathrm{hered}}^\mathrm{SO}= \dot{e_t}_{3 \mathrm{PN}}^\mathrm{SO},
\end{equation}

\begin{equation}\label{eq:A106}
\begin{aligned}
\dot{e_t}_{3 \mathrm{PN}}^\mathrm{SO} = &\frac{e  \pi x^7 \eta}{51840 (1 - {e_t}^2)^{11/2} M} \left( (64622592 + 238783104 {e_t}^2 + 96887280 {e_t}^4 + 2313613 {e_t}^6) \delta \chi_A + \left( -9216 (-7012 + 4937 \eta) \right. \right.  \\ & \left. \left. - 1152 {e_t}^2 (-207277 + 74954 \eta) - 48 {e_t}^4 (-2018485 + 95561 \eta) + {e_t}^6 (2313613 + 1263046 \eta) \right) \chi_S \right),
\end{aligned}
\end{equation}

\begin{equation}\label{eq:A107}
\dot{e_t}_{\mathrm{inst}}^\mathrm{SS}= \dot{e_t}_{2 \mathrm{PN}}^\mathrm{SS}+\dot{e_t}_{3 \mathrm{PN}}^\mathrm{SS},
\end{equation}

\begin{equation}\label{eq:A108}
\begin{aligned}
\dot{e_t}_{2 \mathrm{PN}}^\mathrm{SS} = & -\frac{{e_t} x^6  \eta}{120 (1 - {e_t}^2)^{9/2} M} \left( 4 (3752 + 5950 {e_t}^2 + 555 {e_t}^4) \delta \kappa_A - 4 (3752 + 5950 {e_t}^2 + 555 {e_t}^4) \kappa_S (-1 + 2 \eta) \right. \\ & + 15368 \chi_A^2 + 24340 {e_t}^2 \chi_A^2 + 2265 {e_t}^4 \chi_A^2 - 60032 \eta \chi_A^2 - 95200 {e_t}^2 \eta \chi_A^2 - 8880 {e_t}^4 \eta \chi_A^2 + 2 (15368  + 24340 {e_t}^2 \\ & \left. + 2265 {e_t}^4) \delta \chi_A \chi_S + 15368 \chi_S^2 + 24340 {e_t}^2 \chi_S^2 + 2265 {e_t}^4 \chi_S^2 - 1440 \eta \chi_S^2 - 2160 {e_t}^2 \eta \chi_S^2 - 180 {e_t}^4 \eta \chi_S^2 \right) ,
\end{aligned}
\end{equation}

\begin{equation}\label{eq:A109}
\begin{aligned}
\dot{e_t}_{3 \mathrm{PN}}^\mathrm{SS} = &\frac{{e_t} x^7 \eta}{60480 M} \bigg( - \frac{1}{(1 - {e_t}^2)^{11/2}} \left( 6 \kappa_S \left( 9 {e_t}^6 (184470 - 516661 \eta + 288288 \eta^2) + 32 (89793 - 527402 \eta + 404908 \eta^2) \right. \right. \\ & \left. \left. \left. + 48 {e_t}^2 (847619 - 2691660 \eta + 1530550 \eta^2) + 4 {e_t}^4 (8317317 - 23483168 \eta + 12461848 \eta^2) \right) + 88499360 \chi_A^2 \right. \right. \\ & + 477482624 {e_t}^2 \chi_A^2 + 275826708 {e_t}^4 \chi_A^2 + 11414979 {e_t}^6 \chi_A^2 - 399608096 \eta \chi_A^2 - 2115049424 {e_t}^2 \eta \chi_A^2 \\ & - 1236732852 {e_t}^4 \eta \chi_A^2 - 52550856 {e_t}^6 \eta \chi_A^2 + 155484672 \eta^2 \chi_A^2 + 881596800 {e_t}^2 \eta^2 \chi_A^2 + 598168704 {e_t}^4 \eta^2 \chi_A^2 \\ & + 31135104 {e_t}^6 \eta^2 \chi_A^2 + 88499360 \chi_S^2 + 477482624 {e_t}^2 \chi_S^2 + 275826708 {e_t}^4 \chi_S^2 + 11414979 {e_t}^6 \chi_S^2 \\ & - 219613856 \eta \chi_S^2 - 624223376 {e_t}^2 \eta \chi_S^2 - 235260900 {e_t}^4 \eta \chi_S^2 - 8905680 {e_t}^6 \eta \chi_S^2 + 58808960 \eta^2 \chi_S^2 \\ & + 83203904 {e_t}^2 \eta^2 \chi_S^2 + 15245328 {e_t}^4 \eta^2 \chi_S^2 + 707616 {e_t}^6 \eta^2 \chi_S^2 - 2 \delta \left( 3 \kappa_A \left( 9 {e_t}^6 (-184470 + 147721 \eta) \right. \right. \\ & \left. + 32 (-89793 + 347816 \eta) + 48 {e_t}^2 (-847619 + 996422 \eta) + 4 {e_t}^4 (-8317317 + 6848534 \eta) \right) \\ & + \left( 27 {e_t}^6 (-422777 + 292530 \eta) + 32 (-2765605 + 4144133 \eta) + 12 {e_t}^4 (-22985559 + 15361955 \eta) \right. \\ & \left. \left. \left. + 16 {e_t}^2 (-29842664 + 25916947 \eta) \right) \chi_A \chi_S \right) \right) - \frac{672 (3248 + 6891 {e_t}^2 + 61 {e_t}^4)}{(1 - {e_t}^2)^5} \left( \delta \kappa_A (-14 + 5 \eta) \right. \\ & \left. + \kappa_S (-14 + 33 \eta - 6 \eta^2) + 4 \delta (-11 + 9 \eta) \chi_A \chi_S - 2 \left( (11 - 46 \eta + 6 \eta^2) \chi_A^2 + (11 - 16 \eta + 4 \eta^2) \chi_S^2 \right) \right)\bigg). 
\end{aligned}
\end{equation}

The (2,2) mode of the gravitational waveform $h^{22, \mathrm{SP}}$ for the spin-aligned eccentric BBH system, as provided in Ref. \cite{Henry:2023tka}, can be expressed as
\begin{equation}\label{eq:A110}
h^{22,\mathrm{SP}}=\frac{4 G M \eta }{c^4 R} \sqrt{\frac{\pi}{5}} e^{-2 i \phi} H^{22,\mathrm{SP}},
\end{equation}
where the amplitude $H^{22,\mathrm{SP}}$ can be expressed as
\begin{equation}\label{eq:A111}
H^{22,\mathrm{SP}}=H^{22,\mathrm{NS}} + H^{22,\mathrm{SO}} + H^{22,\mathrm{SS}},
\end{equation}
where $H^{22,\mathrm{NS}}$ is the nonspinning portion of the aforementioned waveform. And the specific expressions of $H^{22,\mathrm{SO}}$ and $H^{22,\mathrm{SO}}$ are
\begin{equation}\label{eq:A112}
H^{22,\mathrm{SO}}=H_{1.5 \mathrm{PN}}^{22,\mathrm{SO}} + H_{2.5 \mathrm{PN}}^{22,\mathrm{SO}} + H_{3 \mathrm{PN}}^{22,\mathrm{SO}},
\end{equation}
\begin{equation}\label{eq:A113}
H_{1.5 \mathrm{PN}}^{22,\mathrm{SO}} =\frac{1}{c^3} \frac{2 M^2}{3 r^2} \left( -i \dot{r} \left( 3 \delta \chi_A + (3 - 8 \eta) \chi_S \right) + r \dot{\phi} \left( -3 \delta \chi_A + (-3 + 5 \eta) \chi_S \right) \right),
\end{equation}
\begin{equation}\label{eq:A114}
\begin{aligned}
H_{2.5 \mathrm{PN}}^{22,\mathrm{SO}} = & \frac{1}{c^5} \frac{M^2}{42 r^3} \Bigg( -8 i M \dot{r} \left( \delta (-55 + 19 \eta) \chi_A + (-55 + 100 \eta - 86 \eta^2) \chi_S \right) + 2 i r {\dot{r}}^3 \left( 3 \delta (-9 + 14 \eta) \chi_A \right. \\ & \left. + (-27 + 30 \eta - 4 \eta^2) \chi_S \right) - r^4 {\dot{\phi}}^3 \left( \delta (120 + 83 \eta) \chi_A + 3 (40 - 161 \eta + 78 \eta^2) \chi_S \right) \\ & + r^2 {\dot{r}}^2 \dot{\phi} \left( \delta (18 + 275 \eta) \chi_A + (18 - 315 \eta + 188 \eta^2) \chi_S \right) - 2 i r^3 \dot{r} \dot{\phi}^2 \left( \delta (51 + 62 \eta) \chi_A \right. \\ & \left. + (51 - 264 \eta + 440 \eta^2) \chi_S \right) + M r \dot{\phi} \left( \delta (238 - 141 \eta) \chi_A + (238 - 181 \eta + 474 \eta^2) \chi_S \right) \Bigg) ,
\end{aligned}
\end{equation}
\begin{equation}\label{eq:A115}
H_{3 \mathrm{PN}}^{22,\mathrm{SO}} = -\frac{1}{c^6}\frac{i M^3 (M + r (2 \dot{r}^2 - 10 i r \dot{r} \dot{\phi} + 7 r^2 \dot{\phi}^2)) (\delta \chi_A + \chi_S - 2 \eta \chi_S)}{3 r^4},
\end{equation}
\begin{equation}\label{eq:A116}
H^{22,\mathrm{SS}}=H_{2 \mathrm{PN}}^{22,\mathrm{SS}} + H_{3 \mathrm{PN}}^{22,\mathrm{SS}},
\end{equation}
\begin{equation}\label{eq:A117}
H_{2 \mathrm{PN}}^{22,\mathrm{SS}} = \frac{1}{c^4} \frac{3 M^3 \left( \kappa_S - 2 \kappa_S \eta + \chi_A^2 - 4 \eta \chi_A^2 + \chi_S^2 + \delta (\kappa_A + 2 \chi_A \chi_S) \right)}{2 r^3}
\end{equation}
\begin{equation}\label{eq:A118}
\begin{aligned}
H_{3 \mathrm{PN}}^{22,\mathrm{SS}} = & - \frac{1}{c^6} \frac{M^3}{84 r^4} \left( 2 M \left( \delta \kappa_A (438 + 103 \eta) + \kappa_S (438 - 773 \eta + 108 \eta^2) + 2 (219 - 863 \eta + 108 \eta^2) \chi_A^2 \right. \right. \\ & \left. \left. + 2 \delta (438 + 145 \eta) \chi_A \chi_S + 6 (73 + 44 \eta - 28 \eta^2) \chi_S^2 \right) + r \left( -r^2 \dot{\phi}^2 \left( \delta \kappa_A (153 + 430 \eta) + \kappa_S (153 \right. \right. \right. \\ & + 124 \eta - 804 \eta^2) + 153 \chi_A^2 - 28 \eta \chi_A^2 - 1608 \eta^2 \chi_A^2 + 2 \delta (153 + 38 \eta) \chi_A \chi_S + 153 \chi_S^2 - 508 \eta \chi_S^2 \\ & \left. \left. \left. - 112 \eta^2 \chi_S^2 \right) - 2 i r \dot{r} \dot{\phi} \left( \delta \kappa_A (117 + 211 \eta) + \kappa_S (117 - 23 \eta - 390 \eta^2) + 117 \chi_A^2 - 439 \eta \chi_A^2 \right. \right. \right. \\ & \left. \left. \left. - 780 \eta^2 \chi_A^2 + 2 \delta (117 - 223 \eta) \chi_A \chi_S + 117 \chi_S^2 - 475 \eta \chi_S^2 + 56 \eta^2 \chi_S^2 \right) + \dot{r}^2 \left( \delta \kappa_A (291 - 308 \eta) \right. \right. \right. \\ & + \kappa_S (291 - 890 \eta + 24 \eta^2) + 291 \chi_A^2 - 940 \eta \chi_A^2 + 48 \eta^2 \chi_A^2 + 2 \delta (291 - 364 \eta) \chi_A \chi_S + 291 \chi_S^2 \\ & \left. \left. \left. - 952 \eta \chi_S^2 + 224 \eta^2 \chi_S^2 \right) \right) \right) .
\end{aligned}
\end{equation}
The high-order harmonic mode for spin-aligned BBH can be expressed as
\begin{equation}\label{eq:A119}
h^{\ell m,\mathrm{SP}}=\frac{4 G M \eta}{c^4 R} \sqrt{\frac{\pi}{5}} e^{-i m \phi} H^{\ell m,\mathrm{SP}}.
\end{equation}
At the leading PN order PN order, the magnitudes of the high-order modes (2,1), (3,3), (3,2), (4,4), (4,3), and (5,5) in spin-aligned BBH closely resemble those in the nonspinning scenario. For a more detailed representation at higher PN orders in the context of complete spin alignment, readers can consult Ref. \cite{Henry:2023tka}.

\subsection{Initial eccentricity estimate}
The technique employed to quantify the nonspinning eccentricity originates from the third PN precise generalized quasi-Keplerian parametrization for compact binaries moving in eccentric orbits under harmonic coordinates \cite{Memmesheimer:2004cv}, characterized by the expression:
\begin{equation}\label{eq:A120}
\begin{aligned}
{{{e_t}_{0}}_{\mathrm{KP}}}^2=&1+2 {E_0} {h_0}^2+\frac{(-2 {E_0})}{4 c^2}\left\{-8+8 \eta-\left(-2 {E_0} {h_0}^2\right)(-17+7 \eta)\right\}+\frac{(-2 {E_0})^2}{8 c^4}\bigg\{12+72 \eta+20 \eta^2 \\ & -24 \sqrt{\left(-2 {E_0} {h_0}^2\right)}(-5 +2 \eta) -\left(-2 {E_0} {h_0}^2\right)\left(112-47 \eta+16 \eta^2\right)-\frac{16}{\left(-2 {E_0} {h_0}^2\right)}(-4+7 \eta)+\frac{24}{\sqrt{\left(-2 {E_0} {h_0}^2\right)}}(-5+2 \eta)\bigg\} \\
& +\frac{(-2 {E_0})^3}{6720 c^6}\bigg\{23520-464800 \eta+179760 \eta^2+16800 \eta^3-2520 \sqrt{\left(-2 {E_0} {h_0}^2\right)}\left(265-193 \eta+46 \eta^2\right) \\
& -525\left(-2 {E_0} {h_0}^2\right)\left(-528+200 \eta-77 \eta^2+24 \eta^3\right)-\frac{6}{\left(-2 {E_0} {h_0}^2\right)}\left(73920-260272 \eta+4305 \pi^2 \eta+61040 \eta^2\right) \\
& +\frac{70}{\sqrt{\left(-2 {E_0} {h_0}^2\right)}}\left(16380-19964 \eta+123 \pi^2 \eta+3240 \eta^2\right)+\frac{8}{\left(-2 {E_0} {h_0}^2\right)^2}\left(53760-176024 \eta+4305 \pi^2 \eta\right. \\
&\left. +15120 \eta^2\right)-\frac{70}{\left(-2 {E_0} {h_0}^2\right)^{3 / 2}}\left(10080-13952 \eta+123 \pi^2 \eta+1440 \eta^2\right) \bigg\}
\end{aligned}
\end{equation}
where we introduce ${E_0}$ and ${h_0}$, defined as ${E_0} = E_b / \eta$ and ${h_0} = L / \eta$, where $E_b$ represents the orbital binding energy, as defined by Eq. (\ref{eq:65}), and $L$ denotes the orbital angular momentum. All relevant eccentricity measurement results for numerical relativity simulations can be accessed from the RIT and SXS catalogs.

In the context of spin-aligned configurations, we employ the quasi-Keplerian parameterization method at the 2PN order to determine eccentricity in ADM coordinates referring to Ref. \cite{Tessmer:2010hp}, represented as:

\begin{equation}\label{eq:A121}
\begin{aligned}
e_t^2= & 1-2 {h_0}^2|{E_0}|+|{E_0}|\left({h_0}^2|{E_0}|(17-7 \eta)+4(\eta-1)\right) \\
& +\frac{\delta}{{h_0}} |{E_0}|\left[2(\eta-2)\left(\chi_1+\chi_2\right)-4 \sqrt{1-4 \eta}\left(\chi_1-\chi_2\right)\right] \\
& +\frac{\delta^2|{E_0}|}{{h_0}^2}\left[\left(\chi_1-\chi_2\right)^2\left(\left(\left(\eta-\frac{1}{2}\right)\left(\lambda_1+\lambda_2\right)-\frac{1}{2} \sqrt{1-4 \eta}\left(\lambda_1-\lambda_2\right)\right)- \eta\right)\right. \\
& +\left(\chi_1+\chi_2\right)^2\left( \eta+\left(\left(\eta-\frac{1}{2}\right)\left(\lambda_1+\lambda_2\right)-\frac{1}{2} \sqrt{1-4 \eta}\left(\lambda_1-\lambda_2\right)\right)\right) \\
& \left.+\left(\chi_1+\chi_2\right)\left(\chi_1-\chi_2\right)\left((2 \eta-1)\left(\lambda_1-\lambda_2\right)-\sqrt{1-4 \eta}\left(\lambda_1+\lambda_2\right)\right)\right] \\
& +\frac { | {E_0} | } { {h_0} ^ { 2 } } \left(-11 \eta+17+{h_0}^4|{E_0}|^2\left(-16 \eta^2+47 \eta-112\right)+12 \sqrt{2} {h_0}^3|{E_0}|^{3 / 2}(5-2 \eta)\right. \\
& \left.+2 {h_0}^2|{E_0}|\left(5 \eta^2+\eta+2\right)+6 \sqrt{2} {h_0} \sqrt{|{E_0}|}(2 \eta-5)\right) \\
& +\frac{\delta}{{h_0}} \frac{|{E_0}|}{2 {h_0}^2}\bigg[( \chi _ { 1 } + \chi _ { 2 } ) \left(-16 \sqrt{2} {h_0}^3|{E_0}|^{3 / 2}\left(\eta^2-8 \eta+6\right)+{h_0}^2|{E_0}|\left(32 \eta^2-159 \eta+124\right)\right. \\
& \left.+8 \sqrt{2} \sqrt{|{E_0}|} {h_0}\left(\eta^2-8 \eta+6\right)-8 \eta^2+78 \eta-64\right) \\
& +\sqrt{1-4 \eta}\left(\chi_1-\chi_2\right)\left(32 \sqrt{2} {h_0}^3|{E_0}|^{3 / 2}(\eta-3)+{h_0}^2|{E_0}|(124-59 \eta)\right. \\
& -16 \sqrt{2} {h_0} \sqrt{|{E_0}|}(\eta-3)+18 \eta-64)\bigg]
\end{aligned}
\end{equation}
where ${E_0}$ and ${h_0}$ have the same meaning as in the nonspinning case, and $\lambda_1 = \lambda_2 = -0.5$.

\section{Fitting waveform and parameters}\label{App:B}
This section showcases the NR waveforms utilized, along with the outcomes of parameter fitting achieved through PN least squares fitting applied to the NR waveforms, encompassing data from the RIT and SXS catalogs for both nonspinning and spin-aligned setups. Additionally, it includes the eccentricity results derived from methodology quasi-Keplerian parameterization ${{e_t}_0}_{\text{KP}}$. In Table \ref{table:1}, we present the details of the NR simulations utilized, including the spin configuration, mass ratio, spins $\chi_1$ and $\chi_2$ in the $z$ direction, the fitting time interval, the PN least squares fitting results for parameters $x_0$, $l_0$, ${e_t}_0$, and $\phi_0$, the fitting residuals $Q$, and the eccentricity values ${{e_t}_0}_{\text{KP}}$ determined through the quasi-Keplerian parameterization.

\begin{longtable*}{|c|c|c|c|c|c|c|c|c|c|c|c|}
    \caption{Details of the NR simulations utilized, including the spin configuration, mass ratio, spins $\chi_1$ and $\chi_2$ in the $z$ direction, the fitting time interval, the PN least squares fitting results for parameters $x_0$, $l_0$, ${e_t}_0$, and $\phi_0$, the fitting residuals $Q$, and the eccentricity values ${{e_t}_0}_{\text{KP}}$ determined through the quasi-Keplerian parameterization.}\label{table:1} \\
    \hline
    Simulations & Type & $q$ & $\chi_1$ & $\chi_2$ & Fitting interval & $x_0$ & $l_0$ & ${e_t}_0$ & $\phi_0$ & $Q$ & ${{e_t}_0}_{\text{KP}}$ \\
    \hline
    \endfirsthead
    \multicolumn{2}{c}{\bfseries\small \tablename\ \thetable\ {\textit{(Continued)}}}\\
    \hline
    Simulations & Type & $q$ & $\chi_1$ & $\chi_2$ & Fitting interval & $x_0$ & $l_0$ & ${e_t}_0$ & $\phi_0$ & $Q$ & ${{e_t}_0}_{\text{KP}}$ \\
    \hline
    \endhead
    \hline
    \multicolumn{2}{r}{\textit{(Table continued)}} 
    \endfoot
    \hline
    \endlastfoot    %
        RIT:eBBH:1090 & Nonspinning & 1 & 0 & 0 & [-1194,-300]  & 0.082991  & 3.481579  & 0.005624  & 2.590685  & $1.12\times 10^{-8}$ & 0.07821  \\ \hline
        RIT:eBBH:1091 & Nonspinning & 1 & 0 & 0 & [-1176,-300]  & 0.083244  & 3.847368  & 0.005751  & 2.617879  & $1.13\times 10^{-8}$ & 0.07798  \\ \hline
        RIT:eBBH:1092 & Nonspinning & 1 & 0 & 0 & [-1157,-300]  & 0.083499  & 4.158421  & 0.006537  & 2.645123  & $1.15\times 10^{-8}$ & 0.07785  \\ \hline
        RIT:eBBH:1093 & Nonspinning & 1 & 0 & 0 & [-1139,-300]  & 0.083750  & 4.387895  & 0.007789  & 2.676829  & $1.20\times 10^{-8}$ & 0.07776  \\ \hline
        RIT:eBBH:1094 & Nonspinning & 1 & 0 & 0 & [-1171, -300]  & 0.084006  & 4.547135  & 0.009305  & 2.705703  & $1.26\times 10^{-8}$ & 0.07774  \\ \hline
        RIT:eBBH:1095 & Nonspinning & 1 & 0 & 0 & [-1153,-300]  & 0.084260  & 4.662105  & 0.011032  & 2.736338  & $1.35\times 10^{-8}$ & 0.07776  \\ \hline
        RIT:eBBH:1096 & Nonspinning & 1 & 0 & 0 & [-1109,-300]  & 0.084900  & 4.835789  & 0.015653  & 2.814509  & $1.85\times 10^{-8}$ & 0.07811  \\ \hline
        RIT:eBBH:1097 & Nonspinning & 1 & 0 & 0 & [-1066,-300]  & 0.085562  & 4.938947  & 0.020546  & 2.886214  & $2.37\times 10^{-8}$ & 0.07882  \\ \hline
        RIT:eBBH:1098 & Nonspinning & 1 & 0 & 0 & [-904,-300]  & 0.088251  & 5.154737  & 0.040786  & 3.236675  & $3.05\times 10^{-8}$ & 0.08499  \\ \hline
        RIT:eBBH:1099 & Nonspinning & 1 & 0 & 0 & [-976,-300]  & 0.089646  & 5.221053  & 0.050058  & 3.436083  & $2.46\times 10^{-8}$ & 0.08979  \\ \hline
        RIT:eBBH:1100 & Nonspinning & 1 & 0 & 0 & [-803,-300]  & 0.091115  & 5.265263  & 0.059644  & 3.619988  & $4.93\times 10^{-8}$ & 0.09551  \\ \hline
        RIT:eBBH:1101 & Nonspinning & 1 & 0 & 0 & [-785,-300]  & 0.092666  & 5.302105  & 0.069016  & -2.490113  & $9.10\times 10^{-8}$ & 0.10197  \\ \hline
        RIT:eBBH:1102 & Nonspinning & 1 & 0 & 0 & [-721,-300]  & 0.094231  & 5.334737  & 0.078479  & -2.301955  & $1.61\times 10^{-7}$ & 0.10903  \\ \hline
        RIT:eBBH:1103 & Nonspinning & 1 & 0 & 0 & [-666,-300]  & 0.095789  & 5.384211  & 0.088226  & -2.075571  & $1.44\times 10^{-7}$ & 0.11658  \\ \hline
        RIT:eBBH:1104 & Nonspinning & 1 & 0 & 0 & [-602,-300]  & 0.097444  & 5.431579  & 0.096159  & -1.849961  & $3.85\times 10^{-8}$ & 0.12450  \\ \hline
        RIT:eBBH:1105 & Nonspinning & 1 & 0 & 0 & [-545,-300]  & 0.099288  & 5.463158  & 0.104953  & -1.665337  & $4.08\times 10^{-8}$ & 0.13273  \\ \hline
        RIT:eBBH:1106 & Nonspinning & 1 & 0 & 0 & [-515,-200]  & 0.101167  & 5.484211  & 0.112440  & -1.473961  & $7.47\times 10^{-8}$ & 0.14120  \\ \hline
        RIT:eBBH:1107 & Nonspinning & 1 & 0 & 0 & [-468,-200]  & 0.103230  & 5.508421  & 0.119887  & -1.292614  & $8.39\times 10^{-8}$ & 0.14986  \\ \hline
        RIT:eBBH:1108 & Nonspinning & 1 & 0 & 0 & [-419,-200]  & 0.105457  & 5.531579  & 0.127149  & -1.103378  & $7.13\times 10^{-8}$ & 0.15868  \\ \hline
        RIT:eBBH:1109 & Nonspinning & 1 & 0 & 0 & [-393,-200]  & 0.106579  & 5.537895  & 0.130435  & -0.999022  & $7.09\times 10^{-8}$ & 0.16313  \\ \hline
        RIT:eBBH:1133 & Nonspinning &  1/4 & 0 & 0 & [-1844,-800]  & 0.082499  & 3.649474  & 0.003881  & 2.656878  & $7.75\times 10^{-9}$ & 0.08120  \\ \hline
        RIT:eBBH:1134 & Nonspinning &  1/4 & 0 & 0 & [-1816,-800]  & 0.082742  & 4.152632  & 0.004500  & 2.684012  & $7.82\times 10^{-9}$ & 0.08091  \\ \hline
        RIT:eBBH:1135 & Nonspinning &  1/4 & 0 & 0 & [-1788,-800]  & 0.083013  & 4.469474  & 0.005784  & 2.691516  & $7.41\times 10^{-9}$ & 0.08070  \\ \hline
        RIT:eBBH:1136 & Nonspinning &  1/4 & 0 & 0 & [-1760,-800]  & 0.083233  & 4.669474  & 0.007497  & 2.736034  & $7.65\times 10^{-9}$ & 0.08053  \\ \hline
        RIT:eBBH:1137 & Nonspinning &  1/4 & 0 & 0 & [-1733,-800]  & 0.083483  & 4.785263  & 0.009318  & 2.759675  & $7.80\times 10^{-9}$ & 0.08043  \\ \hline
        RIT:eBBH:1138 & Nonspinning &  1/4 & 0 & 0 & [-1706,-800]  & 0.083767  & 4.853684  & 0.011230  & 2.761049  & $8.48\times 10^{-9}$ & 0.08038  \\ \hline
        RIT:eBBH:1139 & Nonspinning &  1/4 & 0 & 0 & [-1640,-700]  & 0.084392  & 4.978947  & 0.016471  & 2.830922  & $1.19\times 10^{-8}$ & 0.08052  \\ \hline
        RIT:eBBH:1140 & Nonspinning &  1/4 & 0 & 0 & [-1576,-700]  & 0.085026  & 5.048421  & 0.021390  & 2.899352  & $1.48\times 10^{-8}$ & 0.08101  \\ \hline
        RIT:eBBH:1141 & Nonspinning &  1/4 & 0 & 0 & [-1447,-600]  & 0.086355  & 5.127368  & 0.031613  & 3.018197  & $2.67\times 10^{-8}$ & 0.08300  \\ \hline
        RIT:eBBH:1142 & Nonspinning &  1/4 & 0 & 0 & [-1427,-600]  & 0.087708  & 5.180000  & 0.041883  & 3.150058  & $5.37\times 10^{-8}$ & 0.08630  \\ \hline
        RIT:eBBH:1143 & Nonspinning &  1/4 & 0 & 0 & [-1366,-500]  & 0.089116  & 5.220000  & 0.051877  & 3.282279  & $9.74\times 10^{-8}$ & 0.09071  \\ \hline
        RIT:eBBH:1144 & Nonspinning &  1/4 & 0 & 0 & [-1308,-500]  & 0.089815  & 5.238421  & 0.057072  & 3.360127  & $1.27\times 10^{-7}$ & 0.09328  \\ \hline
        RIT:eBBH:1145 & Nonspinning &  1/4 & 0 & 0 & [-1250,-400]  & 0.090576  & 5.252105  & 0.061713  & 3.420018  & $1.75\times 10^{-7}$ & 0.09607  \\ \hline
        RIT:eBBH:1146 & Nonspinning &  1/4 & 0 & 0 & [-1194,-400]  & 0.091315  & 5.265789  & 0.066837  & -2.786966  & $2.39\times 10^{-7}$ & 0.09905  \\ \hline
        RIT:eBBH:1147 & Nonspinning &  1/4 & 0 & 0 & [-1145,-300]  & 0.092119  & 5.276316  & 0.071281  & -2.731379  & $3.30\times 10^{-7}$ & 0.10222  \\ \hline
        RIT:eBBH:1148 & Nonspinning &  1/4 & 0 & 0 & [-1098,-300]  & 0.092899  & 5.285681  & 0.076129  & -2.664751  & $4.69\times 10^{-7}$ & 0.10554  \\ \hline
        RIT:eBBH:1149 & Nonspinning &  1/4 & 0 & 0 & [-1051,-300]  & 0.093649  & 5.299474  & 0.081438  & -2.570056  & $6.15\times 10^{-7}$ & 0.10900  \\ \hline
        RIT:eBBH:1150 & Nonspinning &  1/4 & 0 & 0 & [-1001,-300]  & 0.094385  & 5.326007  & 0.087130  & -2.450216  & $6.05\times 10^{-7}$ & 0.11261  \\ \hline
        RIT:eBBH:1151 & Nonspinning &  1/4 & 0 & 0 & [-948,-300]  & 0.095165  & 5.362632  & 0.091453  & -2.321655  & $2.92\times 10^{-7}$ & 0.11632  \\ \hline
        RIT:eBBH:1152 & Nonspinning &  1/4 & 0 & 0 & [-897,-300]  & 0.095977  & 5.383684  & 0.095080  & -2.202471  & $2.39\times 10^{-7}$ & 0.12014  \\ \hline
        RIT:eBBH:1153 & Nonspinning &  1/4 & 0 & 0 & [-863,-300]  & 0.096886  & 5.391053  & 0.099510  & -2.138225  & $3.41\times 10^{-7}$ & 0.12405  \\ \hline
        RIT:eBBH:1154 & Nonspinning &  1/4 & 0 & 0 & [-823,-300]  & 0.097786  & 5.401579  & 0.103992  & -2.055945  & $4.80\times 10^{-7}$ & 0.12804  \\ \hline
        RIT:eBBH:1155 & Nonspinning &  1/4 & 0 & 0 & [-786,-300]  & 0.098720  & 5.407895  & 0.108161  & -1.980047  & $6.85\times 10^{-7}$ & 0.13211  \\ \hline
        RIT:eBBH:1156 & Nonspinning &  1/4 & 0 & 0 & [-747,-300]  & 0.099635  & 5.414211  & 0.112781  & -1.890690  & $9.69\times 10^{-7}$ & 0.13624  \\ \hline
        RIT:eBBH:1157 & Nonspinning &  1/4 & 0 & 0 & [-713,-300]  & 0.100522  & 5.427368  & 0.117976  & -1.780219  & $1.24\times 10^{-6}$ & 0.14044  \\ \hline
        RIT:eBBH:1158 & Nonspinning &  1/4 & 0 & 0 & [-664,-300]  & 0.101405  & 5.459532  & 0.123142  & -1.640302  & $8.99\times 10^{-7}$ & 0.14468  \\ \hline
        RIT:eBBH:1159 & Nonspinning &  1/4 & 0 & 0 & [-619,-300]  & 0.102351  & 5.496842  & 0.125690  & -1.485783  & $1.34\times 10^{-7}$ & 0.14896  \\ \hline
        RIT:eBBH:1160 & Nonspinning &  1/4 & 0 & 0 & [-582,-300]  & 0.103418  & 5.506842  & 0.129047  & -1.381539  & $1.28\times 10^{-7}$ & 0.15331  \\ \hline
        RIT:eBBH:1161 & Nonspinning &  1/4 & 0 & 0 & [-551,-200]  & 0.104704  & 5.472105  & 0.135665  & -1.372285  & $2.44\times 10^{-6}$ & 0.15769  \\ \hline
        RIT:eBBH:1162 & Nonspinning &  1/4 & 0 & 0 & [-523,-200]  & 0.105795  & 5.505263  & 0.138477  & -1.233183  & $1.07\times 10^{-6}$ & 0.16209  \\ \hline
        RIT:eBBH:1163 & Nonspinning &  1/4 & 0 & 0 & [-495,-200]  & 0.107011  & 5.524211  & 0.139619  & -1.106726  & $2.00\times 10^{-7}$ & 0.16653  \\ \hline
        RIT:eBBH:1164 & Nonspinning &  1/4 & 0 & 0 & [-475,-200]  & 0.108341  & 5.528421  & 0.142046  & -1.004547  & $1.82\times 10^{-7}$ & 0.17100  \\ \hline
        RIT:eBBH:1165 & Nonspinning &  1/4 & 0 & 0 & [-438,-200]  & 0.109782  & 5.533158  & 0.145150  & -0.900833  & $2.09\times 10^{-7}$ & 0.17549  \\ \hline
        RIT:eBBH:1200 & Nonspinning &  1/2 & 0 & 0 & [-1338,-600]  & 0.082828  & 3.492105  & 0.005325  & 2.620816  & $6.44\times 10^{-9}$ & 0.07926  \\ \hline
        RIT:eBBH:1201 & Nonspinning &  1/2 & 0 & 0 & [-1318,-600]  & 0.083049  & 3.894211  & 0.005613  & 2.662174  & $6.83\times 10^{-9}$ & 0.07903  \\ \hline
        RIT:eBBH:1202 & Nonspinning &  1/2 & 0 & 0 & [-1297,-600]  & 0.083307  & 4.224737  & 0.006504  & 2.685870  & $6.80\times 10^{-9}$ & 0.07886  \\ \hline
        RIT:eBBH:1203 & Nonspinning &  1/2 & 0 & 0 & [-1277,-500]  & 0.083584  & 4.428947  & 0.007645  & 2.702922  & $1.00\times 10^{-8}$ & 0.07875  \\ \hline
        RIT:eBBH:1204 & Nonspinning &  1/2 & 0 & 0 & [-1257,-500]  & 0.083840  & 4.588947  & 0.009305  & 2.727978  & $1.06\times 10^{-8}$ & 0.07870  \\ \hline
        RIT:eBBH:1205 & Nonspinning &  1/2 & 0 & 0 & [-1237,-500]  & 0.084098  & 4.701579  & 0.011162  & 2.753254  & $1.15\times 10^{-8}$ & 0.07871  \\ \hline
        RIT:eBBH:1206 & Nonspinning &  1/2 & 0 & 0 & [-1238,-500]  & 0.084736  & 4.875263  & 0.016016  & 2.826787  & $1.36\times 10^{-8}$ & 0.07899  \\ \hline
        RIT:eBBH:1207 & Nonspinning &  1/2 & 0 & 0 & [-1189,-500]  & 0.085381  & 4.982632  & 0.021096  & 2.904696  & $1.39\times 10^{-8}$ & 0.07962  \\ \hline
        RIT:eBBH:1208 & Nonspinning &  1/2 & 0 & 0 & [-1096,-400]  & 0.086747  & 5.081579  & 0.030955  & 3.035802  & $3.46\times 10^{-8}$ & 0.08193  \\ \hline
        RIT:eBBH:1209 & Nonspinning &  1/2 & 0 & 0 & [-1008,-400]  & 0.088090  & 5.172105  & 0.041114  & 3.217989  & $2.45\times 10^{-8}$ & 0.08550  \\ \hline
        RIT:eBBH:1210 & Nonspinning &  1/2 & 0 & 0 & [-1024,-400]  & 0.089480  & 5.230000  & 0.050503  & 3.400687  & $2.46\times 10^{-8}$ & 0.09018  \\ \hline
        RIT:eBBH:1211 & Nonspinning &  1/2 & 0 & 0 & [-940,-400]  & 0.090958  & 5.271053  & 0.060235  & 3.567925  & $4.81\times 10^{-8}$ & 0.09577  \\ \hline
        RIT:eBBH:1212 & Nonspinning &  1/2 & 0 & 0 & [-861,-400]  & 0.092456  & 5.306842  & 0.069921  & -2.533825  & $9.03\times 10^{-8}$ & 0.10213  \\ \hline
        RIT:eBBH:1213 & Nonspinning &  1/2 & 0 & 0 & [-811,-300]  & 0.094081  & 5.337368  & 0.078902  & -2.366889  & $1.57\times 10^{-7}$ & 0.10909  \\ \hline
        RIT:eBBH:1214 & Nonspinning &  1/2 & 0 & 0 & [-743,-300]  & 0.095700  & 5.364737  & 0.088232  & -2.183018  & $2.83\times 10^{-7}$ & 0.11656  \\ \hline
        RIT:eBBH:1215 & Nonspinning &  1/2 & 0 & 0 & [-673,-300]  & 0.097303  & 5.411053  & 0.098138  & -1.955428  & $2.58\times 10^{-7}$ & 0.12442  \\ \hline
        RIT:eBBH:1216 & Nonspinning &  1/2 & 0 & 0 & [-604,-300]  & 0.099024  & 5.460526  & 0.105617  & -1.720755  & $5.21\times 10^{-8}$ & 0.13259  \\ \hline
        RIT:eBBH:1217 & Nonspinning &  1/2 & 0 & 0 & [-546,-200]  & 0.101028  & 5.459474  & 0.115241  & -1.581007  & $6.42\times 10^{-7}$ & 0.14101  \\ \hline
        RIT:eBBH:1218 & Nonspinning &  1/2 & 0 & 0 & [-495,-200]  & 0.102984  & 5.505789  & 0.121395  & -1.349127  & $9.04\times 10^{-8}$ & 0.14963  \\ \hline
        RIT:eBBH:1219 & Nonspinning &  1/2 & 0 & 0 & [-445,-200]  & 0.105203  & 5.526842  & 0.128510  & -1.163278  & $9.81\times 10^{-8}$ & 0.15841  \\ \hline
        RIT:eBBH:1241 & Nonspinning &  3/4 & 0 & 0 & [-1221,-500]  & 0.082896  & 3.491053  & 0.005832  & 2.619953  & $4.84\times 10^{-9}$ & 0.07841  \\ \hline
        RIT:eBBH:1242 & Nonspinning &  3/4 & 0 & 0 & [-1202,-500]  & 0.083140  & 3.860526  & 0.005992  & 2.651699  & $5.51\times 10^{-9}$ & 0.07820  \\ \hline
        RIT:eBBH:1243 & Nonspinning &  3/4 & 0 & 0 & [-1183,-500]  & 0.083396  & 4.175263  & 0.006806  & 2.678351  & $5.66\times 10^{-9}$ & 0.07805  \\ \hline
        RIT:eBBH:1244 & Nonspinning &  3/4 & 0 & 0 & [-1164,-500]  & 0.083646  & 4.408947  & 0.008117  & 2.709243  & $6.25\times 10^{-9}$ & 0.07797  \\ \hline
        RIT:eBBH:1245 & Nonspinning &  3/4 & 0 & 0 & [-1146,-500]  & 0.083906  & 4.573158  & 0.009682  & 2.736100  & $6.08\times 10^{-9}$ & 0.07794  \\ \hline
        RIT:eBBH:1246 & Nonspinning &  3/4 & 0 & 0 & [-1127,-500]  & 0.084161  & 4.693158  & 0.011389  & 2.767391  & $6.17\times 10^{-9}$ & 0.07797  \\ \hline
        RIT:eBBH:1247 & Nonspinning &  3/4 & 0 & 0 & [-1132,-500]  & 0.084804  & 4.877368  & 0.015982  & 2.845872  & $6.04\times 10^{-9}$ & 0.07830  \\ \hline
        RIT:eBBH:1248 & Nonspinning &  3/4 & 0 & 0 & [-1088,-500]  & 0.085459  & 4.979474  & 0.020735  & 2.923306  & $6.27\times 10^{-9}$ & 0.07899  \\ \hline
        RIT:eBBH:1249 & Nonspinning &  3/4 & 0 & 0 & [-1004,-400]  & 0.086814  & 5.094211  & 0.030822  & 3.076327  & $9.91\times 10^{-9}$ & 0.08141  \\ \hline
        RIT:eBBH:1250 & Nonspinning &  3/4 & 0 & 0 & [-973,-400]  & 0.088168  & 5.170000  & 0.040442  & 3.257745  & $1.44\times 10^{-8}$ & 0.08510  \\ \hline
        RIT:eBBH:1251 & Nonspinning &  3/4 & 0 & 0 & [-944,-400]  & 0.089604  & 5.223684  & 0.050270  & 3.427651  & $2.62\times 10^{-8}$ & 0.08988  \\ \hline
        RIT:eBBH:1252 & Nonspinning &  3/4 & 0 & 0 & [-868,-400]  & 0.091069  & 5.267895  & 0.059978  & 3.610003  & $4.87\times 10^{-8}$ & 0.09557  \\ \hline
        RIT:eBBH:1253 & Nonspinning &  3/4 & 0 & 0 & [-809,-400]  & 0.092544  & 5.313158  & 0.069904  & -2.472680  & $6.54\times 10^{-8}$ & 0.10201  \\ \hline
        RIT:eBBH:1254 & Nonspinning &  3/4 & 0 & 0 & [-744,-300]  & 0.094190  & 5.336316  & 0.078536  & -2.311507  & $1.59\times 10^{-7}$ & 0.10906  \\ \hline
        RIT:eBBH:1255 & Nonspinning &  3/4 & 0 & 0 & [-678,-300]  & 0.095752  & 5.379474  & 0.088385  & -2.093602  & $1.85\times 10^{-7}$ & 0.11658  \\ \hline
        RIT:eBBH:1256 & Nonspinning &  3/4 & 0 & 0 & [-613,-300]  & 0.097391  & 5.432105  & 0.096533  & -1.860349  & $3.83\times 10^{-8}$ & 0.12449  \\ \hline
        RIT:eBBH:1257 & Nonspinning &  3/4 & 0 & 0 & [-564,-200]  & 0.099297  & 5.428947  & 0.105767  & -1.720883  & $5.43\times 10^{-7}$ & 0.13271  \\ \hline
        RIT:eBBH:1258 & Nonspinning &  3/4 & 0 & 0 & [-513,-200]  & 0.101127  & 5.481579  & 0.113113  & -1.490849  & $1.23\times 10^{-7}$ & 0.14117  \\ \hline
        RIT:eBBH:1259 & Nonspinning &  3/4 & 0 & 0 & [-465,-200]  & 0.103134  & 5.508947  & 0.120116  & -1.291997  & $9.38\times 10^{-8}$ & 0.14983  \\ \hline
        RIT:eBBH:1260 & Nonspinning &  3/4 & 0 & 0 & [-416,-200]  & 0.105414  & 5.531579  & 0.127444  & -1.114624  & $8.84\times 10^{-8}$ & 0.15863  \\ \hline
        RIT:eBBH:1282 & Nonspinning & 1 & 0 & 0 & [-11536,-1000]  & 0.048243  & 4.290850  & 0.197120  & -2.014567  & $1.17\times 10^{-7}$ & 0.19553  \\ \hline
        RIT:eBBH:1283 & Nonspinning & 1 & 0 & 0 & [-5696,-1000]  & 0.053019  & 4.353711  & 0.281010  & -2.315150  & $2.34\times 10^{-7}$ & 0.27984  \\ \hline
        RIT:eBBH:1284 & Nonspinning & 1 & 0 & 0 & [-3602,-1000]  & 0.056134  & 4.384316  & 0.327042  & -2.443524  & $5.19\times 10^{-7}$ & 0.32785  \\ \hline
        RIT:eBBH:1285 & Nonspinning & 1 & 0 & 0 & [-2690,-500]  & 0.058312  & 4.420732  & 0.359652  & -2.493538  & $9.86\times 10^{-7}$ & 0.35872  \\ \hline
        RIT:eBBH:1287 & Nonspinning & 1 & 0 & 0 & [-1829,-500]  & 0.060909  & 4.419580  & 0.385463  & -2.536189  & $1.87\times 10^{-6}$ & 0.38867  \\ \hline
        RIT:eBBH:1286 & Nonspinning & 1 & 0 & 0 & [-2102,-500]  & 0.061645  & 4.420937  & 0.389594  & -2.487635  & $2.00\times 10^{-6}$ & 0.39602  \\ \hline
        RIT:eBBH:1289 & Nonspinning & 1 & 0 & 0 & [-1913,-400]  & 0.062348  & 4.426132  & 0.396798  & -2.484868  & $2.52\times 10^{-6}$ & 0.40330  \\ \hline
        RIT:eBBH:1288 & Nonspinning & 1 & 0 & 0 & [-1765,-400]  & 0.063176  & 4.414717  & 0.401683  & -2.513374  & $3.36\times 10^{-6}$ & 0.41053  \\ \hline
        RIT:eBBH:1291 & Nonspinning & 1 & 0 & 0 & [-1595,-400]  & 0.063921  & 4.419815  & 0.407769  & -2.491983  & $3.72\times 10^{-6}$ & 0.41770  \\ \hline
        RIT:eBBH:1290 & Nonspinning & 1 & 0 & 0 & [-1452,-400]  & 0.064839  & 4.409618  & 0.410775  & -2.467450  & $4.15\times 10^{-6}$ & 0.42481  \\ \hline
        RIT:eBBH:1293 & Nonspinning & 1 & 0 & 0 & [-1330,-400]  & 0.065787  & 4.399422  & 0.415695  & -2.494619  & $5.66\times 10^{-6}$ & 0.43187  \\ \hline
        RIT:eBBH:1292 & Nonspinning & 1 & 0 & 0 & [-1183,-400]  & 0.066686  & 4.399422  & 0.419951  & -2.475669  & $6.06\times 10^{-6}$ & 0.43887  \\ \hline
        RIT:eBBH:1295 & Nonspinning & 1 & 0 & 0 & [-1065,-300]  & 0.067709  & 4.384127  & 0.427391  & -2.513742  & $8.73\times 10^{-6}$ & 0.44581  \\ \hline
        RIT:eBBH:1294 & Nonspinning & 1 & 0 & 0 & [-971,-300]  & 0.068842  & 4.373021  & 0.425059  & -2.365113  & $7.55\times 10^{-6}$ & 0.45269  \\ \hline
        RIT:eBBH:1297 & Nonspinning & 1 & 0 & 0 & [-859,-200]  & 0.070006  & 4.357820  & 0.431041  & -2.342054  & $1.13\times 10^{-5}$ & 0.45951  \\ \hline
        RIT:eBBH:1296 & Nonspinning & 1 & 0 & 0 & [-743,-200]  & 0.071370  & 4.336834  & 0.431908  & -2.344454  & $1.35\times 10^{-5}$ & 0.46628  \\ \hline
        RIT:eBBH:1330 & Nonspinning &   9/10 & 0 & 0 & [-11724,-1000]  & 0.048147  & 4.305433  & 0.194825  & -1.861335  & $5.40\times 10^{-8}$ & 0.19408  \\ \hline
        RIT:eBBH:1331 & Nonspinning &   9/10 & 0 & 0 & [-2694,-500]  & 0.058225  & 4.415449  & 0.357173  & -2.471125  & $9.11\times 10^{-7}$ & 0.35759  \\ \hline
        RIT:eBBH:1332 & Nonspinning &   9/10 & 0 & 0 & [-1354,-300]  & 0.065569  & 4.406231  & 0.416251  & -2.443202  & $5.15\times 10^{-6}$ & 0.43088  \\ \hline
        RIT:eBBH:1333 & Nonspinning &   9/10 & 0 & 0 & [-879,-200]  & 0.069805  & 4.361599  & 0.433026  & -2.397944  & $1.29\times 10^{-5}$ & 0.45859  \\ \hline
        RIT:eBBH:1353 & Nonspinning &  4/5 & 0 & 0 & [-11835,-1000]  & 0.048140  & 4.305299  & 0.195131  & -1.854330  & $5.32\times 10^{-8}$ & 0.19408  \\ \hline
        RIT:eBBH:1354 & Nonspinning &  4/5 & 0 & 0 & [-3218,-500]  & 0.058213  & 4.420283  & 0.357490  & -2.466165  & $9.09\times 10^{-7}$ & 0.35756  \\ \hline
        RIT:eBBH:1355 & Nonspinning &  4/5 & 0 & 0 & [-1363,-400]  & 0.065598  & 4.405299  & 0.415505  & -2.475990  & $5.45\times 10^{-6}$ & 0.43085  \\ \hline
        RIT:eBBH:1356 & Nonspinning &  4/5 & 0 & 0 & [-886,-300]  & 0.069857  & 4.363058  & 0.427429  & -2.353169  & $9.27\times 10^{-6}$ & 0.45855  \\ \hline
        RIT:eBBH:1376 & Nonspinning &   7/10 & 0 & 0 & [-12064,-1000]  & 0.048131  & 4.310106  & 0.194554  & -1.830161  & $5.28\times 10^{-8}$ & 0.19407  \\ \hline
        RIT:eBBH:1377 & Nonspinning &   7/10 & 0 & 0 & [-3270,-500]  & 0.058223  & 4.420454  & 0.357109  & -2.458103  & $8.97\times 10^{-7}$ & 0.35750  \\ \hline
        RIT:eBBH:1378 & Nonspinning &   7/10 & 0 & 0 & [-1381,-400]  & 0.065585  & 4.407776  & 0.415781  & -2.466174  & $5.39\times 10^{-6}$ & 0.43077  \\ \hline
        RIT:eBBH:1379 & Nonspinning &   7/10 & 0 & 0 & [-899,-300]  & 0.069874  & 4.363024  & 0.427431  & -2.343689  & $9.23\times 10^{-6}$ & 0.45847  \\ \hline
        RIT:eBBH:1399 & Nonspinning &  3/5 & 0 & 0 & [-12481,-1000]  & 0.048114  & 4.318550  & 0.193576  & -1.781341  & $5.14\times 10^{-8}$ & 0.19406  \\ \hline
        RIT:eBBH:1400 & Nonspinning &  3/5 & 0 & 0 & [-3369,-500]  & 0.058197  & 4.426576  & 0.357626  & -2.438717  & $9.05\times 10^{-7}$ & 0.35740  \\ \hline
        RIT:eBBH:1401 & Nonspinning &  3/5 & 0 & 0 & [-1410,-400]  & 0.065587  & 4.410445  & 0.415863  & -2.455433  & $5.31\times 10^{-6}$ & 0.43065  \\ \hline
        RIT:eBBH:1402 & Nonspinning &  3/5 & 0 & 0 & [-920,-300]  & 0.069849  & 4.368388  & 0.428583  & -2.337774  & $9.30\times 10^{-6}$ & 0.45834  \\ \hline
        RIT:eBBH:1422 & Nonspinning &  1/2 & 0 & 0 & [-12846,-1000]  & 0.048242  & 4.298953  & 0.198058  & -1.882699  & $6.29\times 10^{-8}$ & 0.19536  \\ \hline
        RIT:eBBH:1423 & Nonspinning &  1/2 & 0 & 0 & [-3487,-500]  & 0.058316  & 4.419950  & 0.358598  & -2.447685  & $9.50\times 10^{-7}$ & 0.35826  \\ \hline
        RIT:eBBH:1424 & Nonspinning &  1/2 & 0 & 0 & [-1439,-400]  & 0.065719  & 4.411384  & 0.416867  & -2.443388  & $5.49\times 10^{-6}$ & 0.43134  \\ \hline
        RIT:eBBH:1425 & Nonspinning &  1/2 & 0 & 0 & [-938,-300]  & 0.070032  & 4.368712  & 0.428990  & -2.304345  & $9.90\times 10^{-6}$ & 0.45898  \\ \hline
        RIT:eBBH:1445 & Nonspinning &  2/5 & 0 & 0 & [-14012,-1000]  & 0.048164  & 4.307285  & 0.197959  & -1.844830  & $5.75\times 10^{-8}$ & 0.19403  \\ \hline
        RIT:eBBH:1446 & Nonspinning &  2/5 & 0 & 0 & [-3772,-500]  & 0.058219  & 4.424085  & 0.358421  & -2.435551  & $9.45\times 10^{-7}$ & 0.35700  \\ \hline
        RIT:eBBH:1447 & Nonspinning &  2/5 & 0 & 0 & [-1537,-400]  & 0.065575  & 4.417081  & 0.417076  & -2.415085  & $5.33\times 10^{-6}$ & 0.43017  \\ \hline
        RIT:eBBH:1448 & Nonspinning &  2/5 & 0 & 0 & [-1002,-300]  & 0.069843  & 4.377519  & 0.431653  & -2.298384  & $1.14\times 10^{-5}$ & 0.45785  \\ \hline
        RIT:eBBH:1468 & Nonspinning &  1/3 & 0 & 0 & [-14885,-1000]  & 0.048270  & 4.293214  & 0.202685  & -1.947855  & $9.08\times 10^{-8}$ & 0.19516  \\ \hline
        RIT:eBBH:1469 & Nonspinning &  1/3 & 0 & 0 & [-4021,-500]  & 0.058240  & 4.434938  & 0.361682  & -2.457581  & $1.08\times 10^{-6}$ & 0.35768  \\ \hline
        RIT:eBBH:1470 & Nonspinning &  1/3 & 0 & 0 & [-1627,-400]  & 0.065581  & 4.424968  & 0.424166  & -2.521533  & $7.66\times 10^{-6}$ & 0.43069  \\ \hline
        RIT:eBBH:1471 & Nonspinning &  1/3 & 0 & 0 & [-1038,-300]  & 0.070000  & 4.379848  & 0.435683  & -2.335837  & $1.47\times 10^{-5}$ & 0.45831  \\ \hline
        RIT:eBBH:1491 & Nonspinning &  1/4 & 0 & 0 & [-17606,-1000]  & 0.048185  & 4.327294  & 0.197261  & -1.714333  & $4.89\times 10^{-8}$ & 0.19398  \\ \hline
        RIT:eBBH:1492 & Nonspinning &  1/4 & 0 & 0 & [-4650,-500]  & 0.058175  & 4.436953  & 0.361381  & -2.425922  & $1.03\times 10^{-6}$ & 0.35642  \\ \hline
        RIT:eBBH:1493 & Nonspinning &  1/4 & 0 & 0 & [-1874,-400]  & 0.065422  & 4.433595  & 0.424074  & -2.422223  & $7.31\times 10^{-6}$ & 0.42948  \\ \hline
        RIT:eBBH:1494 & Nonspinning &  1/4 & 0 & 0 & [-1141,-300]  & 0.069895  & 4.388408  & 0.443018  & -2.502784  & $1.99\times 10^{-5}$ & 0.45714  \\ \hline
        RIT:eBBH:1740 & Spin-aligned & 1 & -0.5 & -0.5 & [-9966,-1000]  & 0.048861  & 4.243120  & 0.201827  & -2.789773  & $1.95\times 10^{-7}$ & 0.20702  \\ \hline
        RIT:eBBH:1741 & Spin-aligned & 1 & -0.5 & -0.5 & [-2159,-500]  & 0.061191  & 4.356750  & 0.363587  & -2.906665  & $1.64\times 10^{-6}$ & 0.37058  \\ \hline
        RIT:eBBH:1763 & Spin-aligned & 1 & -0.8 & -0.8 & [-8926,-1000]  & 0.049339  & 4.225607  & 0.205185  & -3.202682  & $3.56\times 10^{-7}$ & 0.22918  \\ \hline
        RIT:eBBH:1764 & Spin-aligned & 1 & -0.8 & -0.8 & [-1493,-400]  & 0.064034  & 4.309648  & 0.365244  & -2.991234  & $2.49\times 10^{-6}$ & 0.38297  \\ \hline
        RIT:eBBH:1786 & Spin-aligned & 1 & 0.5 & 0.5 & [-2320,-400]  & 0.061876  & 4.474999  & 0.411076  & -1.991020  & $5.16\times 10^{-6}$ & 0.42191  \\ \hline
        RIT:eBBH:1787 & Spin-aligned & 1 & 0.5 & 0.5 & [-1739,-400]  & 0.064529  & 4.466832  & 0.431725  & -2.085661  & $1.04\times 10^{-5}$ & 0.44865  \\ \hline
        RIT:eBBH:1788 & Spin-aligned & 1 & 0.5 & 0.5 & [-1091,-400]  & 0.069618  & 4.427534  & 0.446566  & -2.095252  & $1.89\times 10^{-5}$ & 0.48712  \\ \hline
        RIT:eBBH:1807 & Spin-aligned & 1 & 0.8 & 0.8 & [-2854,-400]  & 0.060530  & 4.499562  & 0.404775  & -1.604014  & $6.69\times 10^{-6}$ & 0.41899  \\ \hline
        RIT:eBBH:1808 & Spin-aligned & 1 & 0.8 & 0.8 & [-2221,-400]  & 0.062669  & 4.507297  & 0.428760  & -1.757701  & $9.56\times 10^{-6}$ & 0.44437  \\ \hline
        RIT:eBBH:1809 & Spin-aligned & 1 & 0.8 & 0.8 & [-1487,-400]  & 0.066603  & 4.487266  & 0.455576  & -1.942250  & $2.26\times 10^{-5}$ & 0.48103  \\ \hline
        RIT:eBBH:1810 & Spin-aligned & 1 & 0.8 & 0.8 & [-967,-300]  & 0.072278  & 4.415135  & 0.455850  & -1.814238  & $4.39\times 10^{-5}$ & 0.51587  \\ \hline
        RIT:eBBH:1811 & Spin-aligned & 1 & 0.8 & 0.8 & [-710,-200]  & 0.077459  & 4.305993  & 0.454949  & -1.786925  & $9.04\times 10^{-5}$ & 0.53805  \\ \hline
        RIT:eBBH:1828 & Spin-aligned & 1 & 0 & 0.8 & [-2146,-400]  & 0.062397  & 4.467975  & 0.412924  & -2.125679  & $4.96\times 10^{-6}$ & 0.42570  \\ \hline
        RIT:eBBH:1829 & Spin-aligned & 1 & 0 & 0.8 & [-1592,-400]  & 0.065208  & 4.458413  & 0.434177  & -2.202726  & $9.50\times 10^{-6}$ & 0.45243  \\ \hline
        RIT:eBBH:1830 & Spin-aligned & 1 & 0 & 0.8 & [-961,-300]  & 0.071002  & 4.391649  & 0.444828  & -2.092890  & $2.10\times 10^{-5}$ & 0.49102  \\ \hline
        RIT:eBBH:1862 & Spin-aligned &  1/3 & 0 & -0.8 & [-2081,-400]  & 0.062686  & 4.335796  & 0.371205  & -3.068030  & $3.60\times 10^{-6}$ & 0.37980  \\ \hline
        RIT:eBBH:1883 & Spin-aligned &  1/2 & 0 & -0.8 & [-2130,-400]  & 0.061833  & 4.344211  & 0.366412  & -2.335113  & $2.29\times 10^{-6}$ & 0.35251  \\ \hline
        RIT:eBBH:1899 & Spin-aligned & 1 & 0 & -0.8 & [-10460,-1000]  & 0.048638  & 4.281648  & 0.197918  & -2.489198  & $1.05\times 10^{-7}$ & 0.21209  \\ \hline
        RIT:eBBH:1900 & Spin-aligned & 1 & 0 & -0.8 & [-2356,-400]  & 0.060398  & 4.372612  & 0.364363  & -2.894547  & $1.55\times 10^{-6}$ & 0.37087  \\ \hline
        RIT:eBBH:1901 & Spin-aligned & 1 & 0 & -0.8 & [-813,-200]  & 0.071933  & 4.268929  & 0.408480  & -2.712869  & $4.74\times 10^{-6}$ & 0.44535  \\ \hline
        SXS:BBH:1355 & Nonspinning & 1 & 0 & 0 & [-3071,-300] & 0.068757  & 1.994624  & 0.066123  & 0.951318  & $6.80\times 10^{-8}$ & 0.06123  \\ \hline
        SXS:BBH:1356 & Nonspinning & 1 & 0 & 0 & [-6583,-300] & 0.056819  & 5.976205  & 0.119648  & -0.254013  & $5.36\times 10^{-8}$ & 0.14171  \\ \hline
        SXS:BBH:1357 & Nonspinning & 1 & 0 & 0 & [-3420,-300] & 0.066068  & 6.556144  & 0.128740  & 1.853997  & $1.07\times 10^{-7}$ & 0.12760  \\ \hline
        SXS:BBH:1358 & Nonspinning & 1 & 0 & 0 & [-3181,-300] & 0.067207  & 6.635232  & 0.127084  & 2.096370  & $9.19\times 10^{-8}$ & 0.12789  \\ \hline
        SXS:BBH:1359 & Nonspinning & 1 & 0 & 0 & [-3053,-300] & 0.067894  & 0.387847  & 0.125970  & 2.205003  & $9.35\times 10^{-8}$ & 0.15151  \\ \hline
        SXS:BBH:1360 & Nonspinning & 1 & 0 & 0 & [-2896,-300] & 0.067348  & 0.360219  & 0.178906  & 2.633178  & $1.80\times 10^{-7}$ & 0.19791  \\ \hline
        SXS:BBH:1361 & Nonspinning & 1 & 0 & 0 & [-2849,-300] & 0.067572  & 0.358490  & 0.180128  & 2.660157  & $2.41\times 10^{-7}$ & 0.17654  \\ \hline
        SXS:BBH:1362 & Nonspinning & 1 & 0 & 0 & [-2673,-300] & 0.066603  & 0.393420  & 0.238166  & 3.106062  & $4.35\times 10^{-7}$ & 0.24186  \\ \hline
        SXS:BBH:1363 & Nonspinning & 1 & 0 & 0 & [-2635,-300] & 0.066767  & 0.396938  & 0.238877  & 3.163563  & $9.74\times 10^{-7}$ & 0.24513  \\ \hline
        SXS:BBH:1364 & Nonspinning &  1/2 & 0 & 0 & [-3727,-300] & 0.067400  & 0.669219  & 0.056787  & 1.332571  & $5.92\times 10^{-8}$ & 0.05986  \\ \hline
        SXS:BBH:1365 & Nonspinning &  1/2 & 0 & 0 & [-3709,-300] & 0.067254  & 0.418394  & 0.075791  & 1.519396  & $9.08\times 10^{-8}$ & 0.06439  \\ \hline
        SXS:BBH:1366 & Nonspinning &  1/2 & 0 & 0 & [-3602,-300] & 0.066900  & 0.238871  & 0.124168  & 1.871625  & $1.16\times 10^{-7}$ & 0.13724  \\ \hline
        SXS:BBH:1367 & Nonspinning &  1/2 & 0 & 0 & [-3481,-300] & 0.067475  & 0.264768  & 0.122686  & 1.948643  & $1.16\times 10^{-7}$ & 0.11288  \\ \hline
        SXS:BBH:1368 & Nonspinning &  1/2 & 0 & 0 & [-3374,-300] & 0.067988  & 0.298730  & 0.121591  & 2.051288  & $1.43\times 10^{-7}$ & 0.13330  \\ \hline
        SXS:BBH:1369 & Nonspinning &  1/2 & 0 & 0 & [-3150,-300] & 0.065673  & 0.231351  & 0.237012  & 2.555949  & $5.47\times 10^{-7}$ & 0.24024  \\ \hline
        SXS:BBH:1370 & Nonspinning &  1/2 & 0 & 0 & [-2903,-300] & 0.067051  & 0.311412  & 0.230810  & 2.925016  & $6.08\times 10^{-7}$ & 0.24001  \\ \hline
        SXS:BBH:1371 & Nonspinning &  1/3 & 0 & 0 & [-4236,-300] & 0.067573  & 0.295354  & 0.070978  & 1.375396  & $1.02\times 10^{-7}$ & 0.10062  \\ \hline
        SXS:BBH:1372 & Nonspinning &  1/3 & 0 & 0 & [-4094,-300] & 0.067292  & 0.126380  & 0.119649  & 1.731706  & $2.12\times 10^{-7}$ & 0.11868  \\ \hline
        SXS:BBH:1373 & Nonspinning &  1/3 & 0 & 0 & [-3979,-300] & 0.067743  & 0.155603  & 0.118981  & 1.809612  & $1.86\times 10^{-7}$ & 0.13512  \\ \hline
        SXS:BBH:1374 & Nonspinning &  1/3 & 0 & 0 & [-3547,-300] & 0.066333  & 0.138994  & 0.230359  & 2.324821  & $7.63\times 10^{-7}$ & 0.24115 \\ \hline
\end{longtable*}
\end{widetext}
\end{document}